\documentclass[12pt]{article}
\usepackage{a4wide}
\usepackage{latexsym}
\usepackage{amsmath}
\usepackage{amsfonts}
\usepackage{amscd}
\usepackage{cite}
\usepackage{graphicx}
\usepackage{axodraw}
\usepackage{float}

\usepackage{pslatex}
\usepackage[latin1,utf8]{inputenc}
\usepackage[OT2,T1]{fontenc}

\newcommand{\bq}{\begin{eqnarray}}
\newcommand{\eq}{\end{eqnarray}}
\newcommand{\eps}{\varepsilon}

\begin{document}

\thispagestyle{empty}

\begin{flushright}
  MITP/16-048
\end{flushright}

\vspace{1.5cm}

\begin{center}
  {\Large\bf Numerical integration of subtraction terms\\
  }
  \vspace{1cm}
  {\large Satyajit Seth and Stefan Weinzierl\\
\vspace{2mm}
      {\small \em PRISMA Cluster of Excellence, Institut f{\"u}r Physik, }\\
      {\small \em Johannes Gutenberg-Universit{\"a}t Mainz,}\\
      {\small \em D - 55099 Mainz, Germany}\\
  } 
\end{center}

\vspace{2cm}

\begin{abstract}\noindent
  {
Numerical approaches to higher-order calculations often employ subtraction terms, both for the real emission
and the virtual corrections. These subtraction terms have to be added back.
In this paper we show that at NLO 
the real subtraction terms, 
the virtual subtraction terms, 
the integral representations of the field renormalisation constants 
and -- in the case of initial-state partons -- the integral representation for the collinear counterterm 
can be grouped together to give finite
integrals, which can be evaluated numerically.
This is useful for an extension towards NNLO.
   }
\end{abstract}

\vspace*{\fill}

\newpage

\section{Introduction}
\label{sect:intro}

Numerical methods are a promising path to higher-order corrections. 
The higher-order corrections are required for precision calculations in high energy physics.
Within the numerical approach one subtracts suitable approximation terms from the real emission contribution and the virtual contribution.
The subtraction terms for the real emission contribution at next-to-leading order (NLO) are well 
established \cite{Kunszt:1994mc,Frixione:1996ms,Catani:1997vz,Catani:1997vzerr,Dittmaier:1999mb,Phaf:2001gc,Catani:2002hc,Weinzierl:2005dd,Frederix:2009yq,Frixione:2011kh,Kosower:1998zr,Campbell:1998nn,Gehrmann-DeRidder:2005cm,Daleo:2006xa,Somogyi:2006cz,Nagy:2007ty,Chung:2010fx,Dittmaier:2008md,Czakon:2009ss,Gotz:2012zz}.
More recently, it has become possible to use the subtraction method for the virtual part at NLO
as well \cite{Nagy:2003qn,Gong:2008ww,Assadsolimani:2009cz,Assadsolimani:2010ka,Becker:2010ng,Becker:2011vg,Becker:2012aq,Becker:2012nk,Becker:2012bi,Goetz:2014lla}.
The subtracted real emission contribution and the subtracted virtual contribution can be evaluated separately by numerical methods.
The subtracted approximation terms have to be added back and give a finite contribution to the final result.
Due to the universality of the singular limits, the approximation terms can be chosen as a sum of process-independent building blocks.
When adding the approximation terms back, the integrations over the virtual loop momentum (for the virtual approximation terms)
and the unresolved phase space (for the real approximation terms) are independent of the process-dependent kinematics.
At NLO the corresponding integrals are rather simple and the integration over the loop momentum/unresolved phase space can
be performed analytically once and for all.

The situation changes at NNLO: An analytic integration of local subtraction terms is 
highly non-trivial \cite{Somogyi:2005xz,Somogyi:2006da,Somogyi:2006db,Aglietti:2008fe,Somogyi:2008fc,Bolzoni:2010bt,DelDuca:2016csb}.
It is therefore a natural question to ask, if the integration over the loop momentum/unresolved phase space can be done
numerically. 
The computational costs for the numerical integration of the subtraction terms will be small against the costs for the numerical
integration of the subtracted real emission contribution or the subtracted virtual contribution.
In this paper we will study the issue at NLO.
Let us stress that our motivation is to lay the foundations for an extension towards NNLO.
If analytically integrated results for the approximation terms are available (as they are for NLO) it is more efficient 
to use these in actual NLO computations.
Our focus is therefore more on the principles and practicalities of the cancellations of singularities.
We will soon see that due to some subtleties it is worth the effort to study these issues at NLO.

Taken separately, the integrations over the virtual approximation terms and the real approximations terms are divergent in four-dimensional space-time.
When manipulating divergent integrals, we will always use dimensional regularisation with $D=4-2\eps$ space-time dimensions.
Our final expressions will be finite and the limit $\eps \rightarrow 0$ can be taken safely.
Integrating the approximation terms numerically will therefore require 
a map between the $D$-dimensional loop momentum space and the $(D-1)$-dimensional unresolved
phase space.
The loop-tree duality method \cite{Catani:2008xa,Bierenbaum:2012th,Buchta:2014dfa,Hernandez-Pinto:2015ysa,Buchta:2015wna,Sborlini:2016gbr}
provides a technique to handle this situation.

In the past there have been attempts to combine directly the virtual corrections 
with the real corrections \cite{Soper:1998ye,Soper:1999xk,Soper:2001hu,Kramer:2002cd}.
This has the disadvantage that one deals at all stages with kinematics of an $2 \rightarrow n$ process.
Our approach first subtracts one set of approximation terms from the virtual corrections
and a different set from the real emission.
We only combine the virtual approximation terms with the real approximation terms.
The approximation terms have a much simpler kinematic structure. 
At NLO this limits us to one-loop three-point functions (in the virtual case) and three external momenta (in the real emission case),
independently of the number of hard particles in the scattering process.

Let us now discuss the subtleties of combining the virtual approximation terms with the real approximation terms.
Our main interest is higher-order corrections in QCD. Therefore we deal with massless gauge bosons and massless or massive
fermions.
Now let us consider a collinear singularity from the real emission contribution. 
The two collinear particles will have transverse polarisations.
On the other hand, for a collinear singularity in the virtual part one of the involved particles will have
a longitudinal polarisation.
These two pieces will not match.
A second manifestation of the same problem is obtained by considering the $g \rightarrow q \bar{q}$ splitting.
In the collinear limit this gives a singular contribution in the real emission part, 
however the corresponding limit in the virtual part is finite.
The solution to both problems is to take the field renormalisation constants into account 
in the form of un-integrated expressions.
For massless fields, the $\alpha_s$-contributions to the field renormalisation constants are zero, however this zero
comes from a cancellation between ultraviolet and infrared regions.
Effectively, the field renormalisation constants reshuffle ultraviolet with infrared transverse/longitudinal
singularities and are needed for a local cancellation of singularities at the integrand level.
We will explain these mechanisms in detail.

If initial-state partons are present a further subtlety arises:
The region for the collinear singularity from the virtual part does not match with the region for
the collinear singularity from the real part.
The solution comes in the form of the collinear counterterm, which has to be included.
In integrated form this counterterm has to parts: An $x$-dependent piece, leading to a convolution in $x$,
and an end-point contribution, proportional to $\delta(1-x)$.
We derive an integral representation for both parts, such that 
on the one hand the integrand corresponding to the convolution part
combines with the real part 
and on the other hand the integrand corresponding to the end-point contribution combines with the virtual part.
In this way we achieve a local cancellation of singularities.

This paper is organised as follows:
In section~\ref{sect:setup} we introduce the setup and the notation and review known results.
Sections~\ref{sect:real_subtraction_terms}-\ref{sect:factorisation} give the integral representations
of all required ingredients:
We start in section~\ref{sect:real_subtraction_terms} with the real approximation terms,
followed by the virtual subtraction terms in section~\ref{sect:virtual_subtraction_terms}.
Section~\ref{sect:renormalisation} is devoted to the integral representation of the renormalisation
constants.
Section~\ref{sect:factorisation} discusses the collinear counterterm for initial-state partons.
Having defined all ingredients, we show in section~\ref{sect:cancellations} that the ingredients
can be grouped together to give locally integrable expressions.
However, local integrability does not mean that all contributions can be integrated along the real axes.
In the virtual approximation terms there can be thresholds, which are avoided by a deformation into the complex
plane. Section~\ref{sect:contour_deformation} discusses therefore contour deformation.
Finally, our conclusions are given in section~\ref{sect:conclusions}.
Various technical details are collected in the appendix.

\section{Notation and review of known results}
\label{sect:setup}

\subsection{Setup}

Let us consider a $2 \rightarrow n$ process.
The contributions at leading and next-to-leading order are written in a condensed notation as
\bq
\langle O \rangle^{\mathrm{LO}} =  
 \int\limits_n O_n d\sigma^{\mathrm{B}},
 & &
\langle O \rangle^{\mathrm{NLO}} =  
 \int\limits_{n+1} O_{n+1} d\sigma^{\mathrm{R}} + \int\limits_{n + \mathrm{loop}} O_n d\sigma^{\mathrm{V}} 
 + \int\limits_n O_n d\sigma^{\mathrm{C}}.
\eq
Here, $d\sigma^{\mathrm{B}}$ denotes the Born contribution,
whose matrix elements are given by the square of the Born amplitudes with $(n+2)$ partons
$| {\mathcal A}^{(0)}_{n+2} |^2$, summed over spins and colours.
Similarly, $d\sigma^{\mathrm{R}}$ denotes the real emission contribution,
whose matrix elements are given by the square of the Born amplitudes with $(n+3)$ partons
$| {\mathcal A}^{(0)}_{n+3} |^2$.
The term $d\sigma^{\mathrm{V}}$ gives the virtual contribution, whose matrix elements are given by the interference term
of the renormalised one-loop amplitude ${\mathcal A}^{(1)}_{n+2}$, with $(n+2)$ partons, with the corresponding
Born amplitude ${\mathcal A}^{(0)}_{n+2}$.
The renormalised one-loop amplitude is given as the sum of the bare one-loop amplitude and the ultraviolet counterterm.
We write
\bq
 d\sigma^{\mathrm{V}} 
 & = & 
 d\sigma^{\mathrm{V}}_{\mathrm{bare}} + d\sigma^{\mathrm{V}}_{\mathrm{CT}}.
\eq
Finally, $d\sigma^{\mathrm{C}}$ denotes a collinear counterterm, which subtracts the initial state collinear
singularities.
Taken separately, the individual contributions at next-to-leading order 
are divergent and only their sum is finite.
Within the numerical approach, one adds and subtracts suitably chosen pieces 
to be able to perform the phase space integrations and the loop integration by Monte Carlo methods:
\bq
 \langle O \rangle^{\mathrm{NLO}} 
 & = &
 \int\limits_{n+1} \left( O_{n+1} d\sigma^{\mathrm{R}} - O_{n} d\sigma^{\mathrm{A}}_{\mathrm{R}} \right)
 +
 \int\limits_{n+\mathrm{loop}} \left( O_{n} d\sigma_{\mathrm{bare}}^{\mathrm{V}} - O_{n} d\sigma^{\mathrm{A}}_{\mathrm{V}} \right) 
 \nonumber \\
 & &
 +
 \int\limits_{n} 
 \left[ 
        O_{n} d\sigma^{\mathrm{C}} 
      + O_{n} \int\limits_{1} d\sigma^{\mathrm{A}}_{\mathrm{R}} 
      + O_{n} \int\limits_{\mathrm{loop}} d\sigma^{\mathrm{A}}_{\mathrm{V}} 
      + O_{n} d\sigma_{\mathrm{CT}}^{\mathrm{V}} 
 \right].
\eq
The approximation term for the real emission part is denoted by $d\sigma^{\mathrm{A}}_{\mathrm{R}}$, the approximation term
for the virtual part by $d\sigma^{\mathrm{A}}_{\mathrm{V}}$.
By construction, the expressions
\bq
 \int\limits_{n+1} \left( O_{n+1} d\sigma^{\mathrm{R}} - O_{n} d\sigma^{\mathrm{A}}_{\mathrm{R}} \right)
 & \mbox{and} &
 \int\limits_{n+\mathrm{loop}} \left( O_{n} d\sigma_{\mathrm{bare}}^{\mathrm{V}} - O_{n} d\sigma^{\mathrm{A}}_{\mathrm{V}} \right) 
\eq
are numerically integrable.
In this paper we are interested in the third term
\bq
\label{integrated_subtraction_terms}
\langle O \rangle^{\mathrm{NLO}}_{{\bf I}+{\bf L}} 
 & = & 
 \int\limits_{n} 
 \left[ 
        O_{n} d\sigma^{\mathrm{C}} 
      + O_{n} \int\limits_{1} d\sigma^{\mathrm{A}}_{\mathrm{R}} 
      + O_{n} \int\limits_{\mathrm{loop}} d\sigma^{\mathrm{A}}_{\mathrm{V}} 
      + O_{n} d\sigma_{\mathrm{CT}}^{\mathrm{V}} 
 \right].
\eq
In particular we show that this term can be integrated numerically as well.
We will separate this term into an ultraviolet part and an infrared part. 
The numerical integration of the former part is un-problematic and our focus lies on the numerical integration of the latter part.
As already indicated by the notation, the integration over the phase space of $n$ hard particles will be common to all
terms in eq.~(\ref{integrated_subtraction_terms}).
However, $d\sigma^{\mathrm{A}}_{\mathrm{V}}$ involves an integration over the $D$-dimensional loop momentum space, whereas $d\sigma^{\mathrm{A}}_{\mathrm{R}}$
involves an extra integration over the $(D-1)$-dimensional unresolved
phase space.
As these two terms are individually divergent, this requires a mapping between the loop momentum space and the unresolved phase space, 
such that non-integrable singularities cancel locally in the combination.

Let us now go into more details:
We denote the phase space measure for $n$ final-state particles by
\bq
 d\phi_n(p_a+p_b \rightarrow p_1,...,p_n)
 = 
 (2 \pi)^D \delta^D\left( p_a + p_b - \sum\limits_{i=1}^n p_i \right)
 \;
 \prod\limits_{i=1}^n \frac{d^Dp_i}{(2\pi)^{D-1}} \theta(p_i^0) \delta(p_i^2-m_i^2).
\eq
We have
\bq
 d\sigma^{\mathrm{B}}
 & = &
 \left| {\mathcal A}^{(0)}_{n+2} \right|^2 
 d\phi_n.
\eq
In order to keep the notation simple, we use the convention that the integral symbol includes
the flux factor, the averaging factors for the spin and colour degrees of freedom
of the initial-state particles, the symmetry factor for final-state particles
and (in hadronic collisions) the integration over the parton distribution functions.
With this convention we have for example for hadronic collisions
\bq
 \int\limits_n O_n d\sigma^{\mathrm{B}}
 & = &
 \sum\limits_{a,b}
 \int dx_1 f_a(x_1) \int dx_2 f_b(x_2) 
             \frac{1}{2 \hat{s} n_s(1) n_s(2) n_c(1) n_c(2)}
 \frac{1}{S}
             \int d\phi_{n}
             O_n
             \left| {\mathcal A}^{(0)}_{n+2} \right|^2.
\eq
The symmetry factor $S$ is given by a product of factors $(n_j!)$, where $n_j$ denotes the number 
of identical particles of type $j$ in the final state.
The number of colour degrees of freedom of a particle $a$ is denoted by $n_c(a)$. We have
\bq
 n_c(q) \;\ = \; n_c(\bar{q}) \; = \; 3,
 & &
 n_c(g) \; = \; 8.
\eq
The number of spin degrees of freedom of a particle $a$ is denoted by $n_s(a)$. 
In $D=4-2\eps$ space-time dimensions we have within conventional dimensional regularisation
\bq
 n_s(q) \;\ = \; n_s(\bar{q}) \; = \; 2,
 & &
 n_s(g) \; = \; D-2.
\eq
As long as we are dealing with finite quantities we may take the limit $D\rightarrow 4$, yielding two spin degrees
of freedom for a gluon in four space-time dimensions.
We may write the phase space measure for the real emission part as
\bq
 d\phi_{n+1} & = &
 d\phi_n \; d\phi_{\mathrm{unresolved}}.
\eq
There is some freedom in defining the real approximation terms. 
In this paper we consider for concreteness
dipole subtraction terms \cite{Catani:1997vz,Catani:1997vzerr,Dittmaier:1999mb,Phaf:2001gc,Catani:2002hc,Dittmaier:2008md,Czakon:2009ss,Gotz:2012zz}, although our results can easily be translated to all other local real subtraction schemes.
In this case, $d\sigma^{\mathrm{A}}_{\mathrm{R}}$ is given as a sum over dipoles:
\bq
\lefteqn{
 d\sigma^{\mathrm{A}}_{\mathrm{R}}
 = } & & 
 \\
 & &
 \left(
 \sum\limits_{(i',j')} \sum\limits_{k' \neq i',j'} {\mathcal D}_{i'j',k'}
 +
 \sum\limits_{(i',j')} \sum\limits_{a'} {\mathcal D}_{i'j'}^{a'}
 +
 \sum\limits_{(a',j')} \sum\limits_{k' \neq j'} {\mathcal D}^{a'j'}_{k'}
 +
 \sum\limits_{(a',j')} \sum\limits_{b' \neq a'} {\mathcal D}^{a'j',b'}
 \right) d\phi_n \; d\phi_{\mathrm{unresolved}}.
 \nonumber
\eq
In this paper we use the convention that particles corresponding to a real emission event are denoted
with primes.
The requirement of local subtraction terms implies that in general the dipole subtraction terms are matrices
in spin and colour space.
This is due to the fact that in the factorisation of the matrix elements squared spin correlations survive in the collinear limit,
while colour correlations survive in the soft limit.
At NLO, the integration over the
unresolved one-particle phase space is easily performed analytically in $(D-1)$ dimensions.
In a compact notation the result of this integration is often written as
\bq
 d\sigma^{\mathrm{C}} + \int\limits_1 d\sigma^{\mathrm{A}}_{\mathrm{R}}  
 & = & {\bf I} \otimes d\sigma^{\mathrm{B}} + {\bf K} \otimes d\sigma^{\mathrm{B}} + {\bf P} \otimes d\sigma^{\mathrm{B}}.
\eq
After integration all spin-correlations average out, but colour correlations still remain, indicated by the 
notation $\otimes$.
The terms with the insertion operators ${\bf K}$ and ${\bf P}$ do not have any poles in the dimensional
regularisation parameter $\eps$.
All explicit poles in the dimensional regularisation parameter are contained in the term
${\bf I} \otimes d\sigma^{\mathrm{B}}$.

Let us now turn our attention to the virtual part.
$d\sigma^{\mathrm{V}}$ is given by
\bq
 d\sigma^{\mathrm{V}} & = & 
 2 \; \mbox{Re}\; \left(\left.{\mathcal A}^{(0)}\right.^\ast {\mathcal A}^{(1)} \right) d\phi_n.
\eq
${\mathcal A}^{(1)}$ denotes the renormalised one-loop amplitude. It is related to the bare amplitude by
\bq
\label{eq_one_loop}
 {\mathcal A}^{(1)} & = & {\mathcal A}^{(1)}_{\mathrm{bare}} + {\mathcal A}^{(1)}_{\mathrm{CT}}.
\eq
${\mathcal A}^{(1)}_{\mathrm{CT}}$ denotes the ultraviolet counterterm from renormalisation.
The bare one-loop amplitude involves the loop integration
\bq
\label{integrand_one_loop}
{\mathcal A}^{(1)}_{\mathrm{bare}} & = & \int \frac{d^Dk}{(2\pi)^D} {\mathcal G}^{(1)}_{\mathrm{bare}},
\eq
where ${\mathcal G}^{(1)}_{\mathrm{bare}}$ denotes the integrand of the bare one-loop amplitude.
Within the numerical approach also the one-loop amplitude ${\mathcal A}^{(1)}$ can be calculated numerically.
In order to avoid singularities in the integrand, the subtraction method is used again:
\bq
\label{basic_subtraction_loop}
 {\mathcal A}_{\mathrm{bare}}^{(1)} + {\mathcal A}_{\mathrm{CT}}^{(1)} 
 & = & 
 \left( {\mathcal A}_{\mathrm{bare}}^{(1)} - {\mathcal A}_{\mathrm{soft}}^{(1)} - {\mathcal A}_{\mathrm{coll}}^{(1)} - {\mathcal A}_{\mathrm{UV}}^{(1)} \right)
 + \left( {\mathcal A}_{\mathrm{CT}}^{(1)}  
 + {\mathcal A}_{\mathrm{soft}}^{(1)} + {\mathcal A}_{\mathrm{coll}}^{(1)} + {\mathcal A}_{\mathrm{UV}}^{(1)} \right).
\eq
The subtraction terms ${\mathcal A}_{\mathrm{soft}}^{(1)}$, ${\mathcal A}_{\mathrm{coll}}^{(1)}$ and ${\mathcal A}_{\mathrm{UV}}^{(1)}$
are chosen such that they match locally the singular behaviour of the integrand of ${\mathcal A}_{\mathrm{bare}}^{(1)}$ in $D$ dimensions.
The term ${\mathcal A}_{\mathrm{soft}}^{(1)}$ approximates the soft singularities, 
${\mathcal A}_{\mathrm{coll}}^{(1)}$ approximates
the collinear singularities and 
the term ${\mathcal A}_{\mathrm{UV}}^{(1)}$ approximates the ultraviolet singularities.
These subtraction terms have a local form similar to eq.~(\ref{integrand_one_loop}):
\bq
{\mathcal A}^{(1)}_{\mathrm{soft}} = \int \frac{d^Dk}{(2\pi)^D} {\mathcal G}^{(1)}_{\mathrm{soft}},
\;\;\;\;\;\;
{\mathcal A}^{(1)}_{\mathrm{coll}} = \int \frac{d^Dk}{(2\pi)^D} {\mathcal G}^{(1)}_{\mathrm{coll}},
\;\;\;\;\;\;
{\mathcal A}^{(1)}_{\mathrm{UV}} = \int \frac{d^Dk}{(2\pi)^D} {\mathcal G}^{(1)}_{\mathrm{UV}}.
\eq
Again, there is some freedom in defining these approximation terms. We use the approximation terms
given in \cite{Assadsolimani:2009cz,Assadsolimani:2010ka,Becker:2010ng,Becker:2011vg,Becker:2012aq}.
The approximation term $d\sigma^{\mathrm{A}}_{\mathrm{V}}$ is given by
\bq
 d\sigma^{\mathrm{A}}_{\mathrm{V}}
 & = &
 \frac{d^Dk}{(2\pi)^D} 
 \; 2 \; \mbox{Re}\; \left[ \left.{\mathcal A}^{(0)}\right.^\ast \left( {\mathcal G}^{(1)}_{\mathrm{soft}} + {\mathcal G}^{(1)}_{\mathrm{coll}} + {\mathcal G}^{(1)}_{\mathrm{UV}} \right) \right] d\phi_n.
\eq
At NLO the loop integration for the approximation term $d\sigma^{\mathrm{A}}_{\mathrm{V}}$ is easily performed analytically in $D$ dimensions.
One obtains
\bq
 d\sigma^{\mathrm{V}}_{\mathrm{CT}}
 +
 \int\limits_{\mathrm{loop}} d\sigma^{\mathrm{A}}_{\mathrm{V}}
 & = &
 {\bf L} \otimes d\sigma^{\mathrm{B}}.
\eq
The operator ${\bf L}$ contains, as does the operator ${\bf I}$, colour correlations due to soft gluons.
In addition, the insertion operator ${\bf L}$ contains explicit poles in the dimensional regularisation parameter 
$\eps$ related to the infrared singularities of the one-loop amplitude.
These poles cancel when combined with the insertion operator ${\bf I}$:
\bq
\label{I_plus_L}
 \left( {\bf I} + {\bf L} \right) \otimes d\sigma^{\mathrm{B}} & = &
 \mbox{finite}.
\eq
Eq.~(\ref{I_plus_L}) is a statement on the cancellation of singularities after
the integration over the unresolved phase space and the loop momentum space, respectively.
In this paper we would like to achieve a cancellation of singularities before these integrations.

\subsection{Colour}

The amplitudes are vectors in colour space. It is convenient to define colour charge operators acting on the colour indices
of the amplitudes as follows:
The colour charge operators ${\bf T}_i$ for the emission of a gluon from a quark, gluon or antiquark in the final state are defined by
\bq
\label{colour_charge_operator_final_state}
 \mbox{quark :} 
 & & 
 {\bf T}_{q \rightarrow q g} {\mathcal A}\left(  ... q_j ... \right) 
 =
 \left( T_{ij}^a \right) {\mathcal A}\left(  ... q_j ... \right), 
 \nonumber \\
 \mbox{gluon :} 
 & & 
 {\bf T}_{g \rightarrow g g}{\mathcal A}\left(  ... g^b ... \right) 
 =
 \left( i f^{cab} \right) {\mathcal A}\left(  ... g^b ... \right), 
 \nonumber \\
 \mbox{antiquark :} 
 & & 
 {\bf T}_{\bar{q} \rightarrow \bar{q} g} {\mathcal A}\left(  ... \bar{q}_j ... \right) 
 =
 \left( - T_{ji}^a \right) {\mathcal A}\left(  ... \bar{q}_j ... \right).
\eq
The minus sign for the antiquark has its origin in the fact that for an outgoing antiquark the (outgoing) 
momentum flow is opposite to the flow of the fermion line.
The corresponding colour charge operators for the emission of a gluon from a quark, gluon or antiquark in the initial state are
\bq
\label{colour_charge_operator_initial_state}
 \mbox{quark :} 
 & & 
 {\bf T}_{\bar{q} \rightarrow \bar{q} g} {\mathcal A}\left(  ... \bar{q}_j ... \right) 
 =
 \left( - T_{ji}^a \right) {\mathcal A}\left(  ... \bar{q}_j ... \right), 
 \nonumber \\
 \mbox{gluon :} 
 & & 
 {\bf T}_{g \rightarrow g g} {\mathcal A}\left(  ... g^b ... \right) 
 =
 \left( i f^{cab} \right) {\mathcal A}\left(  ... g^b ... \right), 
 \nonumber \\
 \mbox{antiquark :} 
 & & 
 {\bf T}_{q \rightarrow q g}{\mathcal A}\left(  ... q_j ... \right) 
 =
 \left( T_{ij}^a \right) {\mathcal A}\left(  ... q_j ... \right). 
\eq
In the amplitude an incoming quark is denoted as an outgoing antiquark and vice versa.
For the squares of the colour charge operators one has
\bq
 {\bf T}_{q \rightarrow q g}^2 = C_F,
 & &
 {\bf T}_{g \rightarrow g g}^2 = C_A.
\eq
We also define the colour charge operator for the emission of a quark-antiquark pair from a gluon by
\bq
 {\bf T}_{g \rightarrow q \bar{q}} {\mathcal A}\left(  ... g^b ... \right) 
 & = &
 \left( T^b_{ij} \right) {\mathcal A}\left(  ... g^b ... \right)
\eq
and
\bq
 {\bf T}_{g \rightarrow q \bar{q}}^2 = T_R.
\eq
$C_A$, $C_F$ and $T_R$ are the usual $SU(N_c)$ colour factors, given by
\bq
 C_A = N_c,
 \;\;\;\;\;\;
 C_F = \frac{N_c^2-1}{2 N_c},
 \;\;\;\;\;\;
 T_R = \frac{1}{2}.
\eq 
In squaring an amplitude we obtain terms proportional to ${\bf T}_i \cdot {\bf T}_k$ (with $k \neq i$) and terms proportional to ${\bf T}_i^2$.
We may re-express ${\bf T}_i^2$ as a combination of
terms involving only ${\bf T}_i \cdot {\bf T}_k$ with $k \neq i$.
This can be done using colour conservation. We write for $i \in \{q, g, \bar{q}\}$
\bq
 {\bf T}_i^2 & = & 
 - \sum\limits_{k \neq i}
 {\bf T}_i \cdot {\bf T}_k,
\eq
where the sum runs over all external coloured partons $k$ excluding parton $i$.
For the splitting $g \rightarrow q \bar{q}$ we write
\bq
 {\bf T}_{g \rightarrow q \bar{q}}^2 & = & 
 - \sum\limits_{k \neq i}
 \frac{{\bf T}_{g \rightarrow q \bar{q}}^2}{{\bf T}_{i}^2}
 {\bf T}_i \cdot {\bf T}_k.
\eq
We further denote by $\beta_0$ the first coefficient of the QCD $\beta$-function,
\bq
 \beta_0 & = & \frac{11}{3} C_A - \frac{4}{3} T_R N_f,
\eq
and introduce for later convenience the constants
\bq
 \gamma_q = \gamma_{\bar q} = \frac{3}{2} C_F, 
 & &
 \gamma_g = \frac{1}{2} \beta_0.
\eq
In the real emission part there can be approximation terms corresponding to initial-state singularities
with a flavour transition $q\rightarrow g$ or $g \rightarrow q$.
The averaging factor for the number of colour degrees of freedom for the initial-state particle $a'$
is determined from the real emission matrix element with $(n+3)$ particles.
When adding the real approximation terms back, it is within the dipole formalism common practice 
to take as averaging factor the number of colour degrees of freedom for the particle $a$ in the
Born amplitude with $(n+2)$ particles.
This introduces a compensation factor in the integrated approximation terms.
We have
\bq
 \frac{n_c(q)}{n_c(g)} C_F \; = \; T_R,
 & &
 \frac{n_c(g)}{n_c(q)} T_R \; = \; C_F.
\eq
In this paper we will not use this convention. We are interested in the local cancellation of singularities
at the integrand level. It is therefore natural to work in the phase space of $(n+1)$-final state particles
and we simply keep the averaging factor corresponding to $a'$.

\subsection{Spin}

The amplitudes are vectors in spin space as well. 
It is advantageous to set-up the subtraction method locally in spin space.
This allows the use of optimisation techniques like helicity sampling \cite{Czakon:2009ss,Dittmaier:2008md}.
In QCD, both quarks and gluons have two independent spin states, which we can label
by ``$+$'' and ``$-$''.
The polarisations of an external gluon are described by two polarisation vectors $\eps_\mu^\pm$, 
the polarisations of an outgoing quark are described by the two spinors $\bar{u}_\alpha^\pm$, 
the ones of an incoming quark by $u_\alpha^\pm$.
The polarisations of an outgoing antiquark are described by $v_\alpha^\pm$, 
the ones of an incoming antiquark by $\bar{v}_\alpha^\pm$.
For the convenience of the reader we have listed explicit expressions for all polarisation vectors and polarisation spinors 
in appendix~\ref{appendix:spinors}.

Let us further denote by ${\mathcal A}^{\xi}(...,i,...)$ the amplitude, where the polarisation vector of particle $i$ has been removed.
If particle $i$ is a gluon, $\xi$ is a Lorentz index, while in the case where particle $i$ is a quark $\xi$ corresponds to a Dirac index.

\subsection{The loop-tree duality method}

Let us consider a one-loop integral with $n$ external momenta $\{p_1, ..., p_n\}$.
\begin{figure}
\begin{center}
\includegraphics[scale=0.8]{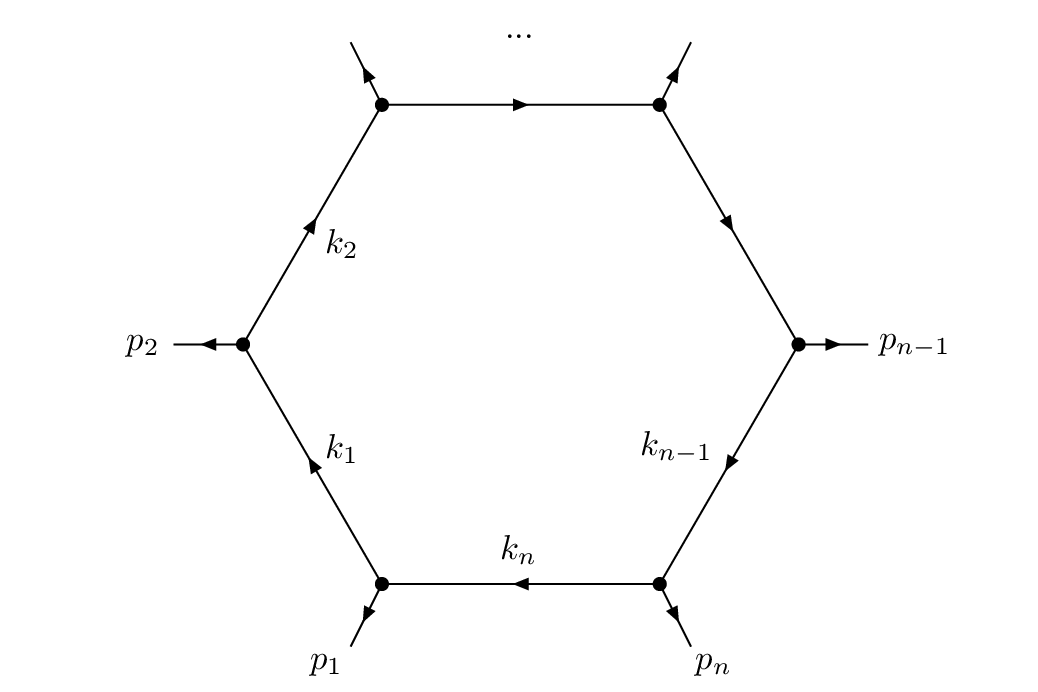}
\caption{\label{figure_momenta_one_loop}
The labelling of the momenta for a generic one-loop integral. The arrows denote the momentum flow.}
\end{center}
\end{figure}
In this sub-section it will be convenient to take all particles as outgoing.
Then, the momenta of the incoming particles will have negative energy components.
We further assume without loss of generality that the cyclic order of the external
momenta is $p_1,p_2,...,p_n$.
If this is not the case, a simple re-labelling of the momenta will achieve this.
With the notation as in fig.~(\ref{figure_momenta_one_loop}) we define
\bq
 k_j & = & k - q_j,
 \;\;\; q_j = \sum\limits_{l=1}^j p_l.
\eq
A generic one-loop integral can be written as
\bq
 I_n 
 & = & 
 \int \frac{d^Dk}{(2\pi)^D} \frac{P(k) }{\prod\limits_{j=1}^n \left( k_j^2 - m_j^2 + i \delta \right)}.
\eq
$P(k)$ is a polynomial in the loop momentum $k$.
The $+i \delta$-prescription in the propagators indicates into which direction the poles of
the propagators should be avoided.
The loop-tree duality technique allows us to replace the integration over the $D$-dimensional loop momentum space
by $n$ integrations over the $(D-1)$-dimensional forward hyperboloids \cite{Catani:2008xa}:
\bq
\label{loop_tree_forward}
 I_n 
 & = &
 - i \sum\limits_{i=1}^n
 \int \frac{d^{D-1}k}{(2\pi)^{D-1} \; 2 k_{i,0}} 
 \frac{P(k) }{\prod\limits_{\stackrel{j=1}{j \neq i}}^n \left[ k_j^2 - m_j^2 - i \delta \eta \left(k_j-k_i\right)\right]}
 \left.
 \vphantom{
 \frac{P(k) }{\prod \left[ k_j^2 - m_j^2 - i \delta \eta \left(k_j-k_i\right)\right]}
 }
 \right|_{k_{i,0}=\sqrt{\vec{k}_i^2+m_i^2}},
\eq
where $\eta$ is a vector with $\eta_0 > 0$ and $\eta^2 \ge 0$.
Alternatively, we may integrate over the backward hyperboloids:
\bq
\label{loop_tree_backward}
 I_n 
 & = &
  i \sum\limits_{i=1}^n
 \int \frac{d^{D-1}k}{(2\pi)^{D-1} \; 2 k_{i,0}} 
 \frac{P(k) }{\prod\limits_{\stackrel{j=1}{j \neq i}}^n \left[ k_j^2 - m_j^2 + i \delta \eta \left(k_j-k_i\right)\right]}
 \left.
 \vphantom{
 \frac{P(k) }{\prod \left[ k_j^2 - m_j^2 + i \delta \eta \left(k_j-k_i\right)\right]}
 }
 \right|_{k_{i,0}=-\sqrt{\vec{k}_i^2+m_i^2}}.
\eq
Note the sign change in the $i \delta \eta (k_j-k_i)$-term.

Typical ultraviolet subtraction terms are of the form
\bq
\label{typical_UV_subtraction}
 I_r^{\mathrm{UV}} 
 & = & 
 \int \frac{d^Dk}{(2\pi)^D} \frac{P\left(\bar{k}\right) }{\left( \bar{k}^2 - \mu_{\mathrm{UV}}^2 + i \delta \right)^r},
\eq
with $\bar{k}=k-Q$ and $\mu_{\mathrm{UV}}$ an arbitrary mass. 
$Q$ is an arbitrary four-vector independent of the loop momentum $k$.
The quantity $P(\bar{k})$ is again a polynomial in $\bar{k}$.
In eq.~(\ref{typical_UV_subtraction}) there is only a single propagator, but this propagator may be raised to the power $r$.
Again, we may use the residue theorem to replace the integration over the $D$-dimensional loop momentum space
by an integration over the $(D-1)$-dimensional forward hyperboloid \cite{Bierenbaum:2012th}:
\bq
\label{loop_tree_duality_UV}
 I_r^{\mathrm{UV}} 
 & = & 
 - i \int \frac{d^{D-1}k}{(2\pi)^{D-1}} 
 \;
 \frac{1}{\left(r-1\right)!}
 \left( \frac{d}{d\bar{k}_0} \right)^{r-1}
 \frac{P\left(\bar{k}\right) }{\left( \bar{k}_0 + \sqrt{\vec{\bar{k}} + \mu_{\mathrm{UV}}^2} \right)^r}
 \left.
 \vphantom{
 \frac{P\left(\bar{k}\right) }{\bar{k}_0 + \sqrt{\vec{\bar{k}} + \mu_{\mathrm{UV}}^2}}
 }
 \right|_{\bar{k}_0=\sqrt{\vec{\bar{k}} + \mu_{\mathrm{UV}}^2}}.
\eq
There are only a finite number of ultraviolet subtraction terms. The differentiations with respect to $\bar{k}_0$
in eq.~(\ref{loop_tree_duality_UV}) may be carried analytically once and for all.
Note that we may take in eq.~(\ref{loop_tree_duality_UV}) the parameter $\mu_{\mathrm{UV}}^2$ to be complex.
Alternatively, we may integrate over the
backward hyperboloid:
\bq
\label{loop_tree_duality_UV_backward}
 I_r^{\mathrm{UV}} 
 & = & 
 i \int \frac{d^{D-1}k}{(2\pi)^{D-1}} 
 \;
 \frac{1}{\left(r-1\right)!}
 \left( \frac{d}{d\bar{k}_0} \right)^{r-1}
 \frac{P\left(\bar{k}\right) }{\left( \bar{k}_0 - \sqrt{\vec{\bar{k}} + \mu_{\mathrm{UV}}^2} \right)^r}
 \left.
 \vphantom{
 \frac{P\left(\bar{k}\right) }{\bar{k}_0 - \sqrt{\vec{\bar{k}} + \mu_{\mathrm{UV}}^2}}
 }
 \right|_{\bar{k}_0=-\sqrt{\vec{\bar{k}} + \mu_{\mathrm{UV}}^2}}.
\eq

\subsection{Phase space generation}

We recapitulate some basic facts about phase space generation.
Let us start from an $n$-parton configuration. In hadron collisions we have an integral
of the form
\bq
 \int\limits_0^1 dz_1
 \int\limits_0^1 dz_2
 \int d\phi_{n}\left(z_1 P_a + z_2 P_b \rightarrow p_1 + ... p_n\right)
 \frac{f_a\left(z_1\right)}{z_1}
 \frac{f_b\left(z_2\right)}{z_2}
 {\mathcal M}_{n}\left(\left\{ p \right\} \right),
\eq
where we suppressed all factors not relevant to the discussion here.
We denote by $P_a$ and $P_b$ the momenta of the incoming hadrons, the set
$\{p\}$ is given by $\{z_1P_a, z_2P_b, p_1, ..., p_n\}$.
Given $P_a$ and $P_b$, we first generate the momentum fractions $z_1$ and $z_2$ and
then the final state momenta $\{p_1,...,p_n\}$.

Now let us look at an $(n+1)$-parton configuration:
\bq
 \int\limits_0^1 dz_1
 \int\limits_0^1 dz_2
 \int d\phi_{n+1}\left(z_1 P_a + z_2 P_b \rightarrow p_1' + ... p_{n+1}'\right)
 \frac{f_a\left(z_1\right)}{z_1}
 \frac{f_b\left(z_2\right)}{z_2}
 {\mathcal M}_{n+1}\left(\left\{ p' \right\} \right),
\eq
with $\{p'\}=\{z_1 P_a, z_2 P_b, p_1', ..., p_{n+1}'\}$.
We would like to re-write this integral as an $n$-parton phase space integral plus some
additional integrations.
Using the phase space factorisation for final-state particles this can be done:
\bq
 \int\limits_0^1 dz_1
 \int\limits_0^1 dz_2
 \int d\phi_{n}\left(z_1 P_a + z_2 P_b \rightarrow p_1 + ... p_n\right)
 \int d\phi_{\mathrm{unres}}
 \frac{f_a\left(z_1\right)}{z_1}
 \frac{f_b\left(z_2\right)}{z_2}
 {\mathcal M}_{n+1}\left(\left\{ p' \right\} \right).
\eq
Thus we first generate the momentum fractions $z_1$ and $z_2$, then
$n$ final-state momenta $\{p_1,...,p_n\}$.
Finally, using $(D-1)$ additional variables, we construct from the set $\{p_1,...,p_n\}$
and the additional variables the final-state momenta $\{p_1',...,p_{n+1}'\}$.

Now let us consider the case, where we use phase space factorisation with initial-state particles.
In this case we obtain a convolution in one variable, which we denote by $x$.
We write
\bq
 d\phi_{\mathrm{unres}}
 & = &
 dx \; d\phi_{\mathrm{unres}}^{\mathrm{red}},
\eq
where $d\phi_{\mathrm{unres}}^{\mathrm{red}}$ is the measure for the remaining $(D-2)$ variables.
We now have
\bq
 \int\limits_0^1 dz_1
 \int\limits_0^1 dz_2
 \int\limits_0^1 dx
 \int d\phi_{\mathrm{unres}}^{\mathrm{red}}
 \int d\phi_{n}\left( x z_1 P_a + z_2 P_b \rightarrow p_1 + ... p_n\right)
 \frac{f_a\left(z_1\right)}{z_1}
 \frac{f_b\left(z_2\right)}{z_2}
 {\mathcal M}_{n+1}\left(\left\{ p' \right\} \right).
\eq
According to this expression, we would first generate the momentum fractions $z_1$ and $z_2$,
then the variables of $d\phi_{\mathrm{unres}}$ (including $x$), then the intermediate 
momenta $\{p\}=\{x z_1 P_a, z_2 P_b, p_1, ..., p_n\}$ and finally the 
momenta $\{p'\}=\{z_1 P_a, z_2 P_b, p_1', ..., p_{n+1}'\}$.
We would like to switch the order and generate $d\phi_n$ before $d\phi_{\mathrm{unres}}$.
We make the change of variables $z_1=z_1'/x$ and obtain
\bq
\lefteqn{
 \int\limits_0^1 dz_1'
 \int\limits_0^1 dz_2
 \int d\phi_{n}\left( z_1' P_a + z_2 P_b \rightarrow p_1 + ... p_n\right)
} \\
 & &
 \times
 \int\limits_0^1 dx
 \int d\phi_{\mathrm{unres}}^{\mathrm{red}}
 \theta\left(x-z_1'\right)
 \frac{f_a\left(\frac{z_1'}{x}\right)}{z_1'}
 \frac{f_b\left(z_2\right)}{z_2}
 {\mathcal M}_{n+1}\left(\left\{ p' \right\} \right).
 \nonumber 
\eq
This allows us to generate $d\phi_{n}$ before $d\phi_{\mathrm{unres}}$.
Consider now the case, where ${\mathcal M}_{n+1}$ factorises as
\bq
 {\mathcal M}_{n+1}\left(\left\{ p' \right\} \right)
 & = &
 \mbox{Sing}\left(\left\{ p' \right\} \right)
 {\mathcal M}_{n}\left(\left\{ p \right\} \right),
\eq
where ${\mathcal M}_{n}$ depends only on $\{p\}$.
We are in particular interested in the case, where the singular function $\mbox{Sing}(\{ p' \} )$ is of the form
\bq
\label{def_sing_form}
 \mbox{Sing}\left(\left\{ p' \right\} \right)
 & = &
 A\left(x\right) 
 - \delta\left(1-x\right) \int\limits_0^1 dy \;B\left(y\right).
\eq
Plugging this in gives
\bq
\label{impl_plus_distr}
\lefteqn{
 \int\limits_0^1 \frac{dz_1'}{z_1'}
 \int\limits_0^1 \frac{dz_2}{z_2}
 \int d\phi_{n}\left( z_1' P_a + z_2 P_b \rightarrow p_1 + ... p_n\right)
 f_b\left(z_2\right)
 {\mathcal M}_{n}\left(\left\{ p \right\} \right)
} \\
 & &
 \times
 \int\limits_0^1 dx
 \int d\phi_{\mathrm{unres}}^{\mathrm{red}}
 \left[ 
  \theta\left(x-z_1'\right)
  f_a\left(\frac{z_1'}{x}\right)
  A\left(x\right) 
  -
  f_a\left(z_1'\right)
  B\left(x\right)
 \right].
 \nonumber 
\eq
Eq.~(\ref{impl_plus_distr}) defines how to implement functions 
of the form of eq.~(\ref{def_sing_form}).
In particular this applies to the cases, where $A(x)$ and $B(x)$ contain the same
singular terms $1/(1-x)$:
\bq
 A\left(x\right)
 & = &
 \frac{c}{1-x} + \mbox{finite terms},
 \nonumber \\
 B\left(x\right)
 & = &
 \frac{c}{1-x} + \mbox{other finite terms}.
\eq

\section{The real approximation terms}
\label{sect:real_subtraction_terms}

In this section we define the real subtraction terms
\bq
\lefteqn{
 d\sigma^{\mathrm{A}}_{\mathrm{R}}
 = } & & 
 \\
 & &
 \left(
 \sum\limits_{(i',j')} \sum\limits_{k' \neq i',j'} {\mathcal D}_{i'j',k'}
 +
 \sum\limits_{(i',j')} \sum\limits_{a'} {\mathcal D}_{i'j'}^{a'}
 +
 \sum\limits_{(a',j')} \sum\limits_{k' \neq j'} {\mathcal D}^{a'j'}_{k'}
 +
 \sum\limits_{(a',j')} \sum\limits_{b' \neq a'} {\mathcal D}^{a'j',b'}
 \right) d\phi_n \; d\phi_{\mathrm{unresolved}}.
 \nonumber
\eq
The definition given here differs -- when summed over the spins of the unobserved particles -- 
from the original dipole subtraction terms \cite{Catani:1997vz} by finite terms.
This is un-problematic as long as we add and subtract exactly the same quantity.
The essential property of the subtraction terms is that they have the same singular behaviour as the matrix
elements squared which they approximate.
If an analytic integration of the subtraction terms is envisaged one may in a second step modify the approximation
terms by finite terms in order to simplify the analytic integration.
However, within the approach based on numerical integration discussed in this paper
the second step is not necessary.
The real approximation terms defined below have the additional pedagocial advantage 
that they show manifestly, that all unresolved particles in the real approximation terms have
transverse polarisations.
This will be important for the cancellation of singularities.

The real approximation terms are obtained from the singular limits of the real emission matrix element squared.
We have to consider soft and collinear limits.
Let us start with the collinear limit. We consider a splitting $i \rightarrow i' + j'$.
The collinear limit occurs only in massless case. However, if the masses of the particles are small against
other invariants of the process, it is advantageous to include approximation terms for the
quasi-collinear limit \cite{Catani:2002hc,Dittmaier:2008md}.
In the quasi-collinear limit we parametrise the momenta of the two quasi-collinear final-state partons $i'$ and $j'$
as
\bq
 p_i' & = & z p + k_\perp - \frac{k_\perp^2+z^2 m_i^2 - m_i'{}^2}{z} \frac{n}{2 p \cdot n},
 \nonumber \\
 p_j' & = & (1-z)  p - k_\perp - \frac{k_\perp^2+(1-z)^2m_i^2-m_j'{}^2}{1-z} \frac{n}{2 p \cdot n }.
\eq
Here $n$ is a massless four-vector and the transverse component $k_\perp$ satisfies
$2pk_\perp = 2n k_\perp =0$.
The four-vectors $p$, $p_i'$ and $p_j'$ are on-shell:
\bq
 p^2 = m_i^2,
 \;\;\;\;\;\;
 p_i'{}^2 = m_i'{}^2,
 \;\;\;\;\;\;
 p_j'{}^2 = m_j'{}^2.
\eq
In the quasi-collinear limit we take terms of the order ${\mathcal O}(k_\perp)$, 
${\mathcal O}(m_i)$, ${\mathcal O}(m_i')$ 
and ${\mathcal O}(m_j')$ to be of the same order.
The collinear limit is a special case of the quasi-collinear limit, obtained by setting $m_i=m_i'=m_j'=0$.
If the emitting particle is in the initial state, the collinear limit is defined as
\bq
 p_a' & = & p,
 \nonumber \\
 p_j' & = & \left(1-x\right) p + k_\perp - \frac{k_\perp^2}{1-x} \frac{n}{2 p \cdot n},
 \nonumber \\
 p_a & = & x p - k_\perp - \frac{k_\perp^2}{x} \frac{n}{2 p \cdot n}.
\eq
Here, all particles are massless. In this paper we restrict ourselves to massless incoming partons, therefore
we do not have to consider the generalisation to the massive quasi-collinear case for initial-state partons.

In the quasi-collinear limit we have to consider terms of order ${\mathcal O}(k_\perp^{-2})$.
In this limit the Born amplitude factorises according to
\bq
\lefteqn{
 \lim\limits_{p_i' || p_j'}
 {\mathcal A}_{n+1}^{(0)}\left(...,p_i',...,p_j',...\right) 
 = } & &
 \nonumber \\
 & & 
 g \mu^\eps 
 S_\eps^{-\frac{1}{2}}
 \sum\limits_{\lambda_i} 
 \; 
 \mbox{Split}_{i \rightarrow i'+j'}^{\lambda_i}(p_i,p_i',p_j',\lambda_i',\lambda_j') 
 \; 
 {\bf T}_{i \rightarrow i'+j'} 
 \; 
 {\mathcal A}_{n}^{(0)}\left(...,p_i,\lambda_i,...\right).
\eq
where the sum is over all polarisations of the intermediate particle.
The quantity
\bq
 S_\eps 
 & = & 
 \left( 4 \pi \right)^\eps e^{-\eps\gamma_E}
\eq
is the typical phase space volume factor in $D =4-2\eps$ dimensions and
$\gamma_E$ is Euler's constant.
The variables $\lambda_i'$ and $\lambda_j'$ denote the polarisations of the particles $i'$ and $j'$, respectively.
The splitting functions $\mbox{Split}$ are given by
\bq
\label{def_split}
 \mbox{Split}^{\lambda_i}_{q \rightarrow q g}\left(p_i,p_i',p_j',\lambda_i',\lambda_j'\right) 
 & = &
 \frac{1}{(p_i' + p_j')^2 - m_i^2} \bar{u}^{\lambda_i'}(p_i') \eps\!\!\!/^{\lambda_j'}(p_j') u^{\lambda_i}(p_i),
\nonumber \\
 \mbox{Split}^{\lambda_i}_{g \rightarrow g g}\left(p_i,p_i',p_j',\lambda_i',\lambda_j'\right) 
 & = &
 \frac{2}{2 p_i' \cdot p_j'} \left[
    \eps^{\lambda_i'}(p_i') \cdot \eps^{\lambda_j'}(p_j') \; p_i' \cdot \left. \eps^{\lambda_i}(p_i) \right.^\ast
 \right. \nonumber \\
 & & \left.
  + \eps^{\lambda_j'}(p_j') \cdot \left. \eps^{\lambda_i}(p_i) \right.^\ast \; p_j' \cdot \eps^{\lambda_i'}(p_i')
  - \eps^{\lambda_i'}(p_i') \cdot \left. \eps^{\lambda_i}(p_i) \right.^\ast \; p_i' \cdot \eps^{\lambda_j'}(p_j')
 \right],
\nonumber \\
 \mbox{Split}^{\lambda_i}_{g \rightarrow q \bar{q}}\left(p_i,p_i',p_j',\lambda_i',\lambda_j'\right) 
 & = &
 \frac{1}{2 p_i' \cdot p_j'} \bar{u}^{\lambda_i'}(p_i') \left. \eps\!\!\!/^{\lambda_i}(p_i) \right.^\ast v^{\lambda_j'}(p_j').
\eq
Here we used the notation $\left. \eps\!\!\!/^\lambda(p) \right.^\ast = \eps_\mu^\lambda(p)^\ast \; \gamma^\mu$, i.e. complex conjugation is only with respect to the polarisation vector.
We define the squares of the splitting amplitudes by
\bq
\label{def_P}
 \left[ P_{i \rightarrow i'+j'}\left(p_i,p_i',p_j',\lambda_i',\lambda_j'\right) \right]_{\alpha\beta} 
 & = &  
 \sum\limits_{\lambda,\lambda'}
 u_\alpha^\lambda(p_i) \left. \; \mbox{Split}^\lambda \right.^\ast \mbox{Split}^{\lambda'} \;\bar{u}_\beta^{\lambda'}(p_i)
 \;\;\;\;\mbox{for quarks,}
 \nonumber \\
 \left[ P_{i \rightarrow i'+j'}\left(p_i,p_i',p_j',\lambda_i',\lambda_j'\right)\right]_{\mu\nu} 
 & = &  
 \sum\limits_{\lambda,\lambda'}
 \left. \eps_\mu^\lambda(p_i) \right.^\ast \;
                \left. \mbox{Split}^\lambda \right.^\ast \mbox{Split}^{\lambda'} \; \eps_\nu^{\lambda'}(p_i)
 \;\;\;\;\mbox{for gluons.}
\eq
The squared amplitude factorises in the (quasi-) collinear limit as
\bq
\label{amplitude_squared_coll}
 \lim\limits_{p_i' || p_j'}
 \left|{\mathcal A}_{n+1}^{(0)} \right|^2 & = & 
 4 \pi \alpha_s S_\eps^{-1} \mu^{2\eps} 
 {{\mathcal A}_{n}^{\xi \;(0)}}^\ast
 {\bf T}_{i \rightarrow i'+j'}^2
 \left[ P_{i \rightarrow i'+j'}\left(p_i,p_i',p_j',\lambda_i',\lambda_j'\right) \right]_{\xi \xi'}
 {\mathcal A}_{n}^{\xi' \;(0)}.
\eq
Let us now consider the soft limit.
We consider the case where particle $j'$ becomes soft.
In the soft limit we parametrise the momentum of the soft parton $p_j'$ as
\bq
 p_j' & = & \lambda q
\eq
and consider contributions to $|{\cal A}_{n+1}^{(0)}|^2$ of the order $\lambda^{-2}$.
Contributions to $|{\cal A}_{n+1}^{(0)}|^2$ which are less singular than $\lambda^{-2}$ are integrable in the soft limit.
In the soft limit a Born amplitude ${\cal A}_{n+1}^{(0)}$ with $(n+1)$ partons behaves as
\bq
 \lim\limits_{p_j' \rightarrow 0} {\cal A}_{n+1}^{(0)}
 & = & 
 g S_\eps^{-\frac{1}{2}} \mu^\eps \eps_\mu(p_j') {\bf J}^\mu
 {\cal A}_{n}^{(0)}.
\eq
The eikonal current is given by
\bq
 {\bf J}^\mu & = & \sum\limits_{i \neq j} {\bf T}_i \frac{p_i'{}^\mu}{p_i' \cdot p_j'}.
\eq
The sum is over the remaining $n$ hard momenta $p_i'$.
The quasi-collinear splittings $q \rightarrow q g$ and $ g \rightarrow g g$ 
have non-vanishing soft limits and a part of the soft limit is already approximated by these terms.
In addition we will need the terms which are singular in the soft limit, but not in the (quasi)-collinear limit.
To this aim we set
\bq
\label{def_S}
 \left[ S_{q \rightarrow q g}\left(p_i,p_i',p_j',p_k',\lambda_i',\lambda_j'\right) \right]_{\alpha \beta} & = &  
 -
 \frac{\left( p_i' \cdot \eps_j'{}^\ast \right) \left( p_k' \cdot \eps_j' \right) + \left( p_k' \cdot \eps_j'{}^\ast \right) \left(p_i' \cdot \eps_j' \right)}
      { \left( p_i' \cdot p_j' \right) \left( p_i' \cdot p_j' + p_j' \cdot p_k'\right)}
 \;
 u_\alpha^{\lambda_i'}(p_i') \bar{u}_\beta^{\lambda_i'}(p_i'),
 \nonumber \\
 \left[ S_{g \rightarrow g g}\left(p_i,p_i',p_j',p_k',\lambda_i',\lambda_j'\right) \right]_{\mu\nu} & = &  
 -
 \frac{\left( p_i' \cdot \eps_j'{}^\ast \right) \left( p_k' \cdot \eps_j' \right) + \left( p_k' \cdot \eps_j'{}^\ast \right) \left(p_i' \cdot \eps_j' \right)}
      { \left( p_i' \cdot p_j' \right) \left( p_i' \cdot p_j' + p_j' \cdot p_k'\right)}
 \;
 \left. \eps_\mu^{\lambda_i'}(p_i') \right.^\ast \eps_\nu^{\lambda_i'}(p_i')
 \nonumber \\
 & &
 -
 \frac{\left( p_j' \cdot \eps_i'{}^\ast \right) \left( p_k' \cdot \eps_i' \right) + \left( p_k' \cdot \eps_i'{}^\ast \right) \left(p_j' \cdot \eps_i' \right)}
      { \left( p_i' \cdot p_j' \right) \left( p_i' \cdot p_j' + p_i' \cdot p_k'\right)}
 \;
 \left. \eps_\mu^{\lambda_j'}(p_j') \right.^\ast \eps_\nu^{\lambda_j'}(p_j'),
 \nonumber \\
\eq
where we used the abbreviation $\eps_l' = \eps^{\lambda_l'}(p_l')$ for $l=i,j$.
In connection with crossing symmetry it is useful to define the following operation
\bq
\label{def_C_operation}
 {\mathcal C}_i & : & 
 \eps_i \leftrightarrow \eps^\ast_i,
 \nonumber \\
 & & \bar{u}_i \leftrightarrow \bar{v}_i,
 \nonumber \\
 & & u_i \leftrightarrow v_i,
\eq
which adjusts the polarisation vector or spinor of the $i$-th particle from the final to the initial state and vice versa.
We may now list the dipole subtraction terms.

\subsection{Final-state emitter and final-state spectator}

If both the emitter and the spectator are in the final state, the dipole approximation terms are given by
\bq
\lefteqn{
{\mathcal D}_{i'j',k'}
 = 
 - 4 \pi \alpha_s S_\eps^{-1} \mu^{2\eps} 
 } & & 
 \\
 & &
 {{\mathcal A}^{\xi \;(0)}}\left(...,p_i,...,p_k,...\right)^\ast
 \;\;
 \frac{{\bf T}_i \cdot {\bf T}_k}{{\bf T}_i^2}
 \left[ V_{i'j',k'}\left(p_i,p_i',p_j',p_k',\lambda_i',\lambda_j'\right) \right]_{\xi \xi'}
 \;\;
 {\mathcal A}^{\xi' \;(0)}\left(...,p_i,...,p_k,...\right).
 \nonumber 
\eq
The functions $V_{i'j',k'}$ are given for the various splittings by
\bq
 V_{i'_q j'_g,k'}\left(p_i,p_i',p_j',p_k',\lambda_i',\lambda_j'\right)
 & = & 
 C_F
 \left[
       P_{q \rightarrow q g}\left(p_i,p_i',p_j',\lambda_i',\lambda_j'\right)
     + 
       S_{q \rightarrow q g}\left(p_i,p_i',p_j',p_k',\lambda_i',\lambda_j'\right)
 \right],
 \nonumber \\
 V_{i'_g j'_g,k'}\left(p_i,p_i',p_j',p_k',\lambda_i',\lambda_j'\right)
 & = & 
 C_A
 \left[
       P_{g \rightarrow g g}\left(p_i,p_i',p_j',\lambda_i',\lambda_j'\right)
     + 
       S_{g \rightarrow g g}\left(p_i,p_i',p_j',p_k',\lambda_i',\lambda_j'\right)
 \right],
 \nonumber \\
 V_{i'_q j'_{\bar{q}},k'}\left(p_i,p_i',p_j',p_k',\lambda_i',\lambda_j'\right)
 & = & 
 T_R
 \left[
       P_{g \rightarrow q \bar{q}}\left(p_i,p_i',p_j',\lambda_i',\lambda_j'\right)
 \right].
\eq
The mapped momenta $p_i$ and $p_k$ are defined in the massless case by
\bq
\label{map_massless}
 p_i = p_i' + p_j' - \frac{y}{1-y} p_k',
 \;\;\;\;\;\;
 p_k = \frac{1}{1-y} p_k',
 \;\;\;\;\;\;
 y = \frac{p_i' \cdot p_j'}{p_i' \cdot p_j' + p_i' \cdot p_k' + p_j' \cdot p_k'}.
\eq
In the massive case we use
\bq
\label{map_massive}
 p_k
 & = & 
 \frac{\sqrt{\lambda(Q^2,m_i^2,m_k^2)}}{\sqrt{\lambda(Q^2,(p_i'+p_j')^2,m_k^2)}}
 \left( p_k' - \frac{Q \cdot p_k'}{Q^2} Q \right) 
 + \frac{Q^2+m_k^2-m_i^2}{2Q^2} Q,
 \nonumber \\
 p_i & = & Q - p_k,
\eq
where $Q=p_i'+p_j'+p_k'$ and $\lambda$ is the K\"allen function
\bq
 \lambda(x,y,z) & = & x^2 + y^2 + z^2 - 2 x y - 2 y z - 2 z x.
\eq
Note that the particle type of the spectator is not changed and therefore $m_k'=m_k$.
Eq.~(\ref{map_massive}) reduces in the massless limit to eq.~(\ref{map_massless}).

\subsection{Final-state emitter and initial-state spectator}

If the emitter is in the final state and the spectator in the initial state, 
the dipole approximation terms are given by
\bq
\lefteqn{
{\mathcal D}_{i'j'}^{a'}
 = 
 - 4 \pi \alpha_s S_\eps^{-1} \mu^{2\eps} 
 } & & 
 \\
 & &
 {{\mathcal A}^{\xi \;(0)}}\left(...,p_i,...,p_a,...\right)^\ast
 \;\;
 \frac{{\bf T}_i \cdot {\bf T}_a}{{\bf T}_i^2}
 \left[ V_{i'j'}^{a'}\left(p_i,p_i',p_j',p_a',\lambda_i',\lambda_j'\right) \right]_{\xi \xi'}
 \;\;
 {\mathcal A}^{\xi' \;(0)}\left(...,p_i,...,p_a,...\right).
 \nonumber 
\eq
The dipole splitting function is related by crossing to the final-final case:
\bq
 V_{i'j'}^{a'}\left(p_i,p_i',p_j',p_a',\lambda_i',\lambda_j'\right)
 & = &
 V_{i'j',a'}\left(p_i,p_i',p_j',-p_a',\lambda_i',\lambda_j'\right).
\eq
The mapped momenta $p_i$ and $p_a$ are defined by
\bq
\label{map_massless_fi}
 p_i = p_i' + p_j' - (1-x) p_a',
 \;\;\;\;\;\;
 p_a = x p_a'.
\eq
The variable $x$ is given by
\bq
 x = \frac{p_i' \cdot p_a' + p_j' \cdot p_a' - p_i' \cdot p_j' + \frac{1}{2}\left(m_i^2-m_i'{}^2-m_j'{}^2\right)}{p_i' \cdot p_a' + p_j' \cdot p_a'}.
\eq
In the massless case and in the case where $m_i'=m_i$ and $m_j'=0$ this reduces to
\bq
 x & = & \frac{p_i' \cdot p_a' + p_j' \cdot p_a' - p_i' \cdot p_j'}{p_i' \cdot p_a' + p_j' \cdot p_a'}.
\eq

\subsection{Initial-state emitter and final-state spectator}
\label{sect_dipole_initial_final}

If the emitter is in the initial state and the spectator in the final state, 
the dipole approximation terms are given by
\bq
\lefteqn{
{\mathcal D}_{k'}^{a'j'}
 = 
 - 4 \pi \alpha_s S_\eps^{-1} \mu^{2\eps} 
 } & & 
 \\
 & &
 {{\mathcal A}^{\xi \;(0)}}\left(...,p_a,...,p_k,...\right)^\ast
 \;\;
 \frac{{\bf T}_a \cdot {\bf T}_k}{{\bf T}_a^2}
 \left[ V_{k'}^{a'j'}\left(p_a,p_a',p_j',p_k',\lambda_a',\lambda_j'\right) \right]_{\xi \xi'}
 \;\;
 {\mathcal A}^{\xi' \;(0)}\left(...,p_a,...,p_k,...\right).
 \nonumber 
\eq
The dipole splitting function is related by crossing to the final-final case:
\bq
 V_{k'}^{a'j'}\left(p_a,p_a',p_j',p_k',\lambda_a',\lambda_j'\right)
 & = &
 {\mathcal C}_{(a',a)} V_{a'j',k'}\left(-p_a,-p_a',p_j',p_k',\lambda_a',\lambda_j'\right).
\eq
The operation ${\mathcal C}$ is defined in eq.~(\ref{def_C_operation}).
The mapped momenta $p_a$ and $p_k$ are defined by
\bq
 p_a = x p_a',
 \;\;\;\;\;\;
 p_k = p_k' + p_j' - (1-x) p_a',
 \;\;\;\;\;\;
 x = \frac{p_k' \cdot p_a' + p_j' \cdot p_a' - p_j' \cdot p_k'}{p_k' \cdot p_a' + p_j' \cdot p_a'}.
\eq
Note that we restrict ourselves to massless initial-state particles. This implies that the
masses of the particles $a$, $a'$ and $j'$ are zero.

\subsection{Initial-state emitter and initial-state spectator}
\label{sect_dipole_initial_initial}

If both the emitter and the spectator are in the initial state, 
the dipole approximation terms are given by
\bq
\lefteqn{
{\mathcal D}^{a'j',b'}
 = 
 - 4 \pi \alpha_s S_\eps^{-1} \mu^{2\eps} 
 } & & 
 \\
 & &
 {{\mathcal A}^{\xi \;(0)}}\left(...,p_a,...,p_b,...\right)^\ast
 \;\;
 \frac{{\bf T}_a \cdot {\bf T}_b}{{\bf T}_a^2}
 \left[ V^{a'j',b'}\left(p_a,p_a',p_j',p_b',\lambda_a',\lambda_j'\right) \right]_{\xi \xi'}
 \;\;
 {\mathcal A}^{\xi' \;(0)}\left(...,p_a,...,p_b,...\right).
 \nonumber 
\eq
The dipole splitting function is related by crossing to the final-final case:
\bq
 V^{a'j',b'}\left(p_a,p_a',p_j',p_b',\lambda_a',\lambda_j'\right)
 & = &
 {\mathcal C}_{(a',a)} V_{a'j',b'}\left(-p_a,-p_a',p_j',-p_b',\lambda_a',\lambda_j'\right).
\eq
In this case the mapped momenta are defined as follows:
\bq
 p_a = x p_a',
 \;\;\;\;\;\;
 p_b = p_b',
 \;\;\;\;\;\;
 x = \frac{p_a' \cdot p_b' - p_j' \cdot p_a' - p_j' \cdot p_b'}{p_a' \cdot p_b'},
\eq
and all final state momenta are transformed as
\bq
 p_l & = & \Lambda p_l',
\eq
where $\Lambda$ is a Lorentz transformation defined by
\bq
\label{def_Lorentz_trafo}
 \Lambda^\mu_{\;\;\nu}
 & = & 
 g^\mu_{\;\;\nu}
 -2 \frac{\left(K^\mu+\tilde{K}^\mu\right) \left(K_\nu+\tilde{K}_\nu\right)}{\left(K+\tilde{K}\right)^2}
 + 2 \frac{\tilde{K}^\mu K_\nu}{K^2},
 \nonumber \\
 & &
 K = p_a' + p_b' - p_j',
 \;\;\;\;\;\;
 \tilde{K} = p_a + p_b.
\eq
Again we consider only the case of massless initial-state particles. 
Therefore the 
masses of the particles $a$, $a'$, $b'$ and $j'$ are zero.

\section{The virtual approximation terms}
\label{sect:virtual_subtraction_terms}

In this section we give the virtual subtraction terms, which we split into an infrared part and an ultraviolet part:
\bq
 d\sigma^{\mathrm{A}}_{\mathrm{V}}
 & = &
 d\sigma^{\mathrm{A}}_{\mathrm{V},\mathrm{IR}}
 +
 d\sigma^{\mathrm{A}}_{\mathrm{V},\mathrm{UV}},
\eq
with
\bq
 d\sigma^{\mathrm{A}}_{\mathrm{V},\mathrm{IR}}
 & = &
 \frac{d^Dk}{(2\pi)^D} 
 \; 2 \; \mbox{Re}\; \left[ \left.{\mathcal A}^{(0)}\right.^\ast \left( {\mathcal G}^{(1)}_{\mathrm{soft}} + {\mathcal G}^{(1)}_{\mathrm{coll}} \right) \right] d\phi_n,
 \nonumber \\
 d\sigma^{\mathrm{A}}_{\mathrm{V},\mathrm{UV}}
 & = &
 \frac{d^Dk}{(2\pi)^D} 
 \; 2 \; \mbox{Re}\; \left[ \left.{\mathcal A}^{(0)}\right.^\ast {\mathcal G}^{(1)}_{\mathrm{UV}} \right] d\phi_n.
\eq
The approximation terms are not unique and may be modified by adding finite terms.
This freedom is advantageous and can be used to improve the numerical stability when integrating over the subtracted virtual
part \cite{Becker:2012aq}.
In this paper our focus lies on the basic principles of the cancellation of singularities.
In order to keep all formulae to a minimal length we quote the original approximation terms from \cite{Becker:2010ng}.
In this section we use the convention to take all particles as outgoing.

\subsection{The virtual infrared approximation terms}

We may write 
the virtual infrared approximation terms as
\bq
 d\sigma^{\mathrm{A}}_{\mathrm{V},\mathrm{IR}}
 & = &
 \left(
 \sum\limits_{i} \sum\limits_{k \neq i} {\mathcal E}_{i,k}
 \right) \frac{d^Dk}{(2\pi)^D} \; d\phi_n,
\eq
with
\bq
{\mathcal E}_{i,k}
 & = &
 - 4 \pi \alpha_s S_\eps^{-1} \mu^{2\eps} 
 \;
 2 \;\mbox{Re}\; 
 {{\mathcal A}^{(0)}}\left(...,p_i,...,p_k,...\right)^\ast
 \;
 \frac{{\bf T}_i \cdot {\bf T}_k}{{\bf T}_i^2}
 \;
 W_i\left(p_i,p_k,k_i\right)
 \;
 {\mathcal A}^{(0)}\left(...,p_i,...,p_k,...\right)
 \nonumber \\
\eq
and
\bq
\label{def_W_virtual}
\lefteqn{
 W_i\left(p_i,p_k,k_i\right)
 = 
} & & \\
& &
2 i \; 
 {\bf T}_i^2
 \left\{
  \frac{p_{i} \cdot p_{k}}{\left[ \left(k_i+p_i\right)^2 -m_i^2 \right] k_{i}^2 \left[ \left(k_i-p_k\right)^2 - m_k^2 \right]}
  - \frac{S_{i}}{\left[ \left(k_i+p_i\right)^2 -m_i^2 \right] k_{i}^2}
  + \frac{S_{i}}{\left( \bar{k}^2-\mu_{\mathrm{UV}}^2 \right)^2}
 \right\}.
 \nonumber 
\eq
$m_i$ and $m_k$ are the masses of the external particles $i$ and $k$, respectively.
\begin{figure}
\begin{center}
\includegraphics[scale=0.8]{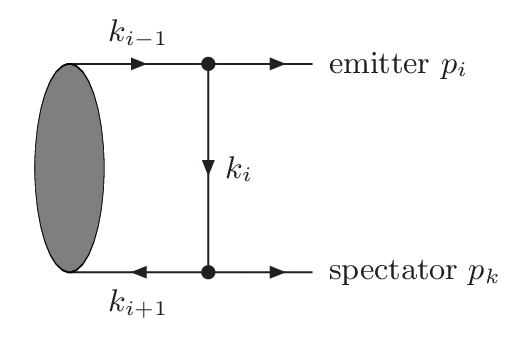}
\caption{\label{figure_virtual_momentum_flow}
The momentum flow for the virtual infrared approximation terms.
The kinematics is specified by the three momenta $p_i$, $p_k$ and $k_i$.}
\end{center}
\end{figure}
Furthermore, $S_i=1$ if the external line $i$ corresponds to a quark 
and $S_i=1/2$ if it corresponds to a gluon. 
The first term in eq.~(\ref{def_W_virtual}) approximates a soft singularity, the second term a (quasi-) collinear singularity.
The third term ensures that the expression is ultraviolet finite.
Note that the loop integrals in the virtual infrared approximation terms are three-point functions at most.
The kinematic configuration is illustrated in fig.~(\ref{figure_virtual_momentum_flow}), with
$k_{i-1}=k_i+p_i$ and $k_{i+1}=k_i-p_k$.

Dimensionally regulated scalar loop integrals are invariant under Lorentz transformations and a shift of the loop momentum.
This applies to the integral over the virtual infrared approximation terms.
For the subtracted one-loop amplitude the loop momenta in the approximation terms has to match the appropriate loop
momenta in the one-loop amplitude. This is best achieved by decomposing the one-loop amplitude into primitive
one-loop amplitudes with a definite cyclic ordering of the external legs and by matching the loop momenta in the
approximation terms for each primitive amplitude \cite{Ellis:2008qc,Ellis:2011cr,Ita:2011ar,Badger:2012pg,Reuschle:2013qna,Schuster:2013aya}.
In adding the approximation terms back, we are in principle free to shift the loop momentum or to do a Lorentz transformation.
Thus, we may choose the relation between $k_i$ and $k$ to be
\bq
\label{poincare_invariance}
 k_i^\mu & = & \Lambda^\mu_{\;\;\nu} k^\nu + a^\mu.
\eq
We may use this freedom for a cancellation of the divergences
with the real emission part.

\subsection{The virtual ultraviolet approximation terms}

We briefly comment on the virtual ultraviolet approximation terms:
\bq
 d\sigma^{\mathrm{A}}_{\mathrm{V},\mathrm{UV}}
 & = &
 \frac{d^Dk}{(2\pi)^D} 
 \; 2 \; \mbox{Re}\; \left[ \left.{\mathcal A}^{(0)}\right.^\ast {\mathcal G}^{(1)}_{\mathrm{UV}} \right] d\phi_n.
\eq
The function ${\mathcal G}^{(1)}_{\mathrm{UV}}$ can be obtained from the Feynman diagrams for ${\mathcal A}^{(0)}$
by replacing in a Feynman diagram exactly one vertex or one propagator by the corresponding one-loop ultraviolet
approximation term, summing over all replacement possibilities and over all Feynman diagrams.
The basic approximation terms for vertices and propagators can be found in \cite{Becker:2010ng,Becker:2012aq}.
In practice, it is advantageous to compute ${\mathcal G}^{(1)}_{\mathrm{UV}}$ not from Feynman diagrams, but to
use recurrence relations \cite{Berends:1987me,Becker:2010ng,Becker:2012aq}.
The virtual ultraviolet approximation terms are of the form
\bq
\label{typical_UV_subtraction_repeat}
 I_r^{\mathrm{UV}} 
 & = & 
 \int \frac{d^Dk}{(2\pi)^D} \frac{P\left(\bar{k}\right) }{\left( \bar{k}^2 - \mu_{\mathrm{UV}}^2 + i \delta \right)^r},
\eq
with $\bar{k}=k-Q$, $Q$ an arbitrary vector and $\mu_{\mathrm{UV}}$ an arbitrary mass. 
The quantity $P(\bar{k})$ is a polynomial in $\bar{k}$.
Note that the integration in eq.~(\ref{typical_UV_subtraction_repeat}) corresponds to a simple tadpole integral.

\section{Renormalisation}
\label{sect:renormalisation}

In the $\overline{\mbox{MS}}$-scheme the relation between
the bare coupling $g_{\mathrm{bare}}$ and the renormalised coupling $g$
is given by
\bq
 g_{\mathrm{bare}} & = & Z_g S_\eps^{-\frac{1}{2}} \mu^\eps g.
\eq
The renormalisation constant $Z_g$ is given by
\bq
 Z_g & = & 1 + \frac{\alpha_s}{4\pi} \left( - \frac{\beta_0}{2} \right) \frac{1}{\eps} + {\cal O}(\alpha_s^2),
\eq
where $\alpha_s=g^2/(4\pi)$.
The scattering amplitudes are calculated from amputated Green functions.
Let us first consider in massless QCD 
an amplitude with $n_q$ external quarks, $n_{\bar q}$ external anti-quarks and $n_g$ external
gluons. We set $n=n_q+n_{\bar q}+n_g$.
Amplitudes with massive quarks will be discussed later.
The relation between the renormalised and the bare amplitude is given by
\bq
\label{LSZ}
 {\mathcal A}(p_1,...,p_n,g)
 & = & 
 \left(Z_2^{1/2} \right)^{n_q+n_{\bar q}} \left( Z_3^{1/2} \right)^{n_g}
 {\mathcal A}_{\mathrm{bare}}(p_1,...,p_n,g_{\mathrm{bare}}).
\eq
$Z_2$ is the quark field renormalisation constant and $Z_3$ is the gluon field renormalisation constant.
The Lehmann-Symanzik-Zimmermann (LSZ) 
reduction formula instructs us to take for the field renormalisation constants the residue
of the propagators at the pole.
In dimensional regularisation and for massless particles 
this residue is $1$ and in an analytic calculation 
it is sufficient to renormalise the coupling:
\bq
 {\mathcal A}(p_1,...,p_n,g)
 & = & 
 {\mathcal A}_{\mathrm{bare}}\left(p_1,...,p_n,Z_g S_\eps^{-\frac{1}{2}} \mu^\eps g\right).
\eq
However $Z_2=Z_3=1$ is due to a cancellation between ultraviolet and infrared divergences.
Keeping track of the ultraviolet or infrared origin of the $1/\eps$-poles one
finds in Feynman gauge
\bq
\label{def_Z2_Z3}
 Z_2 
 & = & 
 1 
 + \frac{\alpha_s}{4 \pi} C_F 
    \left( 
          \frac{1}{\eps_{\mathrm{IR}}} - \frac{1}{\eps_{\mathrm{UV}}}
    \right) 
 + {\mathcal O}(\alpha_s^2),
\nonumber \\
 Z_3 
 & = & 
 1 
 + \frac{\alpha_s}{4 \pi}
     \left( 2 C_A - \beta_0 \right) \left( \frac{1}{\eps_{\mathrm{IR}}} - \frac{1}{\eps_{\mathrm{UV}}} \right) 
 + {\mathcal O}(\alpha_s^2).
\eq
In order to unify the notation we will write in the following ${\mathcal Z}_{\;i}$ for the field renormalisation constants,
with the convention that ${\mathcal Z}_{\;i}=Z_2$ if particle $i$ is a massless quark and ${\mathcal Z}_{\;i}=Z_3$ if particle $i$ is a gluon.
We further write ${\mathcal Z}_{\;i}^{(1)}$ for the ${\mathcal O}(\alpha_s)$-term:
\bq
 {\mathcal Z}_{\;i} & = &
 1 + {\mathcal Z}_{\;i}^{(1)} 
 + {\mathcal O}(\alpha_s^2).
\eq
Thus
\bq
 {\mathcal Z}_{\;q}^{(1)} 
 = 
 \frac{\alpha_s}{4 \pi} C_F 
    \left( 
          \frac{1}{\eps_{\mathrm{IR}}} - \frac{1}{\eps_{\mathrm{UV}}}
    \right),
 & &
 {\mathcal Z}_{\;g}^{(1)} 
 = 
 \frac{\alpha_s}{4 \pi}
     \left( 2 C_A - \beta_0 \right) \left( \frac{1}{\eps_{\mathrm{IR}}} - \frac{1}{\eps_{\mathrm{UV}}} \right). 
\eq
In massless QCD we may write the ultraviolet counterterm as
\bq
\label{massless_UV_counterterm}
 {\mathcal A}^{(1)}_{\mathrm{CT}}
 & = &
 \left[
       - \frac{\alpha_s}{4\pi} \frac{(n-2)}{2} \frac{\beta_0}{\eps_{\mathrm{UV}}}
       - \frac{1}{2} \sum\limits_i \sum\limits_{k \neq i} \frac{{\bf T}_i \cdot {\bf T}_k}{{\bf T}_i^2} {\mathcal Z}_{\;i}^{(1)} 
 \right]{\mathcal A}^{(0)},
\eq
where we used colour conservation in the terms involving ${\mathcal Z}_{\;i}^{(1)}$.

Let us now turn to the massive case.
It is sufficient to consider the case of QCD amplitudes with one heavy flavour, 
the generalisation to several heavy flavours is straightforward.
There are a few modifications.
We have to take into account the heavy quark field renormalisation constant, which is given 
in conventional dimensional regularisation by
\bq
 Z_{2,Q} 
 & = & 
 1 
 + \frac{\alpha_s}{4\pi} C_F 
   \left( 
          -\frac{1}{\eps_{\mathrm{UV}}} - \frac{2}{\eps_{\mathrm{IR}}} - 4 + 3 \ln \frac{m^2}{\mu^2} 
   \right)
 + {\mathcal O}\left(\alpha_s^2\right).
\eq
We write ${\mathcal Z}_{\;i}=Z_{2,Q}$ if particle $i$ is a massive quark.
In this case we also set
\bq 
 {\mathcal Z}_{\;i}^{(1)}
 & = &
 \frac{\alpha_s}{4\pi} C_F 
 \left( 
  -\frac{1}{\eps_{\mathrm{UV}}} -\frac{2}{\eps_{\mathrm{IR}}} - 4 + 3 \ln \frac{m^2}{\mu^2} 
 \right).
\eq
Secondly, the mass of the heavy quark is renormalised.
For the heavy quark mass we have to choose a renormalisation scheme.
In the on-shell scheme the mass renormalisation constant is given 
in conventional dimensional regularisation by
\bq
\label{def_Z_m_onshell}
 Z_{m,\mathrm{on-shell}} 
 & = & 
 1 
 + \frac{\alpha_s}{4\pi} C_F 
   \left( -\frac{3}{\eps_{\mathrm{UV}}} - 4 + 3 \ln \frac{m^2}{\mu^2} \right)
 + {\mathcal O}\left(\alpha_s^2\right).
\eq
In the $\overline{\mathrm{MS}}$-scheme the mass renormalisation constant is simply given by
\bq
 Z_{m,\overline{\mathrm{MS}}} 
 & = & 
 1 
 + \frac{\alpha_s}{4\pi} C_F \left( -\frac{3}{\eps_{\mathrm{UV}}} \right)
 + {\mathcal O}\left(\alpha_s^2\right).
\eq
Again, we write $Z_{m,\mathrm{scheme}}^{(1)}$ for the ${\mathcal O}(\alpha_s)$-term of the mass
renormalisation constant.
In order to present the generalisation of eq.~(\ref{massless_UV_counterterm}) to the massive
case it is convenient to define
the quantity ${\mathcal B}^{(0)}(p_1,...,p_n,g,m)$ through
\bq
 {\mathcal A}^{(0)}\left(p_1,...,p_n,g,m+\delta m\right)
 =  
 {\mathcal A}^{(0)}\left(p_1,...,p_n,g,m\right)
 + \frac{\delta m}{m}
 {\mathcal B}^{(0)}\left(p_1,...,p_n,g,m \right)
 + 
 {\mathcal O}\left( (\delta m)^2 \right).
 \nonumber
\eq
Then
\bq
\label{massive_UV_counterterm}
 {\mathcal A}^{(1)}_{\mathrm{CT}}
 & = &
 \left[
       - \frac{\alpha_s}{4\pi} \frac{(n-2)}{2} \frac{\beta_0}{\eps_{\mathrm{UV}}}
       - \frac{1}{2} \sum\limits_i \sum\limits_{k \neq i} \frac{{\bf T}_i \cdot {\bf T}_k}{{\bf T}_i^2} {\mathcal Z}_{\;i}^{(1)} 
 \right]{\mathcal A}^{(0)}
 +
 Z_{m,\mathrm{scheme}}^{(1)} {\mathcal B}^{(0)}.
\eq
We may group the renormalisation constants into two groups, depending on whether or not they contain in addition to ultraviolet 
divergences also infrared divergences.
The field renormalisation constants belong to the first group, these contain infrared divergences.
The mass renormalisation constants and coupling renormalisation constants belong to the second group,
these do not contain infrared divergences.

We now introduce an integral representation for the counterterm from renormalisation.
It is convenient to separate $d\sigma^{\mathrm{V}}_{\mathrm{CT}}$ into two parts:
\bq
 d\sigma^{\mathrm{V}}_{\mathrm{CT}}
 & = &
 \int\limits_{\mathrm{loop}}
 \left(
 d\sigma^{\mathrm{V}}_{\mathrm{CT},\mathrm{IR}}
 +
 d\sigma^{\mathrm{V}}_{\mathrm{CT},\mathrm{UV}}
 \right).
\eq
This separation is done as follows:
$d\sigma^{\mathrm{V}}_{\mathrm{CT},\mathrm{UV}}$ contains for all renormalisation constants 
(field renormalisation, coupling renormalisation and mass renormalisation) the terms, 
which lead exactly to the $1/\eps_{\mathrm{UV}}$ divergences.
In addition, $d\sigma^{\mathrm{V}}_{\mathrm{CT},\mathrm{UV}}$ contains finite terms from coupling renormalisation and mass renormalisation,
if for these parameters a renormalisation scheme different from the $\overline{\mbox{MS}}$-scheme is used.
On the other hand, $d\sigma^{\mathrm{V}}_{\mathrm{CT},\mathrm{IR}}$ contains for the field renormalisation constants the terms,
which lead to the $1/\eps_{\mathrm{IR}}$ divergences or finite terms.
The splitting of the finite terms is of course arbitrary, but a convenient choice.
We may re-write $d\sigma^{\mathrm{V}}_{\mathrm{CT},\mathrm{IR}}$ as
\bq
 d\sigma^{\mathrm{V}}_{\mathrm{CT},\mathrm{IR}}
 & = &
 \left(
 \sum\limits_{i} \sum\limits_{k \neq i} {\mathcal F}_{i,k}
 \right) \frac{d^{D-1}k}{(2\pi)^{D-1}} \; d\phi_n,
\eq
with
\bq
\lefteqn{
{\mathcal F}_{i,k}
 =
 - 4 \pi \alpha_s S_\eps^{-1} \mu^{2\eps} 
} & &
 \\
 & &
 \mbox{Re}\; 
 {{\mathcal A}^{\xi \;(0)}}\left(...,p_i,...,p_k,...\right)^\ast
 \;\;
 \frac{{\bf T}_i \cdot {\bf T}_k}{{\bf T}_i^2}
 \left[ X_i\left(p_i,k_i\right) \right]_{\xi \xi'}
 \;\;
 {\mathcal A}^{\xi' \;(0)}\left(...,p_i,...,p_k,...\right).
 \nonumber 
\eq
The quantities $[X_i(p_i,k_i) ]_{\xi \xi'}$ are derived from the
self-energy corrections on the external legs.
However, there is a technical complication: The self-energy on an external leg
is attached through a propagator with momentum $p_i$ to the Born amplitude.
This propagator is exactly on-shell, leading to an $1/0$-singularity.
In order to circumvent this problem we follow refs.~\cite{Soper:1998ye,Soper:beowulf}
and we use a dispersion relation for the self-energy
corrections on the external legs.
The technical details are presented in appendix~\ref{sect:self_energies}.

The term $d\sigma^{\mathrm{V}}_{\mathrm{CT},\mathrm{UV}}$ contains all terms which lead
to ultraviolet divergences. An integral representation for these terms 
can be found in ref.~\cite{Becker:2010ng}.
In addition, $d\sigma^{\mathrm{V}}_{\mathrm{CT},\mathrm{UV}}$ contains by definition finite
terms from coupling renormalisation and mass renormalisation,
if for these parameters a renormalisation scheme different 
from the $\overline{\mbox{MS}}$-scheme is used.
The most relevant application would be the case of a massive quark, where the mass
is renormalised in the on-shell scheme.
We discuss the implementation of the finite terms in more detail in section~(\ref{section_pure_UV}).

\section{Factorisation}
\label{sect:factorisation}

In the $\overline{\mbox{MS}}$-scheme 
the collinear subtraction term is given by
\bq
\label{def_initial_coll_subtr}
 d\sigma^{\mathrm{C}}
 & = &
 \frac{\alpha_s}{4\pi} \frac{e^{\eps\gamma_E}}{\Gamma\left(1-\eps\right)}
 \;
 \sum\limits_{a' \;\mathrm{initial}}
 \;
 \sum\limits_{a \in \{q,g,\bar{q}\}} 
 \;
 \int\limits_0^1 dx_a 
 \;
  \frac{2}{\eps} \left( \frac{\mu_F^2}{\mu^2} \right)^{-\eps}
  P^{a'a}\left(x_a\right) d\sigma^{\mathrm{B}}\left(..., x_a p_a', ... \right).
\eq
The splitting functions $P^{a'a}(x)$ are given by
\bq
\label{def_Altarelli_Parisi}
 P^{gq}
 & = &
 \frac{n_c(q)}{n_c(g)} C_F \left[ x^2 + \left(1-x\right)^2 \right],
 \nonumber \\
 P^{qg}
 & = &
 \frac{n_c(g)}{n_c(q)} T_R \left[ \frac{1+\left(1-x\right)^2}{x} \right],
 \nonumber \\
 P^{qq}
 & = &
 C_F \left[ \left. \frac{2}{1-x} \right|_+ - \left(1+x\right) \right] + \frac{3}{2} C_F \delta\left(1-x\right),
 \nonumber \\
 P^{gg}
 & = &
 2 C_A \left[ \left. \frac{1}{1-x}\right|_+ + \frac{1-x}{x} - 1 + x \left(1-x\right) \right]
 + \frac{\beta_0}{2} \delta\left(1-x\right).
\eq
The splitting functions for anti-quarks are identical to the ones for quarks.
The splitting functions in eq.~(\ref{def_Altarelli_Parisi}) are the spin-averaged splitting functions.
We now look for an integral representation of the collinear subtraction term.
The sought after integral representation has to fulfill two conditions:
Firstly, it should match locally the singularities of the other contributions.
Secondly, it should integrate to produce exactly the same finite parts implied by eq.~(\ref{def_initial_coll_subtr}):
\bq
\label{coll_subtr_finite_part}
 d\sigma^{\mathrm{C}}
 = 
 \frac{\alpha_s}{4\pi}
 \;
 \sum\limits_{a' \;\mathrm{initial}}
 \;
 \sum\limits_{a \in \{q,g,\bar{q}\}} 
 \;
 \int\limits_0^1 dx_a 
 \;
 \left[ 
  \frac{1}{\eps} - \ln \left( \frac{\mu_F^2}{\mu^2} \right)
 \right]
  2 P^{a'a}\left(x_a\right) d\sigma^{\mathrm{B}}\left(..., x_a p_a', ... \right)
 + {\mathcal O}\left(\eps\right).
\eq
Let us discuss the first point in more detail: The singularities have to match
the corresponding singularities of the real approximation term and the counterterm from field renormalisation.
The spin-averaged case in eq.~(\ref{def_Altarelli_Parisi}) gives us some guidance:
The $x$-dependent terms in the square brackets will match with the real approximation terms,
while the end-point contributions proportional to $\gamma_q=3 C_F/2$ in $P^{qq}$ and $\gamma_g=\beta_0/2$ in $P^{gg}$
will match with the counterterm from field renormalisation.
Thus we write
\bq
 d\sigma^{\mathrm{C}}
 & = &
 d\sigma^{\mathrm{C}}_{\mathrm{R}}
 +
 d\sigma^{\mathrm{C}}_{\mathrm{CT}},
\eq
where $d\sigma^{\mathrm{C}}_{\mathrm{R}}$ matches with the real approximation term and
$d\sigma^{\mathrm{C}}_{\mathrm{CT}}$ matches with the counterterm from field renormalisation.
Between $d\sigma^{\mathrm{C}}_{\mathrm{R}}$ and $d\sigma_{\mathrm{R}}^{\mathrm{A}}$ the
collinear singularities cancel, the soft singularity in $d\sigma_{\mathrm{R}}^{\mathrm{A}}$
cancels with the virtual part $d\sigma_{\mathrm{V}}^{\mathrm{A}}$, the soft $1/(1-x)$-singularities
in $d\sigma^{\mathrm{C}}_{\mathrm{R}}$ are softened by the plus-distribution.
Between $d\sigma^{\mathrm{C}}_{\mathrm{CT}}$ and $d\sigma^{\mathrm{V}}_{\mathrm{CT}}$ there is a cancellation of
collinear singularities, where both collinear particles have transverse polarisations.
The self-energies contributing to $d\sigma^{\mathrm{V}}_{\mathrm{CT}}$ lead also to collinear singularities,
where one particle has a longitudinal polarisation.
These singularities cancel with the virtual approximation term.

For $d\sigma^{\mathrm{C}}_{\mathrm{R}}$ we make the ansatz
\bq
\label{def_H_cal}
 d\sigma^{\mathrm{C}}
 & = &
 \left(
 \sum\limits_{(a',j')} \sum\limits_{k' \neq j'} {\mathcal H}^{a'j'}_{k'}
 +
 \sum\limits_{(a',j')} \sum\limits_{b' \neq a'} {\mathcal H}^{a'j',b'}
 \right) d\phi_n \; d\phi_{\mathrm{unresolved}}.
\eq
As we would like to match locally the singularities we have to work with the spin-dependent splitting functions
(as opposed to the spin-averaged splitting functions appearing in eq.~(\ref{def_initial_coll_subtr})).
We may however sum over the polarisations of the unobserved particles $a'$ and $j'$.
In the following we drop the adjustment factors $n_c(g)/n_c(q)$ and $n_c(q)/n_c(g)$
appearing in eq.~(\ref{def_Altarelli_Parisi}) and adhere to the convention that the averaging for the
colour degrees of freedom is performed with respect to $a'$.
The same applies to the averaging with respect to the number of spin degrees of freedom for
initial-state particles.
When integrating eq.~(\ref{def_H_cal}), a factor $1/x$ from the unresolved measure is absorbed by the 
flux factor to produce the correct flux factor for the event with $n$ final-state particles.
The integral representation for ${\mathcal H}^{a'j'}_{k'}$ is given in section~(\ref{section_Y_initial_final}),
the one for ${\mathcal H}^{a'j',b'}$ is given in section~(\ref{section_Y_initial_initial}).

For $d\sigma^{\mathrm{C}}_{\mathrm{CT}}$ we write
\bq
 d\sigma^{\mathrm{C}}_{\mathrm{CT}}
 & = &
 \left(
 \sum\limits_{i\;\mathrm{initial}} \sum\limits_{k \neq i} {\mathcal K}_{i,k}
 \right) \frac{d^{D-1}k_i}{(2\pi)^{D-1} 2 k_i^0} \; d\phi_n,
\eq
with
\bq
\lefteqn{
{\mathcal K}_{i,k}
 =
 - 4 \pi \alpha_s S_\eps^{-1} \mu^{2\eps} 
} & &
 \\
 & &
 \mbox{Re}\; 
 {{\mathcal A}^{\xi \;(0)}}\left(...,p_i,...,p_k,...\right)^\ast
 \;\;
 \frac{{\bf T}_i \cdot {\bf T}_k}{{\bf T}_i^2}
 \left[ Z_i\left(p_i,k_i,p_k\right) \right]_{\xi \xi'}
 \;\;
 {\mathcal A}^{\xi' \;(0)}\left(...,p_i,...,p_k,...\right).
 \nonumber 
\eq
The integral representation for $Z_i\left(p_i,k_i,p_k\right)$ is given in section~(\ref{section_Z_initial}).

\subsection{Initial-state emitter and final-state spectator}
\label{section_Y_initial_final}

We first consider the case of an initial-state emitter and a final-state spectator.
The spectator may be massive ($m_k=m_k'$), all other particles are massless.
We use the variables $P=p_k'+p_j'-p_a'$ and
\bq
 x \; = \; \frac{2p_a'p_j'+2p_a'p_k'-2p_j'p_k'}{2p_a'p_j'+2p_a'p_k'},
 \;\;\;
 u \; = \; \frac{2p_a'p_j'}{2p_a'p_j'+2p_a'p_k'},
 \;\;\;
 w \; = \; \frac{2p_a'p_j' \left(2p_j'p_k' +m_k^2 \right)}{2p_j'p_k' \left(2p_a'p_j'+2p_a'p_k'\right)}.
\eq
If we further set
\bq
 x_0 & = & \frac{P^2}{P^2-m_k^2}, 
\eq
then the variables $u$ and $w$ are related by
\bq
u & = & \frac{1-x}{1-x_0x} w.
\eq
We write
\bq
\label{def_Y_initial_final}
\lefteqn{
{\mathcal H}^{a'j'}_{k'}
 =
 - 4 \pi \alpha_s S_\eps^{-1} \mu^{2\eps} 
} & &
 \\
 & &
 \left\{
 {{\mathcal A}^{\xi \;(0)}}\left(p_a,...,p_k,...\right)^\ast
 \;\;
 \frac{{\bf T}_a \cdot {\bf T}_k}{{\bf T}_a^2}
 \left[ Y^{a'j'}_{k'}\left(x,w\right) \right]_{\xi \xi'}
 \;\;
 {\mathcal A}^{\xi' \;(0)}\left(p_a,...,p_k,...\right)
 \right. \nonumber \\
 & & \left.
 - \delta\left(1-x\right)
 \int\limits_0^1 dy \;
 {{\mathcal A}^{\xi \;(0)}}\left(p_a,...,p_k,...\right)^\ast
 \;\;
 \frac{{\bf T}_a \cdot {\bf T}_k}{{\bf T}_a^2}
 \left[ Y^{a'j'}_{k',\mathrm{end}}\left(y,w\right) \right]_{\xi \xi'}
 \;\;
 {\mathcal A}^{\xi' \;(0)}\left(p_a,...,p_k,...\right)
 \right\}.
 \nonumber 
\eq
The relation between the set of momenta $\{p_a',p_j',p_k'\}$ and the set $\{p_a,p_k\}$
is as in section~(\ref{sect_dipole_initial_final}), in particular we have $p_a=x p_a'$.
The expression in eq.~(\ref{def_Y_initial_final}) is of the form as in eq.~(\ref{def_sing_form})
and can be implemented as in eq.~(\ref{impl_plus_distr}).
In order to present the functions $Y^{a'j'}_{k'}$ we factor out some common prefactors 
and we write
\bq
\label{def_Y_tilde_initial_final}
 Y^{a'j'}_{k'}
 & = &
 - 2
 {\bf T}^2_{a\rightarrow a' j'}
 \frac{1}{\left(-P^2\right)}
 \frac{x_0\left(1-x_0x\right)}{\left(1-x\right)}
 \frac{1}{w}
 \tilde{Y}^{a'j'}_{k'}.
\eq
Then
\bq
 \tilde{Y}^{a'_g j'_{\bar{q}}}_{k'} 
 & = & 
 p\!\!\!/_a
 \left\{
  \left[ 1- \eps - 2 x (1-x) \right]
  \left[ 1 - w \ln\left( \frac{\left(-P^2\right) \left(1-x\right)^2}{\mu_F^2 x x_0 \left(1-x_0x\right)} \right) \right]
  - w
 \right\}, 
 \nonumber \\
 \tilde{Y}^{a'_q j'_q}_{k'} 
 & = & 
 \left[ 
        -g^{\mu\nu} x 
        + 4 \frac{(1-x)}{x} \frac{u(1-u)}{2p_j'p_k'{}^\flat} S^{\mu\nu} 
 \right]
  \left[ 1 - w \ln\left( \frac{\left(-P^2\right) \left(1-x\right)^2}{\mu_F^2 x x_0 \left(1-x_0x\right)} \right) \right]
 - 2 g^{\mu\nu} \frac{1-x}{x} w,
 \nonumber \\
 \tilde{Y}^{a'_q j'_g}_{k'} 
 & = &
 p\!\!\!/_a
 \left\{
 \left[ \frac{2}{1-x} -(1+x) - \eps (1-x) \right]
  \left[ 1 - w \ln\left( \frac{\left(-P^2\right) \left(1-x\right)^2}{\mu_F^2 x x_0 \left(1-x_0x\right)} \right) \right]
 - \left(1-x\right) w
 \right\}, 
 \nonumber \\
 \tilde{Y}^{a'_g j'_g}_{k'} 
 & = & 
 2
 \left[ 
      -g^{\mu\nu} \left( \frac{1}{1-x} -1 + x(1-x) \right) 
      + 2 (1- \eps) \frac{(1-x)}{x} \frac{u(1-u)}{2p_j'p_k'{}^\flat} S^{\mu\nu}
 \right]
 \nonumber \\
 & & 
 \times
  \left[ 1 - w \ln\left( \frac{\left(-P^2\right) \left(1-x\right)^2}{\mu_F^2 x x_0 \left(1-x_0x\right)} \right) \right].
\eq
Here $p_k'{}^\flat$ is a light-like vector defined by
\bq
 p_k'{}^\flat & = & p_k' - \frac{m_k^2}{2p_a'p_k'} p_a'.
\eq
The spin correlation tensor is given by
\bq
S^{\mu\nu} & = & 
\left( \frac{1}{u} {p_j'}^\mu - \frac{1}{1-u} p_k'{}^\flat{}^\mu \right)
\left( \frac{1}{u} {p_j'}^\nu - \frac{1}{1-u} p_k'{}^\flat{}^\nu \right).
\eq
The terms proportional to $w$ ensure that the finite part is exactly as in eq.~(\ref{coll_subtr_finite_part}).
Factorisation schemes different from the $\overline{\mbox{MS}}$-scheme
can be implemented by a suitable modification of the finite terms.

The end-point contributions $Y^{a'j'}_{k',\mathrm{end}}$ are rather simple.
They are zero for flavour off-diagonal splittings:
\bq
 Y^{a'_g j'_{\bar{q}}}_{k',\mathrm{end}} 
 \; = \; 
 0,
 & &
 Y^{a'_q j'_q}_{k',\mathrm{end}} 
 \; = \; 
 0.
\eq
For flavour conserving splittings we write in analogy with eq.~(\ref{def_Y_tilde_initial_final})
\bq
 Y^{a'j'}_{k',\mathrm{end}}
 & = &
 - 2
 {\bf T}^2_{a\rightarrow a' j'}
 \frac{1}{\left(-P^2\right)}
 \frac{x_0\left(1-x_0x\right)}{\left(1-x\right)}
 \frac{1}{w}
 \tilde{Y}^{a'j'}_{k',\mathrm{end}}.
\eq
Then we have
\bq
 \tilde{Y}^{a'_q j'_g}_{k',\mathrm{end}} 
 & = &
 p\!\!\!/_a
 \frac{2}{1-x} 
  \left[ 1 - w \ln\left( \frac{\left(-P^2\right) \left(1-x\right)^2}{\mu_F^2 x x_0 \left(1-x_0x\right)} \right) \right],
 \nonumber \\
 \tilde{Y}^{a'_g j'_g}_{k',\mathrm{end}} 
 & = & 
 \left( -g^{\mu\nu} \right) 
 \frac{2}{1-x} 
  \left[ 1 - w \ln\left( \frac{\left(-P^2\right) \left(1-x\right)^2}{\mu_F^2 x x_0 \left(1-x_0x\right)} \right) \right].
\eq

\subsection{Initial-state emitter and initial-state spectator}
\label{section_Y_initial_initial}

We now consider the case of an initial-state emitter and an initial-state spectator.
We use the variables
\bq
 x \; = \; \frac{2p_a'p_b'-2p_a'p_j'-2p_b'p_j'}{2p_a'p_b'},
 \;\;\;
 v \; = \; \frac{2p_a'p_j'}{2p_a'p_b'},
 \;\;\;
 w \; = \; \frac{2p_a'p_j'}{2p_a'p_j'+2p_b'p_j'}.
\eq
The variables $v$ and $w$ are related by
\bq
 v & = & \left(1-x\right) w.
\eq
We write
\bq
\label{def_Y_initial_initial}
\lefteqn{
{\mathcal H}^{a'j',b'}
 =
 - 4 \pi \alpha_s S_\eps^{-1} \mu^{2\eps} 
} & &
 \nonumber \\
 & &
 \left\{
 {{\mathcal A}^{\xi \;(0)}}\left(p_a,p_b,...\right)^\ast
 \;\;
 \frac{{\bf T}_a \cdot {\bf T}_k}{{\bf T}_a^2}
 \left[ Y^{a'j',b'}\left(x,w\right) \right]_{\xi \xi'}
 \;\;
 {\mathcal A}^{\xi' \;(0)}\left(p_a,p_b,...\right)
 \right. \nonumber \\
 & & \left.
 - \delta\left(1-x\right)
 \int\limits_0^1 dy \;
 {{\mathcal A}^{\xi \;(0)}}\left(p_a,p_b,...\right)^\ast
 \;\;
 \frac{{\bf T}_a \cdot {\bf T}_k}{{\bf T}_a^2}
 \left[ Y^{a'j',b'}_{\mathrm{end}}\left(x,y\right) \right]_{\xi \xi'}
 \;\;
 {\mathcal A}^{\xi' \;(0)}\left(p_a,p_b,...\right)
 \right\}.
 \nonumber 
\eq
The relation between the set of momenta $\{p_a',p_b',p_j'\}$ and the set $\{p_a,p_b\}$
is as in section~(\ref{sect_dipole_initial_initial}), in particular we have $p_a=x p_a'$
and $p_b=p_b'$.
The expression in eq.~(\ref{def_Y_initial_initial}) is of the form as in eq.~(\ref{def_sing_form})
and can be implemented as in eq.~(\ref{impl_plus_distr}).
In order to present the functions $Y^{a'j',b'}$ we factor out some common prefactors 
and we write
\bq
\label{def_Y_tilde_initial_initial}
 Y^{a'j', b'}
 & = &
 - 2
 {\bf T}^2_{a\rightarrow a' j'}
 \frac{1}{2 p_a p_b}
 \frac{1}{\left(1-x\right)}
 \frac{1}{w}
 \tilde{Y}^{a'j', b'}.
\eq
Then
\bq
 \tilde{Y}^{a'_g j'_{\bar{q}}, b'} 
 & = &
 p\!\!\!/_a
 \left\{
 \left[ 1- \eps - 2 x (1-x) \right]
 \left[ 1 - w \ln\left( \frac{2 p_a p_b \left(1-x\right)^2}{\mu_F^2 x} \right) \right]
  - w
 \right\}, 
 \nonumber \\
 \tilde{Y}^{a'_q j'_q, b'}
 & = & 
 \left[ 
        -g^{\mu\nu} x 
        + 4 \frac{(1-x)}{x} \frac{2p_a'p_b'}{2p_j'p_a' \; 2p_j'p_b'} S^{\mu\nu} 
 \right]
 \left[ 1 - w \ln\left( \frac{2 p_a p_b \left(1-x\right)^2}{\mu_F^2 x} \right) \right]
 - 2 g^{\mu\nu} \frac{1-x}{x} w,
 \nonumber \\
 \tilde{Y}^{a'_q j'_g, b'} 
 & = & 
 p\!\!\!/_a
 \left\{
 \left[ \frac{2}{1-x} -(1+x) - \eps (1-x) \right]
 \left[ 1 - w \ln\left( \frac{2 p_a p_b \left(1-x\right)^2}{\mu_F^2 x} \right) \right]
 - \left(1-x\right) w
 \right\}, 
 \nonumber \\
 \tilde{Y}^{a'_g j'_g, b'} 
 & = & 
 2
 \left[ 
      -g^{\mu\nu} \left( \frac{1}{1-x} -1 + x(1-x) \right) 
      + 2 (1- \eps) \frac{(1-x)}{x} \frac{2p_a'p_b'}{2p_j'p_a' \; 2p_j'p_b'}  S^{\mu\nu}
 \right]
 \nonumber \\
 & &
 \times
 \left[ 1 - w \ln\left( \frac{2 p_a p_b \left(1-x\right)^2}{\mu_F^2 x} \right) \right].
\eq
The spin correlation tensor is given by
\bq
 S^{\mu\nu} 
 & = & 
 \left( p_j'{}^\mu - \frac{2p_j'p_a'}{2p_a'p_b'} p_b'{}^\mu \right)
 \left( p_j'{}^\nu - \frac{2p_j'p_a'}{2p_a'p_b'} p_b'{}^\nu \right).
\eq
The terms proportional to $w$ ensure that the finite part is exactly as in eq.~(\ref{coll_subtr_finite_part}).
Factorisation schemes different from the $\overline{\mbox{MS}}$-scheme
can be implemented by a suitable modification of the finite terms.

The end-point contributions $Y^{a'j',b'}_{\mathrm{end}}$ are again rather simple.
\bq
 Y^{a'_g j'_{\bar{q}},b'}_{\mathrm{end}} 
 \; = \; 
 0,
 & &
 Y^{a'_q j'_q,b'}_{\mathrm{end}} 
 \; = \; 
 0. 
\eq
For flavour conserving splittings we write in analogy with eq.~(\ref{def_Y_tilde_initial_initial})
\bq
 Y^{a'j', b'}_{\mathrm{end}}
 & = &
 - 2
 {\bf T}^2_{a\rightarrow a' j'}
 \frac{1}{2 p_a p_b}
 \frac{1}{\left(1-x\right)}
 \frac{1}{w}
 \tilde{Y}^{a'j', b'}_{\mathrm{end}}.
\eq
Then
\bq
 \tilde{Y}^{a'_q j'_g, b'}_{\mathrm{end}} 
 & = & 
 p\!\!\!/_a
 \frac{2}{1-x} 
 \left[ 1 - w \ln\left( \frac{2 p_a p_b \left(1-x\right)^2}{\mu_F^2 x} \right) \right],
 \nonumber \\
 \tilde{Y}^{a'_g j'_g, b'}_{\mathrm{end}} 
 & = & 
 \left( -g^{\mu\nu} \right) 
 \frac{2}{1-x} 
 \left[ 1 - w \ln\left( \frac{2 p_a p_b \left(1-x\right)^2}{\mu_F^2 x} \right) \right].
\eq

\subsection{The virtual end-point contributions}
\label{section_Z_initial}

We now consider $d\sigma^{\mathrm{C}}_{\mathrm{CT}}$, which we write as
\bq
 d\sigma^{\mathrm{C}}_{\mathrm{CT}}
 & = &
 \left(
 \sum\limits_{i\;\mathrm{initial}} \sum\limits_{k \neq i} {\mathcal K}_{i,k}
 \right) \frac{d^{D-1}k_i}{(2\pi)^{D-1} 2k_i^0} \; d\phi_n,
\eq
with
\bq
\lefteqn{
{\mathcal K}_{i,k}
 =
 - 4 \pi \alpha_s S_\eps^{-1} \mu^{2\eps} 
} & &
 \\
 & &
 \mbox{Re}\; 
 {{\mathcal A}^{\xi \;(0)}}\left(...,p_i,...,p_k,...\right)^\ast
 \;\;
 \frac{{\bf T}_i \cdot {\bf T}_k}{{\bf T}_i^2}
 \left[ Z_i\left(p_i,k_i,p_k\right) \right]_{\xi \xi'}
 \;\;
 {\mathcal A}^{\xi' \;(0)}\left(...,p_i,...,p_k,...\right).
 \nonumber 
\eq
The particle $i$ is an initial-state particle and we 
write $p_a=-p_i$ such that $p_a$ has positive energy.
Particle $k$ is the spectator. The spectator can either be in the final-state 
(in which case it can be massive or massless)
or in the initial-state (in which case it is assumed to be always massless).
We will treat all cases simultaneously. 
To this aim we first set
\bq
 p_k^\flat & = & p_k - \frac{p_k^2}{2p_k p_a} p_a.
\eq
$p_k^\flat$ is always a massless momentum. We further define
$l_k=p_k^\flat$, if particle $k$ is in the final state, and $l_k=-p_k^\flat=-p_k=p_b$ if particle $k$ is in the inital-state.
The definition is such that $l_k$ is always a massless momentum with positiv energy.
$d\sigma^{\mathrm{C}}_{\mathrm{CT}}$ has to match the collinear singularities of the self-energy corrections.
These occur when the two propagators in the self-energy loop are on-shell.
We define $l_i'$ and $l_j'$ as the on-shell momenta in the self-energy loop flowing 
in the direction of the hard-scattering process.
In the singular collinear limit both $l_i'$ and $l_j'$ have positive energies.
\begin{figure}
\begin{center}
\includegraphics[scale=0.8]{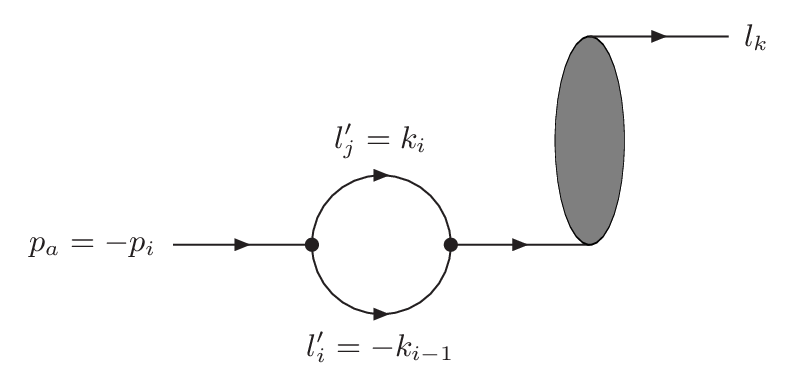}
\caption{\label{figure_self_energy_initial}
The kinematics for self-energy corrections for initial-state particles.
In the collinear limit the momenta $l_i'$ and $l_j'$ are on-shell.
The momenta $l_i'$, $l_j'$ and $l_k$ have positive energy.
}
\end{center}
\end{figure}
The kinematical situation is shown in fig.~(\ref{figure_self_energy_initial}).
Given $p_a$, $l_k$ and $k_i$ we define $l_i'$, $l_j'$ and $l_k'$ by
\bq
\label{mapping_self_energy}
 l_i' & = & p_a - k_i + y l_k,
 \nonumber \\
 l_j' & = & k_i,
 \nonumber \\
 l_k' & = & \left(1-y\right) l_k,
\eq
with 
\bq
\label{def_y_v1}
 y & = & -\frac{\left(p_a-k_i\right)^2}{2 l_k \left(p_a-k_i\right)}.
\eq
We will encounter the mapping in eq.~(\ref{mapping_self_energy}) again
in section~(\ref{sect_final_final_antenna}), where it will be used to relate in the final-final case the virtual approximation
terms to the real approximation terms.
With the definition $l_i=p_a$, the inverse mapping $\{l_i',l_j',l_k'\} \rightarrow \{l_i,l_k,k_i\}$
is just -- when restricted to $\{l_i,l_k\}$ -- the standard Catani-Seymour projection
of eq.~(\ref{map_massless}).
The reason why this mapping is useful for the self-energies related to initial-state particles is
as follows: For a collinear singularity the energy flow across the cut of the self-energy diagrams
has to be in the same direction for both cut propagators.
The momentum $l_k'$ will only be used to define the way the collinear singularity is approached.
Given $l_i'$, $l_j'$ and $l_k'$ we set
\bq
\label{def_y_v2}
 y \;\; = \;\; 
 \frac{2l_i'l_j'}{2l_i'l_j' + 2 l_i' l_k' + 2 l_j' l_k'},
 & &
 z \;\; = \;\;
 \frac{2 l_i' l_k'}{2l_i' l_k' + 2 l_j' l_k'}.
\eq
It is easily checked that the two expressions for the variable $y$ in eq.~(\ref{def_y_v1}) and eq.~(\ref{def_y_v2})
are compatible.
We further set $P=l_i'+l_j'+l_k'=p_a+l_k$.
If the initial-state particle is a quark we have
\bq
\lefteqn{
 Z_i\left(p_i,k_i,p_k\right)_{\alpha\beta}
 = } & & \\
 & = &
 \frac{2 C_F p\!\!\!/_a}{y z P^2} 
 \left\{ \left[ - \left(1+z\right) - \eps \left(1-z\right)\right]
         \left[ 1 - y \ln\left(\frac{P^2 z \left(1-z\right)}{\mu_F^2} \right) \right]
 - \left(1-z\right) y
 \right\}
 \theta\left(E_i'\right) \theta\left(E_k'\right),
 \nonumber 
\eq
in the case where the initial-state particle is a gluon we have
\bq
\lefteqn{
 Z_i\left(p_i,k_i,p_k\right)_{\mu\nu}
 = } & &
 \\
 & = &
 \frac{2 C_A}{y z P^2} 
 \left[ 2 g_{\mu\nu} 
            + \frac{2 \left(1-\eps\right)S_{\mu\nu}}{2 l_i' l_j'}  
     \right]
 \left[ 1 - y \ln\left(\frac{P^2 z \left(1-z\right)}{\mu_F^2} \right) \right]
 \theta\left(E_i'\right) \theta\left(E_k'\right)
 \nonumber \\
 & & 
 + \frac{2 T_R N_f}{y z P^2} 
 \left\{ 
  \left[ - g_{\mu\nu} - \frac{4 S_{\mu\nu} }{2l_i'l_j'} \right] 
  \left[ 1 - y \ln\left(\frac{P^2 z \left(1-z\right)}{\mu_F^2} \right) \right]
  + 2 g_{\mu\nu} z \left(1-z\right) y
 \right\}
 \theta\left(E_i'\right) \theta\left(E_k'\right),
 \nonumber 
\eq
where the spin correlation tensor is given by
\bq
 S^{\mu\nu} & = &
 \left( z l_i'{}^\mu - \left(1-z\right) l_j'{}^\mu \right)
 \left( z l_i'{}^\nu - \left(1-z\right) l_j'{}^\nu \right).
 \nonumber
\eq
It is easily checked that the integrated expression gives
\bq
\lefteqn{
 \int \frac{d^{D-1}k_i}{(2\pi)^{D-1} 2k_i^0} \;
 {\mathcal K}_{i,k}
 = } & & \\
 & &
 - \frac{\alpha_s}{4\pi}
 \frac{2}{\eps} \left( \frac{\mu_F^2}{\mu^2} \right)^{-\eps}
 \gamma_i 
 \;
 {{\mathcal A}^{\xi \;(0)}}\left(...,p_i,...,p_k,...\right)^\ast
 \;\;
 \frac{{\bf T}_i \cdot {\bf T}_k}{{\bf T}_i^2}
 \;\;
 {\mathcal A}^{\xi' \;(0)}\left(...,p_i,...,p_k,...\right)
 + {\mathcal O}\left(\eps\right).
 \nonumber
\eq

\section{Locally integrable combinations}
\label{sect:cancellations}

Our aim is to combine the approximation terms such that they are locally integrable.
In order to achieve this, it is essential to take the field renormalisation constants into account.
The local cancellation of singularities occurs separately for infrared and ultraviolet divergences.
For massless particles the $\alpha_s$-contribution of the field renormalisation constants
is zero after the loop integration.
This does not imply that the integrand is identical to zero, it only implies
that the integrand is a function with possibly ultraviolet and infrared singularities, which
integrates to zero within dimensional regularisation.

Other manifestations, that the contribution from the field renormalisation constants are needed
are:

- The real approximation terms contain a divergent
contribution from the splitting $g \rightarrow q \bar{q}$
of a gluon into massless quarks.
The virtual approximation terms have no such contribution. 
The divergent part from the real approximation terms cancels with the contribution from
the field renormalisation constants.

- In the collinear part of the real approximation terms all unresolved particles have transverse
polarisations.
In the collinear part of the virtual approximation terms one of the two collinear particles
has a longitudinal polarisation.
These two contributions do not match. Again, the cancellation occurs through
the contribution from the field renormalisation constants: 
The longitudinal part from the virtual approximation terms cancels with the longitudinal part
from the field renormalisation constants, the transverse part from the real approximation terms
cancels with the transverse part from the field renormalisation constants.

- It is instructive to look at the explicit poles in $\eps$ of infrared origin in massless QCD.
After integration one has for the various contributions
\bq
 d\sigma_{\mathrm{R}}^{\mathrm{A}}
 & = &
 2 \; \mbox{Re}\; \left\{ \left.{\mathcal A}^{(0)}\right.^\ast 
 \frac{\alpha_s}{4\pi}
 \sum\limits_{i} \sum\limits_{k \neq i}
 {\bf T}_i {\bf T}_k
 \left[ 
        - \frac{1}{\eps_{\mathrm{IR}}^2} \left( \frac{\left|2p_ip_k\right|}{\mu^2} \right)^{-\eps}
        - \frac{\gamma_i}{{\bf T}_i^2} \frac{1}{\eps_{\mathrm{IR}}} 
 \right]
 {\mathcal A}^{(0)}
 \right\}
 d\phi_n
 + ...,
 \nonumber \\
 d\sigma_{\mathrm{V},\mathrm{IR}}^{\mathrm{A}}
 & = &
 2 \; \mbox{Re}\; \left\{ \left.{\mathcal A}^{(0)}\right.^\ast 
 \frac{\alpha_s}{4\pi}
 \sum\limits_{i} \sum\limits_{k \neq i}
 {\bf T}_i {\bf T}_k
 \left[ \frac{1}{\eps_{\mathrm{IR}}^2}  \left( \frac{-2p_ip_k}{\mu^2} \right)^{-\eps}
        + \frac{2 S_i}{\eps_{\mathrm{IR}}}
 \right]
 {\mathcal A}^{(0)}
 \right\}
 d\phi_n
 + ...,
 \nonumber \\
 d\sigma_{\mathrm{CT},\mathrm{IR}}^{\mathrm{V}}
 & = &
 2 \; \mbox{Re}\; \left\{ \left.{\mathcal A}^{(0)}\right.^\ast 
 \frac{\alpha_s}{4\pi}
 \sum\limits_{i} \sum\limits_{k \neq i}
 {\bf T}_i {\bf T}_k
 \left[ \left( - 2 S_i + \frac{\gamma_i}{{\bf T}_i^2} \right) \frac{1}{\eps_{\mathrm{IR}}}
 \right]
 {\mathcal A}^{(0)}
 \right\}
 d\phi_n
 + ...,
\eq
where the dots denote ultraviolet poles and terms of order ${\mathcal O}(\eps^0)$.
The infrared poles cancel in the sum of the three contributions.
However, there is not a complete cancellation between $d\sigma_{\mathrm{R}}^{\mathrm{A}}$
and $d\sigma_{\mathrm{V},\mathrm{IR}}^{\mathrm{A}}$ alone.

We would like to evaluate numerically the expression of eq.~(\ref{integrated_subtraction_terms}):
\bq
 \langle O \rangle^{\mathrm{NLO}}_{{\bf I}+{\bf L}} 
 & = & 
 \int\limits_{n} 
 \left[ 
        O_{n} d\sigma^{\mathrm{C}} 
      + O_{n} \int\limits_{1} d\sigma^{\mathrm{A}}_{\mathrm{R}} 
      + O_{n} \int\limits_{\mathrm{loop}} d\sigma^{\mathrm{A}}_{\mathrm{V}} 
      + O_{n} d\sigma_{\mathrm{CT}}^{\mathrm{V}} 
 \right].
\eq
We split this expression into two parts
\bq
 \langle O \rangle^{\mathrm{NLO}}_{{\bf I}+{\bf L}} 
 & = & 
 \langle O \rangle^{\mathrm{NLO}}_{{\bf I}+{\bf L},\mathrm{IR}} 
 +
 \langle O \rangle^{\mathrm{NLO}}_{{\bf I}+{\bf L},\mathrm{UV}}, 
\eq
with
\bq
\label{separation_pure_UV_rest}
 \langle O \rangle^{\mathrm{NLO}}_{{\bf I}+{\bf L},\mathrm{IR}} 
 & = &
 \int\limits_{n} 
 O_{n} 
 \left[ 
        d\sigma^{\mathrm{C}} 
      + \int\limits_{1} d\sigma^{\mathrm{A}}_{\mathrm{R}} 
      + \int\limits_{\mathrm{loop}} d\sigma^{\mathrm{A}}_{\mathrm{V},\mathrm{IR}} 
      + \int\limits_{\mathrm{loop}} d\sigma_{\mathrm{CT},\mathrm{IR}}^{\mathrm{V}} 
 \right],
 \nonumber \\
 \langle O \rangle^{\mathrm{NLO}}_{{\bf I}+{\bf L},\mathrm{UV}}
 & = &
 \int\limits_{n} 
 O_{n} 
      \int\limits_{\mathrm{loop}} \left( d\sigma^{\mathrm{A}}_{\mathrm{V},\mathrm{UV}}
      + d\sigma_{\mathrm{CT},\mathrm{UV}}^{\mathrm{V}}
 \right).
\eq
The two contributions in eq.~(\ref{separation_pure_UV_rest}) are separately numerically
integrable.
We may break up the term $\langle O \rangle^{\mathrm{NLO}}_{{\bf I}+{\bf L},\mathrm{IR}}$
into even smaller pieces, where an individual piece corresponds to an antenna and
is separately numerically integrable.
This is discussed in section~(\ref{section_antenna}).
The term $\langle O \rangle^{\mathrm{NLO}}_{{\bf I}+{\bf L},\mathrm{UV}}$ is discussed in section~(\ref{section_pure_UV}).

\subsection{The antenna structure}
\label{section_antenna}

In this sub-section we consider
\bq
\label{I_plus_L_IR}
 \langle O \rangle^{\mathrm{NLO}}_{{\bf I}+{\bf L},\mathrm{IR}}.
\eq
All terms contributing to eq.~(\ref{I_plus_L_IR}) can be written as colour dipoles, i.e. they are of the
form
\bq
 - \sum\limits_{i} \sum\limits_{k \neq i} {\bf T}_i {\bf T}_k ...,
\eq
where $i$ denotes the emitter and $k$ denotes the spectator.
We combine the colour dipole with emitter $i$ and spectator $k$
with the colour dipole with emitter $k$ and spectator $i$.
This forms a colour antenna with the hard particles $i$ and $k$ \cite{Kosower:1998zr,Campbell:1998nn,Gehrmann-DeRidder:2005cm,Daleo:2006xa} and we may write 
eq.~(\ref{I_plus_L_IR}) as
\bq
 \langle O \rangle^{\mathrm{NLO}}_{{\bf I}+{\bf L},\mathrm{IR}}
 & = &
 - \sum\limits_{i < k} {\bf T}_i {\bf T}_k
 \;
 \langle O \rangle^{\mathrm{NLO},i,k}_{{\bf I}+{\bf L},\mathrm{IR}}.
\eq
Each antenna contribution is separately numerically integrable.
We have to consider three types of antenna structures.
The two hard particles can either be both in the final-state, 
of mixed type (one in the final-state and the other in the initial-state)
or both in the initial-state.

Let us consider the contribution of $d\sigma^{\mathrm{A}}_{\mathrm{V},\mathrm{IR}}$ to a given antenna, i.e.
a contribution of the form
\bq
\lefteqn{
 {\mathcal E}_{i,k} + {\mathcal E}_{k,i}
 = 
  - 4 \pi \alpha_s S_\eps^{-1} \mu^{2\eps} 
} & &  \\
 & &
 2 \;\mbox{Re}\; 
 {{\mathcal A}^{(0)}}\left(...,p_i,...,p_k,...\right)^\ast
 \;
 {\bf T}_i \cdot {\bf T}_k
 \;
 \left[ \frac{W_i\left(p_i,p_k,k_i\right)}{{\bf T}_i^2} + \frac{W_k\left(p_k,p_i,k_k\right)}{{\bf T}_k^2} \right]
 \;
 {\mathcal A}^{(0)}\left(...,p_i,...,p_k,...\right).
 \nonumber
\eq
The loop integrals are three-point functions (where lower-point functions are considered as three-point functions with
appropriate inverse propagators in the numerator).
\begin{figure}
\begin{center}
\includegraphics[scale=0.8]{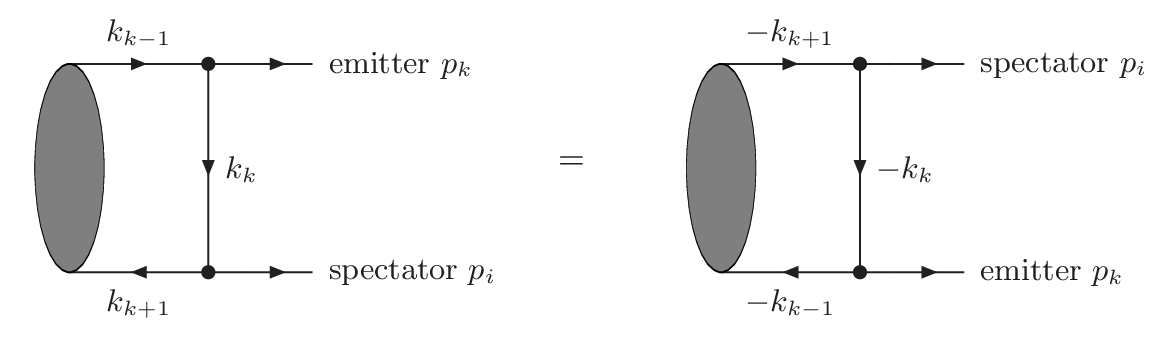}
\caption{\label{figure_virtual_emitter_spectator}
The momentum flow for the virtual infrared approximation terms with emitter $p_k$ and spectator $p_i$.
A comparison with fig.~(\ref{figure_virtual_momentum_flow}) shows $k_{i-1}=-k_{k+1}$, $k_i=-k_k$ and $k_{i+1}=-k_{k-1}$.}
\end{center}
\end{figure}
The momenta flowing through the loop propagators are given for ${\mathcal E}_{i,k}$ by 
\bq
\label{virtual_momenta_i_k}
 k_{i-1} \;\; = \;\; k_i + p_i,
 & &
 k_{i+1} \;\; = \;\; k_i - p_k.
\eq
and shown in fig.~(\ref{figure_virtual_momentum_flow}).
For ${\mathcal E}_{k,i}$ the momenta are given by
\bq
\label{virtual_momenta_k_i}
 k_{k-1} \;\; = \;\; k_k + p_k,
 & &
 k_{k+1} \;\; = \;\; k_k - p_i
\eq
and shown in fig.~(\ref{figure_virtual_emitter_spectator}).
We may use the freedom of Poincar\'e-invariance of the loop integrals of eq.~(\ref{poincare_invariance}) and set
\bq
\label{synchro_E_i_k_and_E_k_i}
 k_i & = & - k_k.
\eq
This implies 
\bq
 k_{i-1} \;\; = \;\; -k_{k+1},
 & &
 k_{i+1} \;\; = \;\; -k_{k-1}.
\eq
Eq.~(\ref{synchro_E_i_k_and_E_k_i}) defines how ${\mathcal E}_{i,k}$ and ${\mathcal E}_{k,i}$ are integrated together.

Our general strategy is as follows: We will write all integrals as integrals
over the spatial components of a momentum:
\bq
 \int \frac{d^{D-1}k}{(2\pi)^{D-1}} ... .
\eq
For the virtual integrals this can be done using the loop-tree duality method.
The loop-tree duality method will put one of the loop propagators on-shell.
The task is to find suitable mappings between the various contributions, such that all
singularities cancel locally in $\vec{k}$-space and the limit $D\rightarrow 4$ can be taken.
This will leave us with a three-dimensional integral, which can be performed numerically.
We will also need the associated Jacobian factors for the various mappings.
The essential ingredient is a mapping between the virtual and real configurations.
Let us denote by $\{p\}$ a set of $(n+2)$ external momenta (including the two initial-state momenta), 
by $\{k\}$ the single-element set of the on-shell loop momentum
and by $\{p'\}$ a set of $(n+3)$ external momenta.
In the next sub-section we will define an invertible  mapping
\bq
 \phi & : &
 \left\{ p \right\} \times \left\{ k \right\} 
 \;\; \rightarrow \;\; 
 \left\{ p' \right\},
\eq
such that the inverse mapping, when restricted to
\bq
 \pi \phi^{-1} & : &
 \left\{ p' \right\}
 \;\; \rightarrow \;\; 
 \left\{ p \right\}
\eq
agrees with the Catani-Seymour projections given in section~(\ref{sect:real_subtraction_terms}),
relating the $(n+1)$-particle phase space to the $n$-particle phase space.
We will use this mapping on the pre-image $\phi^{-1}(\{ p' \})$ to associate 
a real configuration $\{p'\}$
to a virtual configuration specified by $\{p\}$ and $k$.
Note that there are points in $\{ p \} \times \{ k \}$, which do not map to physical points $\{p'\}$.
A typical example would be a loop momentum $k$ in the ultraviolet region, leading to a configuration $\{\tilde{p}'\}$
with final-state particles of negative energy.
This explains the restriction on the pre-image $\phi^{-1}(\{ p' \})$.
In practice, the correct physical region will be implemented by theta-functions.

We will discuss the mappings for the
three cases corresponding to a final-final antenna, a final-initial antenna and an initial-initial
antenny separately in the next three sub-sections.

\subsubsection{Final-final antenna}
\label{sect_final_final_antenna}

We consider the case, where $p_i$ and $p_k$ are final-state momenta, e.g. have positive energy components.
Let us first consider the dipole with emitter $i$ and spectator $k$.
With the kinematics as in fig.~(\ref{figure_virtual_momentum_flow}) we would like to have 
that in the collinear limit the momentum $(-k_i)$ has positive energy as well.
Turned around this means that the momentum $k_i$ has negative energy in the collinear limit.
Thus, we use the loop-tree duality formula for the backward hyperboloids of eq.~(\ref{loop_tree_backward})
to convert the loop integrals into phase space integrals.
This gives three backward hyperboloids with origins at $q_{i-1}=q_i-p_i$, $q_i$ and $q_{i+1}=q_i+p_k$, 
plus an extra backward hyperboloid
with origin at $Q$ corresponding to ultraviolet subtraction terms.
\begin{figure}
\begin{center}
\includegraphics[scale=0.8]{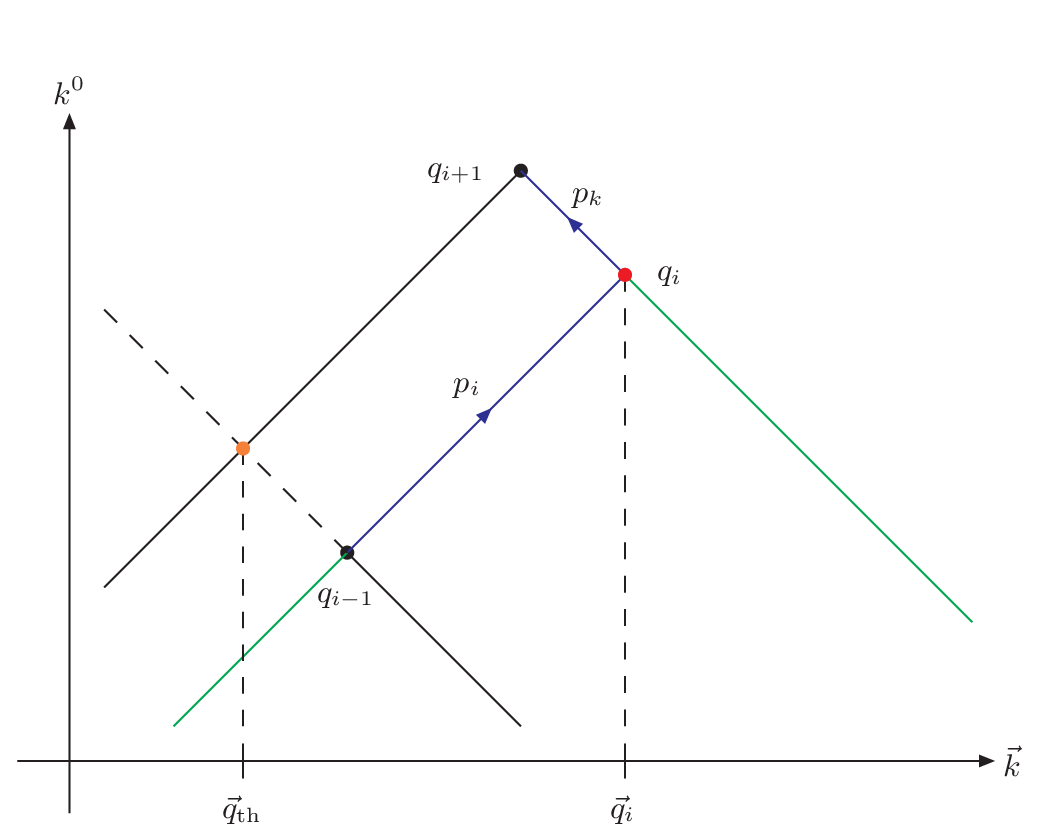}
\\
\includegraphics[scale=0.8]{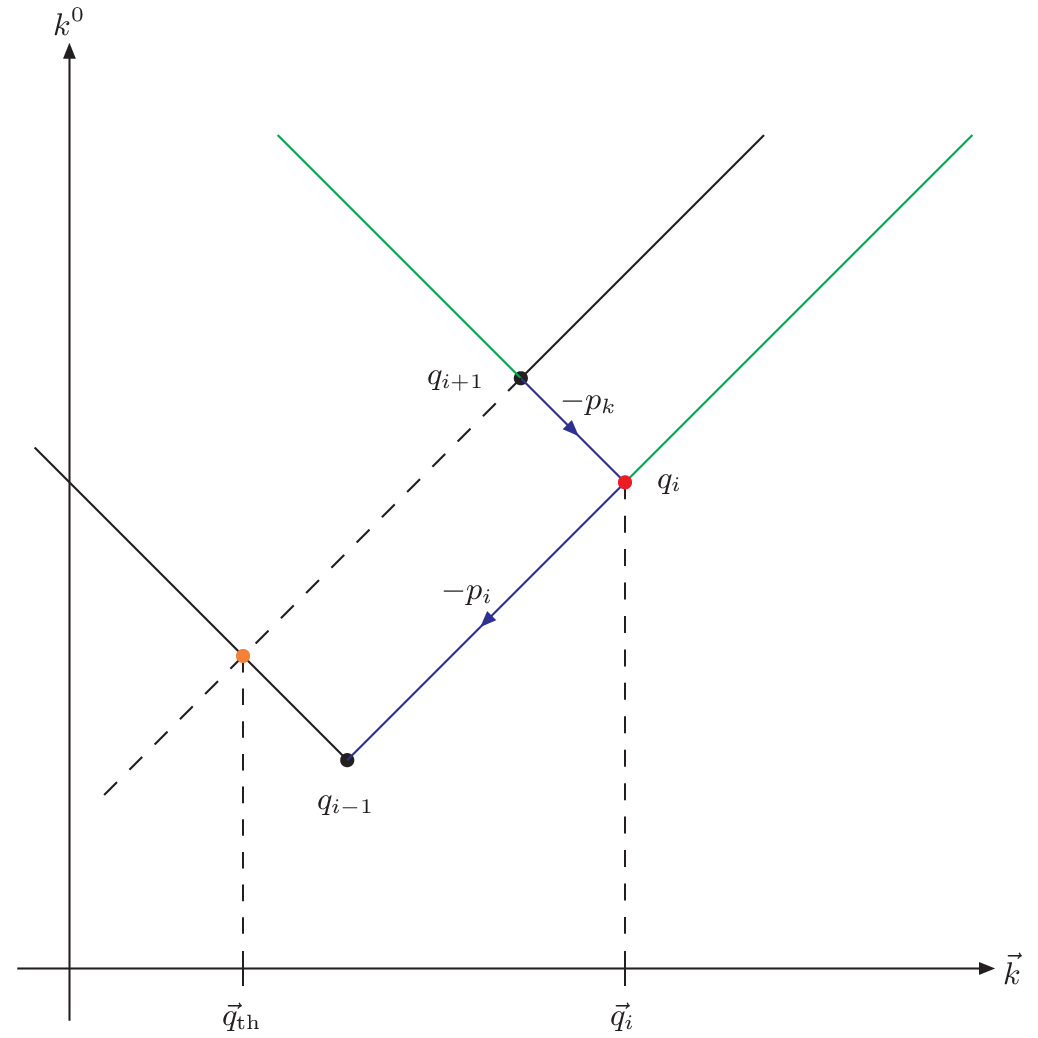}
\caption{\label{figure_integration_cones_ff}
The integration regions for a final-final antenna.
The upper picture corresponds to the dipole with emitter $i$ and spectator $k$,
the lower picture corresponds to the dipole with emitter $k$ and spectator $i$.
}
\end{center}
\end{figure}
The latter is free of infrared singularities.
In order to combine the real approximation terms with the virtual approximation terms we define
a mapping between the set of momenta $\{p_i,p_k,k_i\}$ and $\{p_i',p_j',p_k'\}$.
In the massless case we use
\bq
\label{map_virtual_to_real_massless}
 p_i' & = & p_i + k_i + y p_k,
 \nonumber \\
 p_j' & = & - k_i,
 \nonumber \\
 p_k' & = & \left(1-y\right) p_k,
\eq
with
\bq
 y & = & - \frac{\left(k_i+p_i\right)^2}{2 p_k \left(k_i+p_i\right)}
 \;\; = \;\;
 - \frac{k_{i-1}^2}{2p_k k_{i-1}}.
\eq
Note that the inverse mapping $\{p_i',p_j',p_k'\} \rightarrow \{p_i,p_k,k_i\}$ coincides with the mapping in eq.~(\ref{map_massless})
when restricted to $\{p_i',p_j',p_k'\} \rightarrow \{p_i,p_k\}$.

In the massive case we use
\bq
\label{map_virtual_to_real_massive}
 p_k' & = & \alpha \left( p_k - \frac{Q\cdot p_k}{Q^2} Q \right) + \beta Q,
 \nonumber \\
 p_j' & = & - k_i,
 \nonumber \\
 p_i' & = & Q + k_i - p_k',
\eq
with $Q=p_i+p_k$.
The constants $\alpha$ and $\beta$ are given in appendix~\ref{appendix:virtual_to_real}.
Again, this mapping can be considered to be the inverse of eq.~(\ref{map_massive}) together with supplementary information $p_j'=-k_i$.
Eq.~(\ref{map_virtual_to_real_massive}), which includes in the massless limit eq.~(\ref{map_virtual_to_real_massless}),
defines how the contributions ${\mathcal E}_{i,k}$ and ${\mathcal D}_{i'j',k'}$ are integrated together.

Writing the measure for the unresolved phase space as an integration over $\vec{k_i}$ (or the forward mass hyperboloid for particle $j'$)
introduces a Jacobian factor:
\bq
 d\phi_{\mathrm{unres}}
 & = & 
 \frac{d^{D-1}k_i}{\left(2 \pi\right)^{D-1} \left(-2 k_{i,0}\right)} J,
\eq
with
\bq
 J & = & 
 \frac{\left[\lambda\left(Q^2,m_i^2,m_k^2\right)\right]^{\frac{3-D}{2}} \left[ \lambda\left(Q^2,\left(p_i'+p_j'\right)^2,m_k'{}^2\right) \right]^{\frac{D-1}{2}}}
      { 2 p_i' p_k' \left( 2p_i'p_k' + 2p_j'p_k'\right) - 2 m_k'{}^2 \left( 2p_i'p_j' + 2 m_i'{}^2\right)}
 \theta\left(E_i'\right) \theta\left(E_k'\right).
\eq
Let us now consider the dipole with emitter $k$ and spectator $i$.
With the kinematics as in fig.~(\ref{figure_virtual_emitter_spectator}) we would like to have 
that in the collinear limit the momentum $(-k_k)$ has positive energy.
Since $k_i=-k_k$ this implies that the momentum $k_i$ has positive energy in the collinear limit.
Thus, we use the loop-tree duality formula for the forward hyperboloids of eq.~(\ref{loop_tree_forward})
to convert the loop integrals into phase space integrals.
In the next step we have to relate the real emission integrals to the virtual integrals.
This is straightforward. We may use the same mappings as in eq.~(\ref{map_virtual_to_real_massive}) and
eq.~(\ref{map_virtual_to_real_massless}) with the roles of $i$ and $k$ exchanged.
Thus, the integrations for ${\mathcal E}_{k,i}$ and ${\mathcal D}_{k'j',i'}$ are related in the same way as the
integrations for ${\mathcal E}_{i,k}$ and ${\mathcal D}_{i'j',k'}$.
Taking into account the relation between ${\mathcal E}_{i,k}$ and ${\mathcal E}_{k,i}$ in eq.~(\ref{synchro_E_i_k_and_E_k_i}), we may relate the integration for ${\mathcal D}_{k'j',i'}$ to ${\mathcal E}_{i,k}$ and obtain in the massless case
\bq
 p_i' & = & \left(1-y\right) p_i,
 \nonumber \\
 p_j' & = & k_i,
 \nonumber \\
 p_k' & = & p_k - k_i + y p_i,
\eq
with
\bq
 y & = & 
 \frac{k_{i+1}^2}{2p_i k_{i+1}}.
\eq
The geometric situation for the integration over the on-shell hyperboloids 
is sketched in fig.~(\ref{figure_integration_cones_ff}).
The upper picture shows the contribution from the virtual approximation terms with emitter $i$ and spectator $k$
in the massless case.
The integration is over three backward light-cones with origins at $q_{i-1}$, $q_i$ and $q_{i+1}$.
The soft singularity resides in the integration over the backward light-cone with origin at $q_i$ at the origin $q_i$
and is indicated by a red dot.
The collinear singularities occur on the lines between $q_{i-1}$ and $q_i$ (collinear singularity of $i$) and
between $q_i$ and $q_{i+1}$ (collinear singularity of $k$).
The collinear regions are indicated in blue.
There is a cancellation of singularities within the virtual dual contributions in the regions where two propagators are
on-shell and have the same sign in the energy component. These regions are indicated in green.
There is a threshold singularity (indicated by an orange dot) at $\vec{q}_{\mathrm{th}}$. 
The threshold singularity is avoided by contour deformation.
The integration region for the real approximation term is the backward light-cone with origin at $q_i$.
The collinear singular region for the real approximation term with emitter $i$ is the line segment between $q_{i-1}$ and $q_i$.

The lower picture shows the corresponding integration regions, where the roles of emitter and spectator are exchanged, i.e.
emitter $k$ and spectator $i$.
Note that the soft and collinear singularities occur in the same regions of $D$-dimensional loop momentum space.
The integration region for the real approximation term is the forward light-cone with origin at $q_i$.
The collinear singular region for the real approximation term with emitter $k$ is the line segment between $q_{i}$ and $q_{i+1}$.

\subsubsection{Final-initial antenna}

Let us now consider a final-initial antenna.
Without loss of generality we assume that $p_i$ is a final-state momentum (i.e. a outgoing momentum with positive energy)
and that $p_k$ corresponds to an initial-state momentum. With our conventions $p_k$ is an outgoing momentum with negative
energy.
In order to match the notation of section~\ref{sect:real_subtraction_terms}
we set $p_a=-p_k$. Thus $p_a$ is an incoming momentum with positive energy.
\begin{figure}
\begin{center}
\includegraphics[scale=0.8]{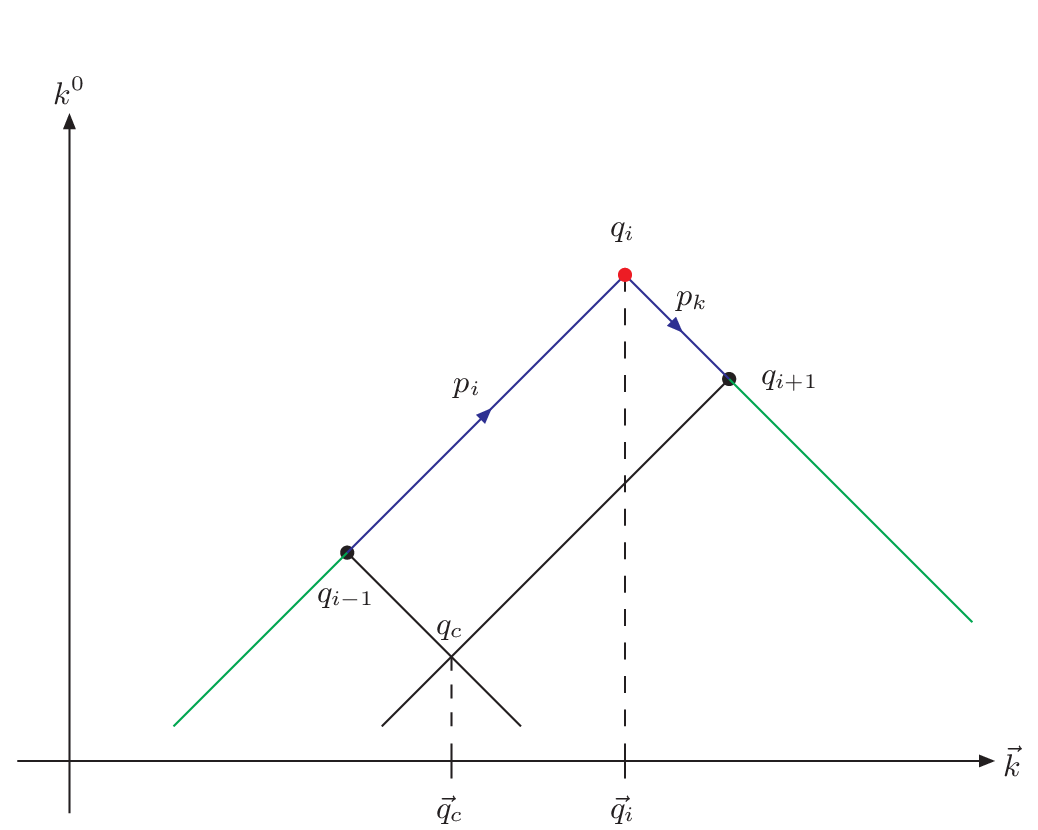}
\\
\includegraphics[scale=0.8]{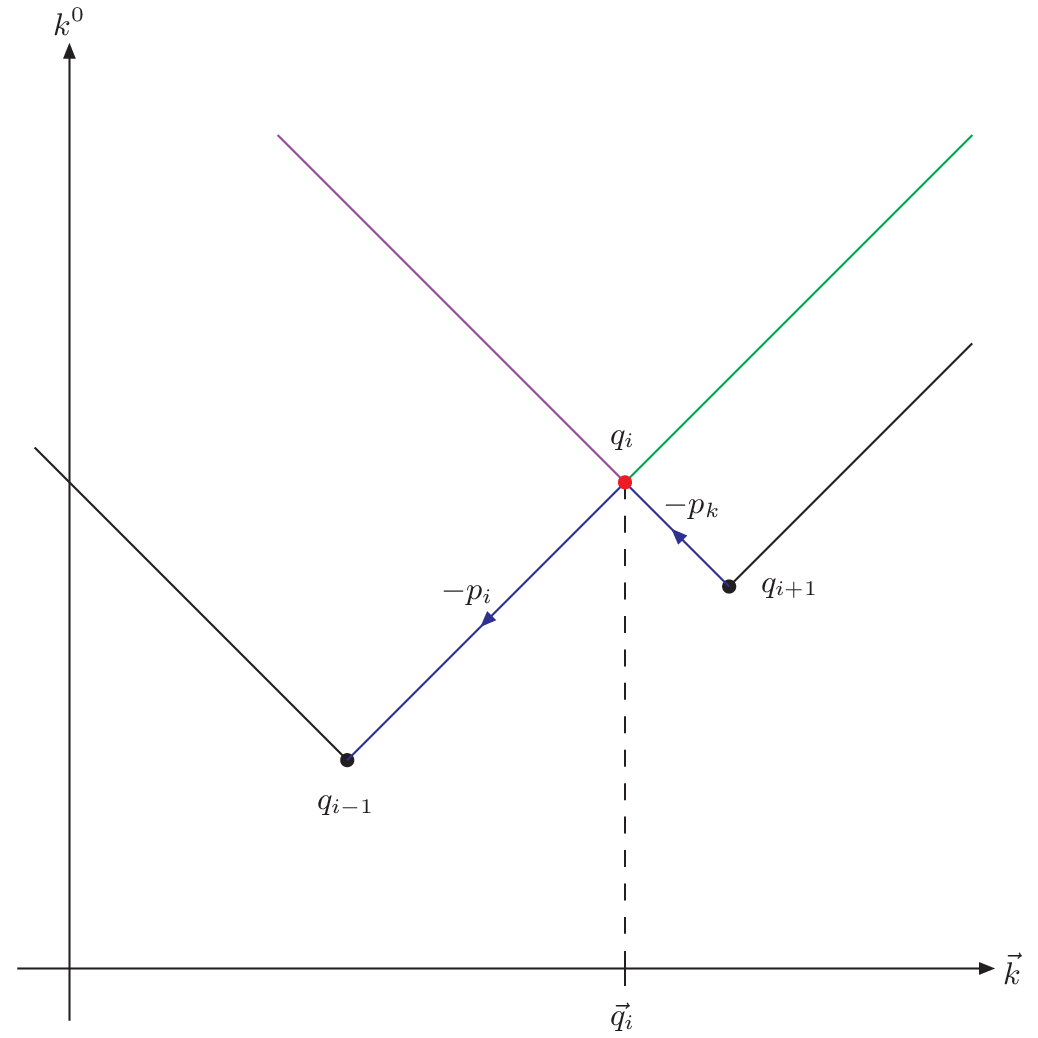}
\caption{\label{figure_integration_cones_fi}
The integration regions for a final-initial antenna
for the case where particle $i$ is in the final state
and particle $k$ is in the initial state.
The upper picture corresponds to the dipole with emitter $i$ and spectator $k$,
the lower picture corresponds to the dipole with emitter $k$ and spectator $i$.
}
\end{center}
\end{figure}
Let us start with the virtual dipole with emitter $i$ and spectator $k$.
As before we would like that in the collinear limit the momentum $(-k_i)$ has positive energy.
Therefore we use the loop-tree duality formula for the backward hyperboloids of eq.~(\ref{loop_tree_backward})
to convert the loop integrals into phase space integrals.
Next, we relate the integration for the real dipole ${\mathcal D}_{i'j'}^{a'}$ to the integration for the virtual dipole ${\mathcal E}_{i,k}$.
We recall that we use the notation $p_k=-p_a$ and $p_k'=-p_a'$.
We define the mapping between the set of momenta $\{p_i,p_k,k_i\}$ and $\{p_i',p_j',p_a'\}$  by
\bq
 p_i' & = & p_i + k_i - \frac{1-x}{x} p_k,
 \nonumber \\
 p_j' & = & - k_i,
 \nonumber \\
 p_a' & = & - \frac{1}{x} p_k,
\eq
with
\bq
 x & = &
 \frac{2p_ip_k + 2p_k k_i}{2p_ip_k + 2p_i k_i + 2p_k k_i + m_i^2 - m_i'{}^2 + m_j'{}^2}.
\eq
Again, the inverse mapping $\{p_i',p_j',p_a'\} \rightarrow \{p_i,p_k,k_i\}$ coincides with the mapping defined
in eq.(~\ref{map_massless_fi}), when restricted to $\{p_i',p_j',p_a'\} \rightarrow \{p_i,p_k\}$.

Expressing the measure for the unresolved phase space as an integration over $\vec{k}_i$ we find
\bq
 d\phi_{\mathrm{unres}}
 & = & 
 \frac{d^{D-1}k_i}{\left(2 \pi\right)^{D-1} \left(-2 k_{i,0}\right)} J,
\eq
with
\bq
 J & = & 
 \frac{2p_ap_i}{2p_a'p_i'}
 \theta\left(E_i'\right)
 \theta\left(x\right) \theta\left(1-x\right).
\eq
Let us now turn to the dipole ${\mathcal E}_{k,i}$ with emitter $k$ and spectator $i$.
In the collinear limit we require that the momentum $k_i$ has positive energy.
Thus, we use the loop-tree duality formula for the forward hyperboloids of eq.~(\ref{loop_tree_forward})
to convert the loop integrals into phase space integrals.
We then relate the real emission integrals to the virtual integrals.
Following the same procedure as in the final-final case we find
\bq
 p_i' & = & p_i - k_i - \frac{1-x}{x} p_k,
 \nonumber \\
 p_j' & = & k_i,
 \nonumber \\
 p_a' & = & - \frac{1}{x} p_k,
\eq
with
\bq
 x & = &
 \frac{2p_ip_k - 2p_k k_i}{2p_ip_k - 2p_i k_i - 2p_k k_i + m_j'{}^2}.
\eq
Note that in this case particle $i$ is the spectator and we have $m_i'=m_i$.

The geometric situation for the integration over the on-shell hyperboloids 
for a final-initial antenna for the case where particle $i$ is in the final state
and particle $k$ is in the initial state
is sketched in fig.~(\ref{figure_integration_cones_fi}).
The upper picture shows the contribution from the virtual approximation terms with emitter $i$ and spectator $k$
in the massless case.
The integration is over three backward light-cones with origins at $q_{i-1}$, $q_i$ and $q_{i+1}$.
The soft singularity resides in the integration over the backward light-cone with origin at $q_i$ at the origin $q_i$
and is indicated by a red dot.
The collinear singularities in the virtual terms occur on the lines between $q_{i-1}$ and $q_i$ (collinear singularity of $i$) and
between $q_i$ and $q_{i+1}$ (collinear singularity of $k$).
The virtual collinear regions are indicated in blue.
There is a cancellation of singularities within the virtual dual contributions in the regions where two propagators are
on-shell and have the same sign in the energy component. These regions are indicated in green.
There is also a cancellation of singularities within the virtual dual contributions at the point $q_c$.
The integration region for the real approximation term is the backward light-cone with origin at $q_i$.
The collinear singular region for the real approximation term with emitter $i$ is the line segment between $q_{i-1}$ and $q_i$
and matches with the corresponding line segment from the virtual term.

The lower picture shows the corresponding integration regions, where the roles of emitter and spectator are exchanged, i.e.
emitter $k$ and spectator $i$.
Note that the soft and the virtual collinear singularities occur in the same regions of $D$-dimensional loop momentum space.
The integration region for the real approximation term is the forward light-cone with origin at $q_i$.
The collinear singular region for the real approximation term with emitter $k$ is now the line segment indicated in purple.
Note that the real collinear singular region (purple line)
does not match with the virtual collinear singular region (blue line segment between $q_i$ and $q_{i+1}$).
This mismatch is compensated by the collinear counterterm for initial-state partons.

\subsubsection{Initial-initial antenna}

We now consider an initial-initial antenna.
The momenta $p_i$ and $p_k$ are outgoing momenta with negative energies.
In order to match the notation of section~\ref{sect:real_subtraction_terms}
we set $p_a=-p_i$ and $p_b=-p_k$. 
Thus $p_a$ and $p_b$ have positive energies.
Let us look at the virtual dipole ${\mathcal E}_{i,k}$.
In the collinear limit we require that the momentum $(-k_i)$ has positive energy.
Therefore we use the loop-tree duality formula for the backward hyperboloids of eq.~(\ref{loop_tree_backward})
to convert the loop integrals into phase space integrals.
\begin{figure}
\begin{center}
\includegraphics[scale=0.8]{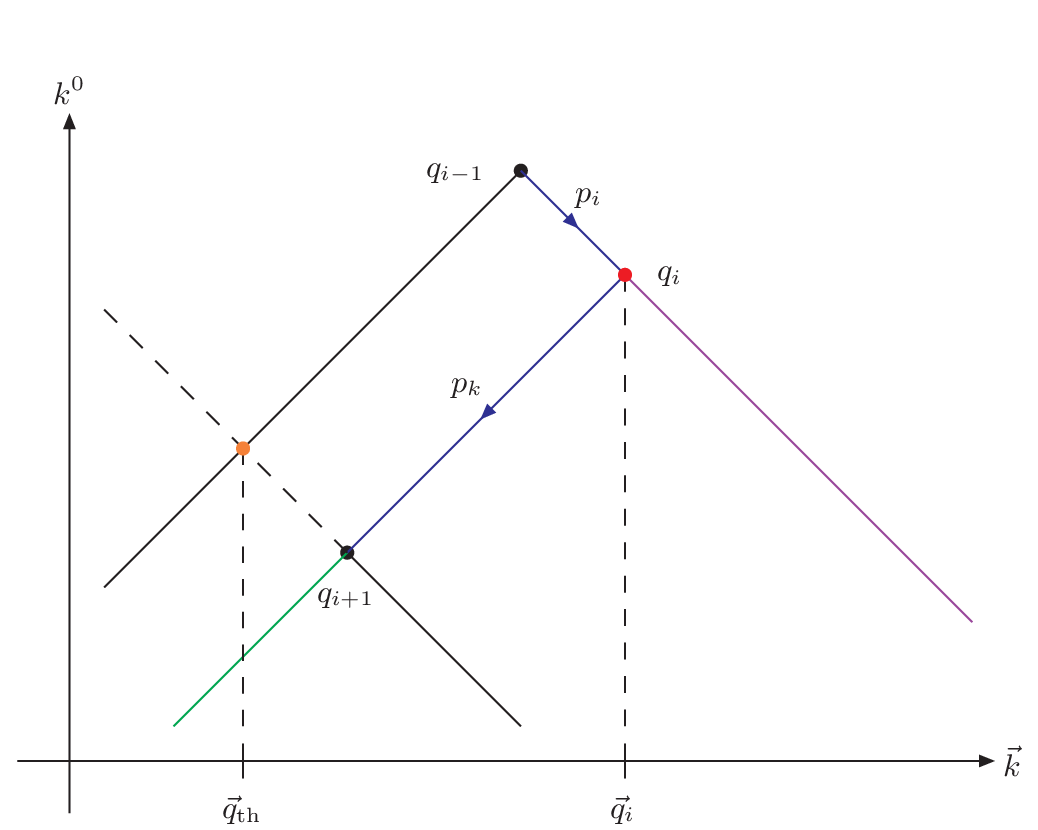}
\\
\includegraphics[scale=0.8]{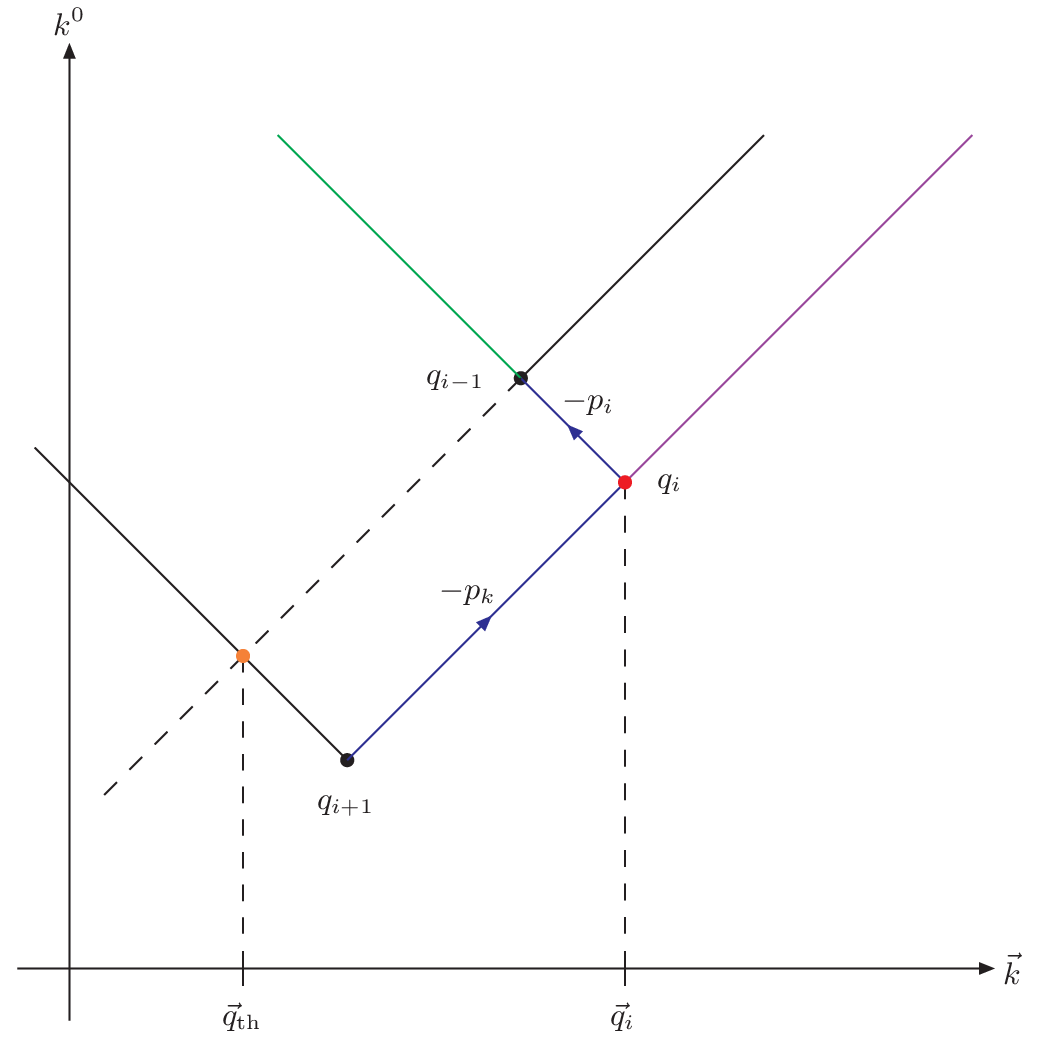}
\caption{\label{figure_integration_cones_ii}
The integration regions for an initial-initial antenna.
The upper picture corresponds to the dipole with emitter $i$ and spectator $k$,
the lower picture corresponds to the dipole with emitter $k$ and spectator $i$.
}
\end{center}
\end{figure}
Next, we relate the integration for the real dipole ${\mathcal D}^{a'j',b'}$ to the integration for the virtual dipole ${\mathcal E}_{i,k}$.
We recall that we use the notation $p_i=-p_a$, $p_i'=-p_a'$, $p_k=-p_b$ and $p_k'=-p_b'$.
We set
\bq
 p_a' & = & -\frac{1}{x} p_i,
 \nonumber \\
 p_j' & = & - k_i,
 \nonumber \\
 p_b' & = & - p_k,
\eq
with
\bq
 x & = &
 \frac{2 p_i p_k - 2 p_i k_i}{2 p_i p_k + 2 p_k k_i}.
\eq
All final state momenta are transformed as
\bq
 p_l' & = &
 \Lambda^{-1} p_l,
\eq
where $\Lambda^{-1}$ is the inverse Lorentz transformation to eq.~(\ref{def_Lorentz_trafo}).
Explicitly we have
\bq
 \left( \Lambda^{-1} \right)^\mu_{\;\;\nu}
 & = &
 g^\mu_{\;\;\nu}
 + a_1 \left(K^\mu+\tilde{K}^\mu\right) \left(K_\nu+\tilde{K}_\nu\right)
 + a_2 \left(K^\mu+\tilde{K}^\mu\right) K_\nu
 + a_3 K^\mu \left(K_\nu+\tilde{K}_\nu\right)
 \nonumber \\
 & &
 + a_4 K^\mu K_\nu.
\eq
The momenta $K$ and $\tilde{K}$ are given by
\bq
 K = p_a' + p_b' - p_j',
 \;\;\;\;\;\;
 \tilde{K} = p_a + p_b.
\eq
The coefficients are
\bq
 a_1 
 \;\; = \;\;
 2 \frac{K^2}{\left(\tilde{K}^2-K^2\right)^2 - \tilde{K}^2 \left(K+\tilde{K}\right)^2},
 & &
 a_2
 \;\; = \;\;
 2 \frac{\tilde{K}^2 - K^2}{\left(\tilde{K}^2-K^2\right)^2 - \tilde{K}^2 \left(K+\tilde{K}\right)^2},
 \nonumber \\
 a_3
 \;\; = \;\;
 2 \frac{\tilde{K}^2 - K^2- \left(K+\tilde{K}\right)^2}{\left(\tilde{K}^2-K^2\right)^2 - \tilde{K}^2 \left(K+\tilde{K}\right)^2},
 & & 
 a_4
 \;\; = \;\;
 2 \frac{\left(K+\tilde{K}\right)^2}{\left(\tilde{K}^2-K^2\right)^2 - \tilde{K}^2 \left(K+\tilde{K}\right)^2}.
\eq
Expressing the measure for the unresolved phase space as an integration over $\vec{k}_i$ we have
\bq
 d\phi_{\mathrm{unres}}
 & = & 
 \frac{d^{D-1}k_i}{\left(2 \pi\right)^{D-1} \left(-2 k_{i,0}\right)} J,
\eq
with
\bq
 J & = & 
 \theta\left(x\right) \theta\left(1-x\right).
\eq
Let us now turn to the dipole ${\mathcal E}_{k,i}$ with emitter $k$ and spectator $i$.
In the collinear limit we require that the momentum $k_i$ has positive energy.
Thus, we use the loop-tree duality formula for the forward hyperboloids of eq.~(\ref{loop_tree_forward})
to convert the loop integrals into phase space integrals.
In relating the real emission integrals to the virtual integrals we set now
\bq
 p_a' & = & - p_i,
 \nonumber \\
 p_j' & = & k_i,
 \nonumber \\
 p_b' & = & - \frac{1}{x} p_k,
\eq
with
\bq
 x & = &
 \frac{2 p_i p_k + 2 p_k k_i}{2 p_i p_k - 2 p_i k_i}.
\eq
All final state momenta are transformed as
\bq
 p_l' & = &
 \Lambda^{-1} p_l.
\eq
The geometric situation for the integration over the on-shell hyperboloids 
for an initial-initial antenna 
is sketched in fig.~(\ref{figure_integration_cones_ii}).
The upper picture shows the contribution from the virtual approximation terms with emitter $i$ and spectator $k$
in the massless case.
The integration is over three backward light-cones with origins at $q_{i-1}$, $q_i$ and $q_{i+1}$.
The soft singularity resides in the integration over the backward light-cone with origin at $q_i$ at the origin $q_i$
and is indicated by a red dot.
The collinear singularities in the virtual terms occur on the lines between $q_{i-1}$ and $q_i$ (collinear singularity of $i$) and
between $q_i$ and $q_{i+1}$ (collinear singularity of $k$).
The virtual collinear regions are indicated in blue.
There is a cancellation of singularities within the virtual dual contributions in the regions where two propagators are
on-shell and have the same sign in the energy component. These regions are indicated in green and purple.
There is a threshold singularity (indicated by an orange dot) at $\vec{q}_{\mathrm{th}}$. 
The threshold singularity is avoided by contour deformation.
The integration region for the real approximation term is the backward light-cone with origin at $q_i$.
The collinear singular region for the real approximation term with emitter $i$ is the line segment indicated in purple.
Note that the real collinear singular region (purple line)
does not match with the virtual collinear singular region (blue line segment between $q_{i-1}$ and $q_{i}$).
This mismatch is compensated by the collinear counterterm for initial-state partons.

The lower picture shows the corresponding integration regions, where the roles of emitter and spectator are exchanged, i.e.
emitter $k$ and spectator $i$.
Note that the soft and the virtual collinear singularities occur in the same regions of $D$-dimensional loop momentum space.
The integration region for the real approximation term is the forward light-cone with origin at $q_i$.
The collinear singular region for the real approximation term with emitter $k$ is the line segment indicated in purple.
Note that the real collinear singular region (purple line)
does not match with the virtual collinear singular region (blue line segment between $q_i$ and $q_{i+1}$).
This mismatch is compensated by the collinear counterterm for initial-state partons.

\subsubsection{Summary on the cancellation within an antenna}

It is worth to summarise how infrared singularities cancel within an antenna.
\begin{figure}
\begin{center}
\includegraphics[scale=1.0]{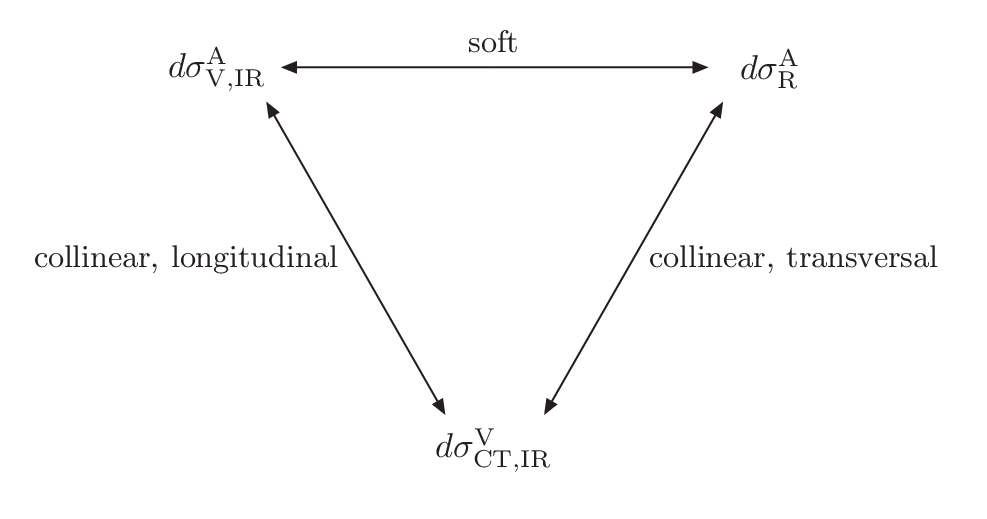}
\caption{\label{figure_summary_final_state}
The cancellation of infrared singularities within an antenna involving only final-state particles.
}
\end{center}
\end{figure}
Let us first consider an antenna involving only final-state particles.
In this case we have contributions from three terms: 
The virtual approximation term $d\sigma^{\mathrm{A}}_{\mathrm{V},\mathrm{IR}}$,
the real approximation term $d\sigma^{\mathrm{A}}_{\mathrm{R}}$
and the infrared part from the field renormalisation constants, given by $d\sigma^{\mathrm{V}}_{\mathrm{CT},\mathrm{IR}}$.
The soft singularity of the antenna cancels between the virtual part $d\sigma^{\mathrm{A}}_{\mathrm{V},\mathrm{IR}}$
and the real part $d\sigma^{\mathrm{A}}_{\mathrm{R}}$.
The contribution from the field renormalisation constants does not contain any soft singularity.
The real part has in addition collinear singularities, where the two particles in the collinear splitting have
transverse polarisations.
These collinear singularities cancel with corresponding singularities from the field renormalisation constants.
On the other hand, the virtual part has collinear singularities, where one of the two particles in the collinear splitting
has a longitudinal polarisation.
These singularities cancel as well with corresponding singularities from the field renormalisation constants.
This mechanism is summarised in fig.~(\ref{figure_summary_final_state}).

Let us now consider an antenna with initial-state particles.
\begin{figure}
\begin{center}
\includegraphics[scale=1.0]{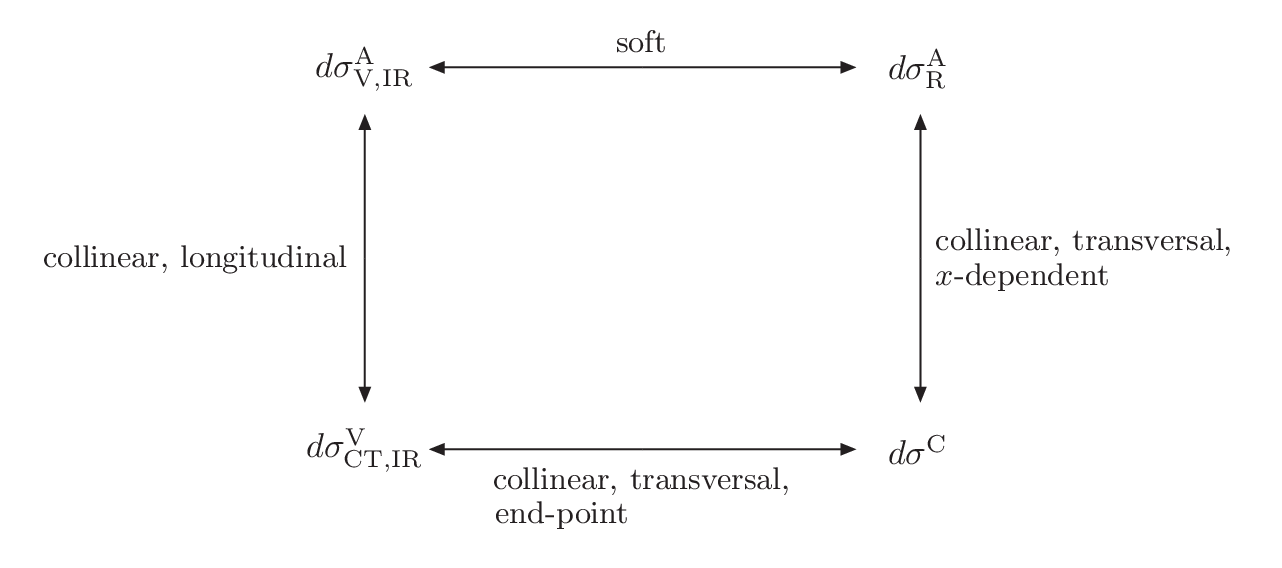}
\caption{\label{figure_summary_initial_state}
The cancellation of infrared singularities within an antenna involving initial-state particles.
}
\end{center}
\end{figure}
We have contributions from four terms:
As before, there are contributions from 
the virtual approximation term $d\sigma^{\mathrm{A}}_{\mathrm{V},\mathrm{IR}}$,
the real approximation term $d\sigma^{\mathrm{A}}_{\mathrm{R}}$
and the infrared part from the field renormalisation constants, given by $d\sigma^{\mathrm{V}}_{\mathrm{CT},\mathrm{IR}}$.
In addition we have a contribution from the collinear subtraction term $d\sigma^{\mathrm{C}}$,
which splits into an $x$-dependent convolution part $d\sigma^{\mathrm{C}}_{\mathrm{R}}$
and an end-point part $d\sigma^{\mathrm{C}}_{\mathrm{CT}}$.
As before, the soft singularity of the antenna cancels between the virtual part $d\sigma^{\mathrm{A}}_{\mathrm{V},\mathrm{IR}}$
and the real part $d\sigma^{\mathrm{A}}_{\mathrm{R}}$.
However the mechanism for initial-state collinear singularities is different:
The collinear singularity with two transverse polarisations 
from the real part $d\sigma^{\mathrm{A}}_{\mathrm{R}}$ cancels with the corresponding singularity from 
the $x$-dependent convolution part $d\sigma^{\mathrm{C}}_{\mathrm{R}}$.
The collinear singularity with two transverse polarisations from the end-point contribution
$d\sigma^{\mathrm{C}}_{\mathrm{CT}}$ cancels with the corresponding singularity
from the field renormalisation constants in $d\sigma^{\mathrm{V}}_{\mathrm{CT},\mathrm{IR}}$,
and finally the collinear singularity, where one of the two particles in the collinear splitting
has a longitudinal polarisation cancels between $d\sigma^{\mathrm{V}}_{\mathrm{CT},\mathrm{IR}}$
and $d\sigma^{\mathrm{A}}_{\mathrm{V},\mathrm{IR}}$.
This is summarised in fig.~(\ref{figure_summary_initial_state}).

\subsection{The pure ultraviolet contribution}
\label{section_pure_UV}

In this sub-section we consider the pure ultraviolet contribution
\bq
 \langle O \rangle^{\mathrm{NLO}}_{{\bf I}+{\bf L},\mathrm{UV}}.
 & = &
 \int\limits_{n} 
 O_{n} 
      \int\limits_{\mathrm{loop}} \left( d\sigma^{\mathrm{A}}_{\mathrm{V},\mathrm{UV}}
      + d\sigma_{\mathrm{CT},\mathrm{UV}}^{\mathrm{V}}
 \right).
\eq
The term
$d\sigma^{\mathrm{A}}_{\mathrm{V},\mathrm{UV}}$ contains the ultraviolet approximation terms
for the vertices and the propagators.
To give an example, let us consider the ultraviolet approximation term for the quark-gluon vertex.
This vertex has a leading colour contribution and a subleading colour contribution.
The subtraction term for the leading colour contribution is given by \cite{Becker:2010ng}
\bq
 V^{\mathrm{V},\mathrm{UV}}_{qqg, \mathrm{lc}}\left(\mu^2,\mu_{\mathrm{UV}}^2\right) & = &   
 i S_\eps^{-1} \mu^{2\eps} \int \frac{d^Dk}{(2\pi)^Di} 
 \left[
 \frac{2 \gamma^\mu}{\left(\bar{k}^2-\mu_{\mathrm{UV}}^2\right)^2}
 +
 \frac{4 \left(1-\eps \right) \bar{k}\!\!\!/ \; \bar{k}^\mu - 2 \mu_{\mathrm{UV}}^2 \gamma^\mu}{\left(\bar{k}^2-\mu_{\mathrm{UV}}^2\right)^3}
 \right].
\eq
The quantity $\mu_{\mathrm{UV}}$ is an arbitrary mass.
For the subleading colour contribution we have
\bq
 V^{\mathrm{V},\mathrm{UV}}_{qqg, \mathrm{sc}}\left(\mu^2,\mu_{\mathrm{UV}}^2\right) & = &   
 i S_\eps^{-1} \mu^{2\eps} \int \frac{d^Dk}{(2\pi)^Di} 
 \left[
 \frac{2 \left(1-\eps \right) \bar{k}\!\!\!/ \gamma^\mu \bar{k}\!\!\!/ + 4 \mu_{\mathrm{UV}}^2 \gamma^\mu}{\left(\bar{k}^2-\mu_{\mathrm{UV}}^2\right)^3}
 \right].
\eq
A complete set of ultraviolet approximation terms can be found in ref.~\cite{Becker:2010ng}.
The terms proportional to $\mu_{\mathrm{UV}}^2$ in the numerator are not divergent, but ensure that the 
finite part of the integrated expression is proportional to the pole part.
For the quark-gluon vertex approximation term this implies
\bq
\label{integrated_UV_approx_terms}
 V^{\mathrm{V},\mathrm{UV}}_{qqg, \mathrm{lc}} & = & 
 i \frac{1}{(4\pi)^2} \gamma^\mu 
 \left( 3 \right) \left( \frac{1}{\eps} - \ln \frac{\mu_{\mathrm{UV}}^2}{\mu^2} \right)
 + {\cal O}(\eps),
 \nonumber \\
 V^{\mathrm{V},\mathrm{UV}}_{qqg, \mathrm{sc}} & = & 
 i \frac{1}{(4\pi)^2} \gamma^\mu 
 \left( -1 \right) \left( \frac{1}{\eps} - \ln \frac{\mu_{\mathrm{UV}}^2}{\mu^2} \right)
 + {\cal O}(\eps).
\eq
Now let us turn to $d\sigma_{\mathrm{CT},\mathrm{UV}}^{\mathrm{V}}$. 
By construction, $d\sigma^{\mathrm{V}}_{\mathrm{CT},\mathrm{UV}}$ contains for all renormalisation constants 
(field renormalisation, coupling renormalisation and mass renormalisation) the terms, 
which lead exactly to the $1/\eps_{\mathrm{UV}}$ divergences.
In addition, $d\sigma^{\mathrm{V}}_{\mathrm{CT},\mathrm{UV}}$ contains finite terms from coupling renormalisation and mass renormalisation.
Let us first focus on the ultraviolet divergent terms.
Note that we may obtain these terms from the renormalisation counterterms for all vertices and propagators.
The set of vertices and propagators for $d\sigma_{\mathrm{CT},\mathrm{UV}}^{\mathrm{V}}$
will correspond exactly to the set of vertices and propagators for $d\sigma^{\mathrm{A}}_{\mathrm{V},\mathrm{UV}}$.
This makes it easy to find an integral representation for $d\sigma_{\mathrm{CT},\mathrm{UV}}^{\mathrm{V}}$.
We may use the integral representation for the ultraviolet approximation terms, substitute $\mu_{\mathrm{UV}} \rightarrow \mu$ and
add a minus sign to obtain the integral representation for the contributions to $d\sigma_{\mathrm{CT},\mathrm{UV}}^{\mathrm{V}}$.
For example
\bq
 V^{\mathrm{CT},\mathrm{UV}}_{qqg, \mathrm{lc}}\left(\mu^2\right) 
 & = &   
 - V^{\mathrm{V},\mathrm{UV}}_{qqg, \mathrm{lc}}\left(\mu^2,\mu^2\right),
 \nonumber \\
 V^{\mathrm{CT},\mathrm{UV}}_{qqg, \mathrm{sc}}\left(\mu^2\right) 
 & = &   
 - V^{\mathrm{V},\mathrm{UV}}_{qqg, \mathrm{sc}}\left(\mu^2,\mu^2\right).
\eq
Choosing $\mu_{\mathrm{UV}} = \mu$ ensures, that the counterterms just subtract out the $1/\eps$-pole.
This can easily be seen from eq.~(\ref{integrated_UV_approx_terms}), yielding
\bq
 V^{\mathrm{CT},\mathrm{UV}}_{qqg, \mathrm{lc}} & = & 
 i \frac{1}{(4\pi)^2} \gamma^\mu 
 \frac{\left( -3 \right) }{\eps} 
 + {\cal O}(\eps),
 \nonumber \\
 V^{\mathrm{CT},\mathrm{UV}}_{qqg, \mathrm{sc}} & = & 
 i \frac{1}{(4\pi)^2} \gamma^\mu 
 \frac{1}{\eps} 
 + {\cal O}(\eps).
\eq
The terms $d\sigma^{\mathrm{A}}_{\mathrm{V},\mathrm{UV}}$
and $d\sigma_{\mathrm{CT},\mathrm{UV}}^{\mathrm{V}}$ contain as far as divergent contributions are concerned
only ultraviolet divergences.
Moreover, the singular behaviour in the ultraviolet is up to a sign exactly equal.
Therefore, the sum of $d\sigma^{\mathrm{A}}_{\mathrm{V},\mathrm{UV}}$ and $d\sigma_{\mathrm{CT},\mathrm{UV}}^{\mathrm{V}}$
is integrable in the ultraviolet region and hence integrable everywhere.

Let us now turn our attention to finite terms from renormalisation of the masses or couplings.
In the $\overline{\mbox{MS}}$-scheme these finite terms are absent.
Therefore, if we take the strong coupling $g$ and the quark masses $m$ in the $\overline{\mbox{MS}}$-scheme
nothing needs to be done.
For the strong coupling the $\overline{\mbox{MS}}$-scheme is the conventional choice.
However, for the quark masses the use of the on-shell scheme is an alternative to the $\overline{\mbox{MS}}$-scheme.
We now discuss how to implement the on-shell scheme for quark masses in our framework.
This will require only minor modifications.
We start with the ultraviolet approximation term for a massive quark propagator:
\bq
\lefteqn{
-i \Sigma^{\mathrm{V},\mathrm{UV}}\left(\mu^2,\mu_{\mathrm{UV}}^2\right) = 
} & & 
 \nonumber \\
 & &
 - i S_\eps^{-1} \mu^{2\eps} \int \frac{d^Dk}{(2\pi)^Di}
 \left[ 
        \frac{-2(1-\eps)\left(Q\!\!\!\!/+\bar{k}\!\!\!/\right)+4\left(1-\frac{1}{2}\eps\right)m}{\left(\bar{k}^2-\mu_{\mathrm{UV}}^2\right)^2}
        - 4 \left(1-\eps\right) \frac{\bar{k}\cdot \left( p - 2 Q \right) \; \bar{k}\!\!\!/}{\left(\bar{k}^2-\mu_{\mathrm{UV}}^2\right)^3} 
 \right. \nonumber \\
 & & \left.
        + \frac{2 \mu_{\mathrm{UV}}^2 \left( p\!\!\!/-2m \right)}{\left(\bar{k}^2-\mu_{\mathrm{UV}}^2\right)^3} 
 \right].
\eq
This approximation term integrates to
\bq
-i \Sigma^{\mathrm{V},\mathrm{UV}}
 & = &
 - i 
 \frac{1}{(4\pi)^2}
 \left( - p\!\!\!/ + 4 m \right) \left( \frac{1}{\eps} - \ln \frac{\mu_{\mathrm{UV}}^2}{\mu^2} \right)
 + {\cal O}(\eps).
\eq
In order to find $\Sigma^{\mathrm{CT},\mathrm{UV}}$ corresponding to a mass definition in the on-shell scheme
one proceeds as before (i.e. adding an extra minus sign and substituting $\mu_{\mathrm{UV}} \rightarrow \mu$)
and one adds an additional finite term, using the fact that
\bq
 S_\eps^{-1} \mu^{2\eps}
 \int \frac{d^Dk}{(2\pi)^D i} \frac{-2 \mu_{\mathrm{UV}}^2}{\left(\bar{k}^2-\mu_{\mathrm{UV}}^2\right)^3}
 \; \; = \;\;
 \frac{1}{\left(4\pi\right)^2}
 e^{\eps \gamma_E} \Gamma\left(1+\eps\right)
 \left( \frac{\mu_{\mathrm{UV}}^2}{\mu^2} \right)^{-\eps}
 \;\; = \;\;
 \frac{1}{\left(4\pi\right)^2} \left[ 1 + \mathcal{O}\left(\eps\right) \right].
 \;\;
\eq
The required finite term is easily found by recalling that the counter-term leads to the Feynman rule
\bq
\begin{picture}(100,10)(0,0)
\ArrowLine(50,5)(10,5)
\Line(90,5)(50,5)
\Line(45,0)(55,10)
\Line(45,10)(55,0)
\end{picture}
 & = &
 i \left[ \left( p\!\!\!/ - m \right) Z_2^{(1)} - m Z_{m,\mathrm{on-shell}}^{(1)} \right],
\eq
with $Z_{m,\mathrm{on-shell}}^{(1)}$ given in eq.~(\ref{def_Z_m_onshell}).
Thus
\bq
\label{def_Sigma_CT_UV}
\lefteqn{
-i \Sigma^{\mathrm{CT},\mathrm{UV}}\left(\mu^2\right) = 
} & & 
 \nonumber \\
 & &
 - i S_\eps^{-1} \mu^{2\eps} \int \frac{d^Dk}{(2\pi)^Di}
 \left[ 
        \frac{2(1-\eps)\left(Q\!\!\!\!/+\bar{k}\!\!\!/\right)-4\left(1-\frac{1}{2}\eps\right)m}{\left(\bar{k}^2-\mu^2\right)^2}
        + 4 \left(1-\eps\right) \frac{\bar{k}\cdot \left( p - 2 Q \right) \; \bar{k}\!\!\!/}{\left(\bar{k}^2-\mu^2\right)^3} 
 \right. \nonumber \\
 & & \left.
        - \frac{2 \mu^2 \left( p\!\!\!/-2m \right)}{\left(\bar{k}^2-\mu^2\right)^3} 
        - \frac{2 \mu^2 m}{\left(\bar{k}^2-\mu^2\right)^3} \left( -4 + 3 \ln \frac{m^2}{\mu^2} \right)
 \right]
\eq
integrates to
\bq
-i \Sigma^{\mathrm{CT},\mathrm{UV}} & = &
 - i 
 \frac{1}{(4\pi)^2}
 \left[
        \left( p\!\!\!/ - 4 m \right) \frac{1}{\eps}
        + m \left( -4 + 3 \ln \frac{m^2}{\mu^2} \right)
 \right]
 + {\cal O}(\eps)
\eq
and implements the on-shell scheme for the quark mass.

\section{Contour deformation}
\label{sect:contour_deformation}

Up to now we have defined integral representations for all contributions and maps between different
contributions such that the combination can be written as a single integral
over
\bq
 \int \frac{d^{D-1}k}{(2\pi)^{D-1}} ... .
\eq
We have achieved that all singularities, which would produce poles in the dimensional regularisation parameter $\eps$
cancel locally at the integrand level.
Therefore we may take the limit $D\rightarrow 4$ and we arrive at a three-dimensional integral
\bq
 \int \frac{d^{3}k}{(2\pi)^{3}} ... .
\eq
However, this does not yet imply that we can simply or safely integrate each of the three components of the loop momentum $\vec{k}$
from minus infinity to plus infinity along the real axis.
In the virtual part there is still the possibility that some of the loop propagators go on-shell 
for real values of the loop momentum.
We have seen examples of these threshold singularities 
in the case of a final-final antenna in fig.~(\ref{figure_integration_cones_ff})
or in the case of an initial-initial antenna in fig.~(\ref{figure_integration_cones_ii}).
In the case of a final-initial antenna we have cancellation between the various dual integrands.
The threshold singularities are avoided by a deformation of the integration contour into the complex plane.
For the loop three-momentum we write
\bq
 \vec{k} & = &
 \vec{\tilde{k}} + i \vec{\kappa}\left(\vec{\tilde{k}}\right),
\eq
where $\vec{\tilde{k}}$ is real and $\vec{\kappa}(\vec{\tilde{k}})$ defines the deformation.
This introduces a Jacobian in integral over the virtual approximation terms:
\bq
 \int \frac{d^{3}k}{(2\pi)^{3}} f\left(\vec{k}\right)
 & = &
 \int \frac{d^{3}\tilde{k}}{(2\pi)^{3}} 
 \left| \frac{\partial k^i}{\partial \tilde{k}^j} \right|
 f\left(\vec{k}\left(\vec{\tilde{k}}\right)\right).
\eq
The deformation has to satisfy three requirements:
\begin{enumerate}
\item The deformation has to match the $i\delta$-prescription of the dual propagators in eq.~(\ref{loop_tree_forward}) and
eq.~(\ref{loop_tree_backward}).
\item The deformation has to respect the ultraviolet power counting.
\item The deformation has to vanish for soft and collinear singularities in order not to spoil the local
cancellation of these singularities.
\end{enumerate}
Algorithms for the contour deformation can be found
in the literature \cite{Gong:2008ww,Becker:2012aq,Becker:2012nk,Becker:2012bi,Buchta:2015wna}.

\section{Conclusions}
\label{sect:conclusions}

In this paper we considered NLO calculations within a numerical approach.
The numerical approach employs subtraction terms both for the real emission contribution and the virtual
contribution,
such that the subtracted real emission contribution and the subtracted virtual contribution
can be integrated numerically.
The subtraction terms have to be added back.
In this paper we showed that the various subtraction terms can be combined to give an integrable
function, which again can be integrated numerically.
Our motivation is not to improve NLO calculations.
At NLO, all subtraction terms are easily integrated analytically and in practical
calculations it is more efficient to use those.
However, the situation is different at NNLO and beyond:
There the task of finding local subtraction terms is manageable, while the analytic integration 
of the local subtraction terms is highly non-trivial.
It is therefore desirable to have at NNLO and beyond a method, which integrates the subtraction
terms numerically.
In order to achieve this, the subtraction terms have to be combined in the right way with appropriate mappings between
them.
There are some subtleties related to field renormalisation and initial-state collinear singularities.
In this paper we studied these subtleties at NLO and obtained a clear picture how all singularities 
cancel at the integrand level.

At a more technical level the new results of this paper include
a mapping between virtual configurations and real configurations
for all relevent cases, including initial-state particles and final-state massive particles.
In addition we derived an integral representation for the collinear subtraction term for initial-state particles,
which matches locally with the singularities of the other contributions.
Furthermore we presented a method on how to implement a mass definition in the on-shell scheme within the numerical
approach.

With the results of this paper we can now split a NLO calculation into three parts, the subtracted virtual part,
the subtracted real part and the combined subtraction terms.
All three parts can now be evaluated numerically.
Does this eliminate the need of any analytic calculation of an integral?
Not quite.
While it is true that infrared singularities cancel between the real and the virtual contributions
at the integrand level
and no integral needs to be computed analytically for this to happen, there are singularities, which
are absorbed into a redefinition of the parameters.
These are the ultraviolet divergences, treated by renormalisation, and initial-state collinear singularities, treated
by a redefinition of the parton distribution functions.
This introduces a scheme-dependence and each scheme has a well-defined prescription which finite terms are absorbed
in a redefinition of the parameters and which not.
The numerical approach has to reproduce the correct finite terms.
This requires to add certain finite terms to the integral representations of some quantities in order to reproduce
the correct finite terms of a given scheme.
In order to find the correct finite terms for the integral representations we have
to perform some simple integrals analytically.
The required integrals are tadpole integrals
\bq
 \int \frac{d^Dk}{(2\pi)^D} \frac{1}{\left( k^2 - m \right)^\nu},
\eq
for the virtual case and Euler beta-function type integrals
\bq
 \int\limits_0^1 dx \; x^{\nu-\eps} \left(1-x\right)^{-\eps}
\eq
in the real case.
These two integrals are significantly simpler than the integrals required to integrate all subtraction terms
analytically and we might expect that this remains true at NNLO and beyond.

\subsection*{Acknowledgements}

This research was supported in part by the National Science Foundation under Grant No. NSF PHY11-25915.


\begin{appendix}

\section{Polarisation vectors and polarisation spinors}
\label{appendix:spinors}

We define the light-cone coordinates of a four-vector $p_\mu$ as
\bq
 p_+ = p_0 + p_3, 
 \;\;\; 
 p_- = p_0 - p_3, 
 \;\;\; 
 p_{\bot} = p_1 + i p_2,
 \;\;\; 
 p_{\bot^\ast} = p_1 - i p_2.
\eq
In terms of the light-cone components of a light-like four-vector, 
the corresponding massless spinors $\langle p \pm |$ and $| p \pm \rangle$ can be chosen as
\bq
\left| p+ \right\rangle = \frac{e^{-i \frac{\phi}{2}}}{\sqrt{\left| p_+ \right|}} \left( \begin{array}{c}
  -p_{\bot^\ast} \\ p_+ \end{array} \right),
 & &
\left| p- \right\rangle = \frac{e^{-i \frac{\phi}{2}}}{\sqrt{\left| p_+ \right|}} \left( \begin{array}{c}
  p_+ \\ p_\bot \end{array} \right),
 \nonumber \\
\left\langle p+ \right| = \frac{e^{-i \frac{\phi}{2}}}{\sqrt{\left| p_+ \right|}} 
 \left( -p_\bot, p_+ \right),
 & &
\left\langle p- \right| = \frac{e^{-i \frac{\phi}{2}}}{\sqrt{\left| p_+ \right|}} 
 \left( p_+, p_{\bot^\ast} \right),
\eq
where the phase $\phi$ is given by
\bq
p_+ & = & \left| p_+ \right| e^{i\phi},
 \;\;\;\;\;
 0 \le \phi < 2 \pi.
\eq
If the Cartesian coordinates $p_0$, $p_1$, $p_2$ and $p_3$ are real numbers, we have
\bq
 \left| p \pm \right\rangle^\dagger = e^{i \phi} \left\langle p \pm \right|,
 & &
 \left\langle p \pm \right|^\dagger = e^{i \phi} \left| p \pm \right\rangle,
 \;\;\;\;\;\;
 e^{i \phi} = \pm 1.
\eq
Spinor products are denoted as
\bq
 \langle p q \rangle = \langle p - | q + \rangle,
 & &
 [ q p ] = \langle q + | p - \rangle.
\eq 
Let $q$ be a light-like four-vector.
We define polarisation vectors for the gluons by
\bq
\eps_{\mu}^{+} = \frac{\langle q-|\sigma_{\mu}|p-\rangle}{\sqrt{2} \langle q p \rangle},
 & &
\eps_{\mu}^{-} = \frac{\langle q+|\bar{\sigma}_{\mu}|p+\rangle}{\sqrt{2} [ p q ]},
\eq
with $\sigma_\mu = ( 1, \vec{\sigma} )$ and $\bar{\sigma}_\mu = ( 1, - \vec{\sigma} )$, where $\vec{\sigma}=(\sigma_1,\sigma_2,\sigma_3)$ are the
Pauli matrices.
The dependence on the reference four-vector $q$ drops out in gauge invariant quantities.
Under complex conjugation we have
\bq
 \left( \eps_\mu^+ \right)^\ast = \eps_\mu^-,
 & &
 \left( \eps_\mu^- \right)^\ast = \eps_\mu^+.
\eq
For the spin sum we have
\bq
 \sum\limits_{\lambda} \left( \eps_\mu^\lambda \right)^\ast \eps_\nu^\lambda
 & = &
 - g_{\mu\nu} + \frac{p_\mu q_\nu + q_\mu p_\nu}{p \cdot  q}.
\eq
The reference four-vector $q$ can be used to project any not necessarily light-like four-vector $P$ 
on a light-like four-vector $P^\flat$:
\bq
\label{projection_null}
 P^\flat & = & P - \frac{P^2}{2 P \cdot q} q.
\eq
The four-vector $P^\flat$ satisfies $(P^\flat)^2=0$.
Let $P$ be a four-vector satisfying $P^2=m^2$. We define the spinors associated to massive fermions by
\bq
\label{eq:spinors} 
 u^\pm = \frac{1}{\langle P^\flat \pm | q \mp \rangle} \left( P\!\!\!/ + m \right) | q \mp \rangle,
 & &
\bar{u}^\pm = \frac{1}{\langle q \mp | P^\flat \pm \rangle} \langle q \mp | \left( P\!\!\!/ + m \right),
 \nonumber \\
 v^\mp = \frac{1}{\langle P^\flat \pm | q \mp \rangle} \left( P\!\!\!/ - m \right) | q \mp \rangle,
 & &
\bar{v}^\mp = \frac{1}{\langle q \mp | P^\flat \pm \rangle} \langle q \mp | \left( P\!\!\!/ - m \right).
\eq
These spinors satisfy the Dirac equations
\bq 
\left( p\!\!\!/ - m \right) u^\lambda = 0, & & 
\bar{u}^\lambda \left( p\!\!\!/ - m \right) = 0,
 \nonumber \\
\left( p\!\!\!/ + m \right) v^\lambda = 0, & & 
\bar{v}^\lambda \left( p\!\!\!/ + m \right) = 0,
\eq
the orthogonality relations
\bq
\bar{u}^{\bar{\lambda}} u^\lambda = 2 m \delta^{\bar{\lambda}\lambda}, 
 & &
\bar{v}^{\bar{\lambda}} v^\lambda = -2 m \delta^{\bar{\lambda}\lambda}, 
\eq
and the completeness relation
\bq
\sum\limits_{\lambda} u^\lambda \bar{u}^\lambda = p\!\!\!/ + m,
 & &
\sum\limits_{\lambda} v^\lambda \bar{v}^\lambda = p\!\!\!/ - m.
\eq
We further have
\bq
 \bar{u}^{\bar{\lambda}} \gamma^\mu u^\lambda & = & 2 p^\mu \delta^{\bar{\lambda} \lambda},
 \nonumber \\
 \bar{v}^{\bar{\lambda}} \gamma^\mu v^\lambda & = & 2 p^\mu \delta^{\bar{\lambda} \lambda}.
\eq
In the massless limit the definition reduces to
\bq
 \bar{u}^\pm = \bar{v}^\mp = \langle p \pm |, 
 & &
 u^\pm = v^\mp = | p \pm \rangle.
\eq

\section{Self-energies and field renormalisation}
\label{sect:self_energies}

\subsection{The gluon self-energy}

We first consider the gluon self-energy. 
With the notation $k_1=k+p/2$, $k_2=k-p/2$ we find
\bq
\lefteqn{
 - i \Pi^{\mu\nu}\left(p,\mu^2\right) =  
 - i g^2 S_\eps^{-1} \mu^{4-D} \int \frac{d^Dk}{(2\pi)^Di} \frac{1}{k_1^2 k_2^2}
 \left\{ 2 C_A \left[ -p^2 g^{\mu\nu} + p^\mu p^\nu - 2 \left( 1 - \eps \right) k^\mu k^\nu 
 \right. \right. } & & \nonumber \\
 & & \left. \left.
                      + \frac{1}{2} \left( 1 - \eps \right) g^{\mu\nu} \left( k_1^2 + k_2^2 \right) \right]
       + 2 T_R N_f \left[ p^2 g^{\mu\nu} - p^\mu p^\nu + 4 k^\mu k^\nu 
                          - g^{\mu\nu} \left( k_1^2 + k_2^2 \right) \right] \right\}.
\eq
The self-energy may be written as
\bq
 - i \Pi^{\mu\nu}
 & = &
 - i 
 \left( -p^2 g^{\mu\nu} + p^\mu p^\nu \right)
 \Pi\left(p^2\right).
\eq
An analytic calculation of $\Pi(p^2)$ gives
\bq
 \Pi\left(p^2\right)
 & = & 
 g^2
 \left[ \beta_0 - 2 C_A + \left( \frac{C_A}{9} + \frac{4}{9} T_R N_f \right) \frac{\eps}{1-\frac{2}{3} \eps} \right] 
 B_0\left(p^2,0,0\right),
\eq
where $B_0(p^2,0,0)$ is the scalar two-point function with masses $m_1=0$ and $m_2=0$, given for $p^2 \neq 0$ by
\bq
 B_0\left(p^2,0,0\right)
 & = & 
 \frac{1}{\left(4\pi\right)^2}
 e^{\eps \gamma_E} 
 \frac{\Gamma\left(\eps\right) \Gamma\left(1-\eps\right)^2}{\Gamma\left(2-2\eps\right)}
 \left( \frac{-p^2}{\mu^2} \right)^{-\eps}.
\eq
For $p^2=0$ one has
\bq
 B_0\left(0,0,0\right)
 & = & 
 0.
\eq
The one-loop contribution to the field renormalisation constant ${\mathcal Z}_{\;g}$ is given by
\bq
 {\mathcal Z}_{\;g}^{(1)} 
 & = &
 \Pi\left(0\right).
\eq
In dimensional regularisation this contribution is zero, due to a cancellation between ultraviolet and infrared parts.
Keeping track of the divergent ultraviolet and infrared parts we may write this zero as 
\bq
 {\mathcal Z}_{\;g}^{(1)}
 & = &
 \frac{\alpha_s}{4 \pi}
 \left( 2 C_A - \beta_0 \right) 
            \left( \frac{1}{\eps_{\mathrm{IR}}} - \frac{1}{\eps_{\mathrm{UV}}} \right).
\eq
Let us denote by $\Pi^{\mu\nu}_{\mathrm{UV}}(p,\mu^2,\mu_{\mathrm{UV}}^2)$ an ultraviolet approximation term to the one-loop self-energy.
$\Pi^{\mu\nu}_{\mathrm{UV}}$ has the integral representation
\bq
 - i \Pi^{\mu\nu}_{\mathrm{UV}}\left(p,\mu^2,\mu_{\mathrm{UV}}^2\right)
 & = &
 - i g^2 S_\eps^{-1} \mu^{4-D} \int \frac{d^Dk}{(2\pi)^Di} \frac{P^{\mu\nu}_{\mathrm{UV}}\left(\bar{k},p,\mu_{\mathrm{UV}}^2\right)}{\left(\bar{k}^2- \mu_{\mathrm{UV}}^2\right)^4},
\eq
where $P^{\mu\nu}_{\mathrm{UV}}$ is a polynomial in $\bar{k}$ and $p$.
The explicit expression can be found in ref.~\cite{Becker:2010ng}.
Here, we will only need the fact that with the choice $\mu_{\mathrm{UV}}=\mu$ the ultraviolet subtraction
term integrates to
\bq
 -i \Pi^{\mu\nu}_{\mathrm{UV}}\left(p,\mu^2,\mu^2\right)
 & = &
 - i 
 \left( -p^2 g^{\mu\nu} + p^\mu p^\nu \right)
 {\mathcal Z}_{\;g,\mathrm{UV}}^{(1)}
 +
 {\mathcal O}\left(\eps\right),
\eq
with
\bq
 {\mathcal Z}_{\;g,\mathrm{UV}}^{(1)}
 & = &
 \frac{\alpha_s}{4 \pi}
 \left( 2 C_A - \beta_0 \right) 
 \left(- \frac{1}{\eps_{\mathrm{UV}}} \right).
\eq
Let us now define the quantity $\hat{\Pi}_{\mu\nu}(p^0,\vec{p})$
\bq
\label{def_hat_Pi_mu_nu}
 \hat{\Pi}_{\mu\nu}\left(p^0,\vec{p}\right)
 = 
 \left( -g_{\mu\rho} + \frac{p_\mu q_\rho + q_\mu p_\rho}{p \cdot q} - \frac{p^2}{\left(p q\right)^2} q_\mu q_\rho \right)
 \left( -i \Pi^{\rho\sigma} +i \Pi^{\rho\sigma}_{\mathrm{UV}} \right) 
 \left( \frac{-ig_{\sigma\nu}}{p^2} \right).
\eq
Here, $q$ is a light-like reference vector ($q^2=0$). On the other hand, we do not require
that $p$ is light-like.
The loop integration inherent in $(\Pi^{\rho\sigma} - \Pi^{\rho\sigma}_{\mathrm{UV}})$ is by definition of $\Pi^{\rho\sigma}_{\mathrm{UV}}$ ultraviolet finite.
Thus $\hat{\Pi}_{\mu\nu}$ is finite for $\eps<0$.
It is also finite for $\eps=0$, provided $p^2 \neq 0$.
For $\eps=0$ and $p^2=0$ we have an infrared singularity from the loop integration and in 
addition an explicit $1/p^2$-pole from the definition in eq.~(\ref{def_hat_Pi_mu_nu}).
The former infrared singularity we would like to combine with corresponding singularities
from other terms in $d\sigma^{\mathrm{NLO}}_{{\bf I}+{\bf L},\mathrm{IR}}$, 
the latter pole requires special treatment.

If we sandwich the analytic expression for $\hat{\Pi}_{\mu\nu}(p^0,\vec{p})$ 
between two amplitudes, where the polarisation vector of the gluon
has been amputated, we find for $\eps<0$ and in the limit $p^2\rightarrow 0$
\bq
 \mbox{Re}\; \left(
                         \left.{\mathcal A}^{\mu \; (0)}\right.^\ast 
                         \hat{\Pi}_{\mu\nu}\left(p^0,\vec{p}\right)
                         {\mathcal A}^{\nu \; (0)} 
                  \right)
 & = &
 \left( {\mathcal Z}_{\;g}^{(1)}  - {\mathcal Z}_{\;g,\mathrm{UV}}^{(1)} \right) \left| {\mathcal A}^{(0)} \right|^2
 + {\mathcal O}\left(\eps\right).
\eq
Thus, this expression contains exactly the terms from the field renormalisation constants, which
lead to the $1/\eps_{\mathrm{IR}}$ divergences or finite terms.
This is the contribution which we would like to include in $d\sigma^{\mathrm{V}}_{\mathrm{CT},\mathrm{IR}}$.
Note that the $1/p^2$-pole cancels after the (analytic) loop integration.
However, we would like to have an expression, where we can take the limit $p^2 \rightarrow 0$
in the integrand before the loop integration.
The expression on the right-hand side of eq.~(\ref{def_hat_Pi_mu_nu}) is not suited for a numerical evaluation, due to
the $1/p^2$-singularity from the propagator.
In order to arrive at an expression suitable for numerical evaluation we will use 
a dispersion relation in the variable $p^0$ for $\hat{\Pi}_{\mu\nu}$ \cite{Soper:1998ye,Soper:beowulf}.
Two properties of $\hat{\Pi}_{\mu\nu}$ are relevant:
First, for $|p^0|<|\vec{p}|$ the function $\hat{\Pi}_{\mu\nu}(p^0,\vec{p})$ is analytic in $p^0$.
Secondly, for large $|p^0|$ the quantity $\hat{\Pi}_{\mu\nu}$ behaves like a constant up to logarithmic corrections.
Therefore we will use a dispersion relation with a subtraction.
The starting point is Cauchy's theorem:
\bq
 \frac{\hat{\Pi}_{\mu\nu}\left(p^0,\vec{p}\right)}{p^2-\mu_{\mathrm{DI}}^2}
 & = &
 \frac{1}{2\pi i}
 \oint d\tilde{p}^0
 \frac{\hat{\Pi}_{\mu\nu}\left(\tilde{p}^0,\vec{p}\right)}{\left(\tilde{p}^0 - p^0\right)\left(\tilde{p}^2-\mu_{\mathrm{DI}}^2\right)},
\eq
with $\tilde{p}=(\tilde{p}^0,\vec{p})$ and
where the contour is a small counter-clockwise circle around $p^0$.
The factor $1/(p^2-\mu_{\mathrm{DI}}^2)$ improves the large $|p^0|$-behaviour.
$\mu_{\mathrm{DI}}^2$ is an arbitrary parameter, which may be complex.
Ignoring the $1/(p\cdot q)$-terms, which will vanish when contracted into the amplitude, we may deform
\begin{figure}
\begin{center}
\includegraphics[scale=0.8]{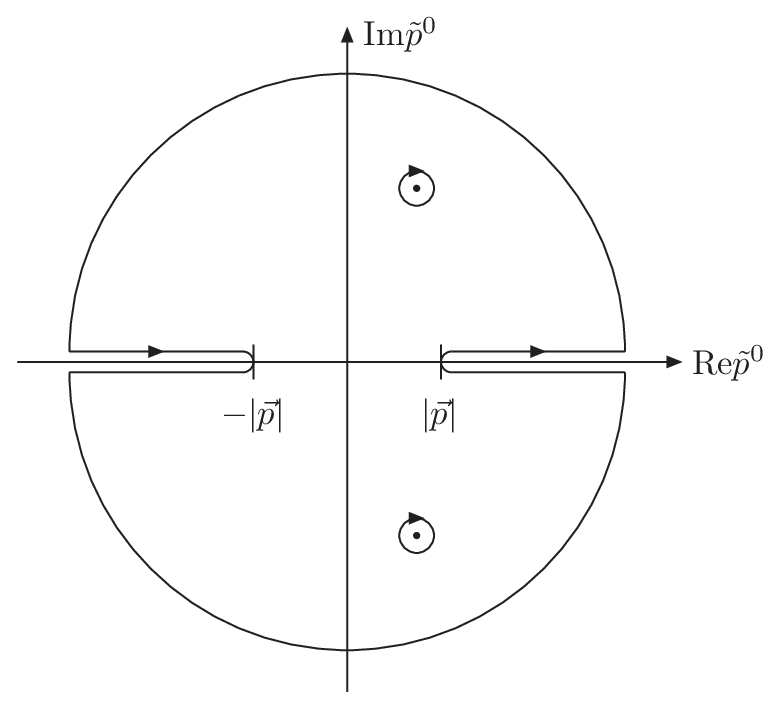}
\caption{\label{figure_dispersion_relation}
The integration contour for the dispersion relation.
The two small circles enclose the poles of $1/(\tilde{p}^2-\mu_{\mathrm{DI}}^2)$.}
\end{center}
\end{figure}
the contour as in fig.~\ref{figure_dispersion_relation}
and obtain
\bq
\lefteqn{
 \hat{\Pi}_{\mu\nu}\left(p^0,\vec{p}\right)
 =
 \frac{p^2-\mu_{\mathrm{DI}}^2}{2\pi i}
 \int\limits_{|\vec{p}|}^\infty  d\tilde{p}^0
 \left[
 \frac{\mathrm{Disc} \; \hat{\Pi}_{\mu\nu}\left(\tilde{p}^0,\vec{p}\right)}{\left( \tilde{p}^0 - p^0 \right) \left(\tilde{p}^2-\mu_{\mathrm{DI}}^2\right)}
 +
 \frac{\mathrm{Disc} \; \hat{\Pi}_{\mu\nu}\left(-\tilde{p}^0,\vec{p}\right)}{\left( -\tilde{p}^0 - p^0 \right) \left(\tilde{p}^2-\mu_{\mathrm{DI}}^2\right)}
 \right]
} \\
 & &
 - 
 \left. \frac{\left( p^2-\mu_{\mathrm{DI}}^2 \right) \hat{\Pi}_{\mu\nu}\left(\tilde{p}^0,\vec{p}\right)}{2 \tilde{p}^0 \left( \tilde{p}^0 - p^0 \right)} \right|_{\tilde{p}^0=\sqrt{|\vec{p}|^2+\mu_{\mathrm{DI}}^2}}
 -
 \left. \frac{\left( p^2-\mu_{\mathrm{DI}}^2 \right) \hat{\Pi}_{\mu\nu}\left(\tilde{p}^0,\vec{p}\right)}{2 \tilde{p}^0 \left( \tilde{p}^0 - p^0 \right)} \right|_{\tilde{p}^0=-\sqrt{|\vec{p}|^2+\mu_{\mathrm{DI}}^2}}.
 \nonumber
\eq
The last two terms subtract the residues at $\tilde{p}^0 = \pm \sqrt{|\vec{p}|^2+\mu_{\mathrm{DI}}^2}$.
The factor $1/(p^2-\mu_{\mathrm{DI}}^2)$ ensures that the half-circles at infinity give a vanishing contribution.
Let us now consider the discontinuity $\mathrm{Disc} \; \hat{\Pi}_{\mu\nu}$. We first note, that the ultraviolet approximation term
$\Pi^{\rho\sigma}_{\mathrm{UV}}$ does not contribute to $\mathrm{Disc} \; \hat{\Pi}_{\mu\nu}$.
The ultraviolet approximation term $\Pi^{\rho\sigma}_{\mathrm{UV}}$ contains only tadpole integrals, which 
are independent of $p^2$.
Let us further denote by $P^{\mu\nu}(k,p,\mu^2)$ the numerator of the integrand of the gluon self-energy,
i.e.
\bq
\lefteqn{
 - i \Pi^{\mu\nu}\left(p,\mu^2\right) =  
 \int \frac{d^Dk}{(2\pi)^Di} \frac{P^{\mu\nu}\left(k,p,\mu^2\right)}{k_1^2 k_2^2},
 } & & \\
\lefteqn{
 P^{\mu\nu}\left(k,p,\mu^2\right)
 =
 - i
 g^2 S_\eps^{-1} \mu^{4-D} 
 \left\{ 2 C_A \left[ -p^2 g^{\mu\nu} + p^\mu p^\nu - 2 \left( 1 - \eps \right) k^\mu k^\nu 
 \right. \right. } & & \nonumber \\
 & & \left. \left.
                      + \frac{1}{2} \left( 1 - \eps \right) g^{\mu\nu} \left( k_1^2 + k_2^2 \right) \right]
       + 2 T_R N_f \left[ p^2 g^{\mu\nu} - p^\mu p^\nu + 4 k^\mu k^\nu 
                          - g^{\mu\nu} \left( k_1^2 + k_2^2 \right) \right] \right\},
 \nonumber
\eq
and define $\hat{N}_{\mu\nu}(p,k,q)$ in analogy with eq.~(\ref{def_hat_Pi_mu_nu}):
\bq
 \hat{N}_{\mu\nu}\left(p,k,q\right)
 = 
 \left( -g_{\mu\rho} + \frac{p_\mu q_\rho + q_\mu p_\rho}{p \cdot q} - \frac{p^2}{\left(p q\right)^2} q_\mu q_\rho \right)
 P^{\rho\sigma}\left(k,p,\mu^2\right) 
 \left( \frac{-ig_{\sigma\nu}}{p^2} \right).
\eq
Working out the discontinuity gives us then an expression which we can evaluate at $p^2=0$:
\bq
\label{eq_Pi_hat_final}
\lefteqn{
 \left. \hat{\Pi}_{\mu\nu}\left(p^0,\vec{p}\right) \right|_{p^2=0}
 = } & &
 \\
 & &
 \frac{1}{\left(2\pi\right)^{D-1}} 
 \int d^D k_1
 \int d^D k_2
 \delta_+\left(k_1^2\right)
 \delta_-\left(k_2^2\right)
 \delta^{D-1}\left(\vec{p}-\vec{k}_1+\vec{k}_2\right)
 \frac{\hat{N}_{\mu\nu}\left(\tilde{p},k,q\right)}{\left(\tilde{p}^0-p^0\right) \left(1-\frac{\tilde{p}^2}{\mu_{\mathrm{DI}}^2}\right)} 
 \nonumber \\
 & &
 +
 \frac{1}{\left(2\pi\right)^{D-1}} 
 \int d^D k_1
 \int d^D k_2
 \delta_-\left(k_1^2\right)
 \delta_+\left(k_2^2\right)
 \delta^{D-1}\left(\vec{p}-\vec{k}_1+\vec{k}_2\right)
 \frac{\hat{N}_{\mu\nu}\left(\tilde{p},k,q\right)}{\left(\tilde{p}^0-p^0\right) \left(1-\frac{\tilde{p}^2}{\mu_{\mathrm{DI}}^2}\right)} 
 \nonumber \\
 & &
 +
 \left. \frac{\mu_{\mathrm{DI}}^2 \hat{\Pi}_{\mu\nu}\left(\tilde{p}^0,\vec{p}\right)}{2 \tilde{p}^0 \left( \tilde{p}^0 - p^0 \right)} \right|_{\tilde{p}^0=\sqrt{|\vec{p}|^2+\mu_{\mathrm{DI}}^2}}
 +
 \left. \frac{\mu_{\mathrm{DI}}^2 \hat{\Pi}_{\mu\nu}\left(\tilde{p}^0,\vec{p}\right)}{2 \tilde{p}^0 \left( \tilde{p}^0 - p^0 \right)} \right|_{\tilde{p}^0=-\sqrt{|\vec{p}|^2+\mu_{\mathrm{DI}}^2}},
 \nonumber 
\eq
with $\tilde{p}=(k_1^0-k_2^0, \vec{p})$ 
and
$k = 1/2 \; (k_1+k_2)$.
Note that the ultraviolet approximation term $\Pi^{\rho\sigma}_{\mathrm{UV}}$
enters in $\hat{\Pi}_{\mu\nu}$, but not $\hat{N}_{\mu\nu}$.
Further note that in the last two terms in eq.~(\ref{eq_Pi_hat_final}) the UV-subtracted self-energy
is evaluated at $\tilde{p}^2 = \mu_{\mathrm{DI}}^2$. These two terms give a finite contribution.
The infrared singularity is contained in the first two terms of eq.~(\ref{eq_Pi_hat_final}),
for $p^0>0$ in the first term, for $p^0<0$ in the second term.

Finally, using the loop-tree duality for the loop integrals in the last two terms, we may re-write
all terms as an integration over $d^{D-1}k_2$ (or alternatively $d^{D-1}k$):
\bq
 \left. \hat{\Pi}_{\mu\nu}\left(p^0,\vec{p}\right) \right|_{p^2=0}
 & = &
 g^2 S_\eps^{-1} \mu^{2\eps}
 \int \frac{d^{D-1}k_2}{\left(2\pi\right)^{D-1}}
 \;
 \left[ X_g\left(p,k_2\right) \right]_{\mu\nu}.
\eq
This defines $[ X_g(p,k_2) ]_{\mu\nu}$. The explicit expression is rather long and not reproduced here.
However, it can be extracted in a straightforward way from eq.~(\ref{eq_Pi_hat_final}).

\subsection{The massless quark self-energy}

The self-energy for a massless quark is given by
\bq
 - i \Sigma^{\alpha\beta} 
 & = & 
 - i g^2 C_F S_\eps^{-1} \mu^{4-D} \int \frac{d^Dk}{(2\pi)^Di} \frac{\left[ -2 \left(1-\eps\right) k\!\!\!/_1^{\alpha\beta} \right]}{k_1^2 k_2^2}.
\eq
We may write
\bq
 - i \Sigma^{\alpha\beta} 
 & = & 
 - i p\!\!\!/^{\alpha\beta} \Sigma'\left(p^2\right).
\eq
An analytic calculation of $\Sigma'(p^2)$ gives
\bq
 \Sigma'\left(p^2\right)
 & = &
 - \frac{\alpha_s}{4\pi} C_F \left(1-\eps\right) B_0\left(p^2,0,0\right).
\eq
The one-loop contribution to the field renormalisation constant ${\mathcal Z}_{\;q}$ for
massless quarks is given by
\bq
 {\mathcal Z}_{\;q}^{(1)} 
 & = &
 \Sigma'\left(0\right).
\eq
Again, this contribution is zero in dimensional regularisation due to a cancellation between
ultraviolet and infrared parts.
Keeping track only of divergent ultraviolet and infrared parts one finds
\bq
 {\mathcal Z}_q^{(1)} 
 & = & \frac{\alpha_s}{4 \pi} C_F \left( \frac{1}{\eps_{\mathrm{IR}}} - \frac{1}{\eps_{\mathrm{UV}}}
                                   \right).
\eq
Let us denote by $\Sigma^{\alpha\beta}_{\mathrm{UV}}(p,\mu^2,\mu_{\mathrm{UV}}^2)$ an ultraviolet approximation term to the one-loop self-energy.
$\Sigma^{\alpha\beta}_{\mathrm{UV}}$ has the integral representation
\bq
 - i \Sigma^{\alpha\beta}_{\mathrm{UV}}\left(p,\mu^2,\mu_{\mathrm{UV}}^2\right)
 & = &
 - i g^2 S_\eps^{-1} \mu^{4-D} \int \frac{d^Dk}{(2\pi)^Di} \frac{P^{\alpha\beta}_{\mathrm{UV}}\left(\bar{k},p,\mu_{\mathrm{UV}}^2\right)}{\left(\bar{k}^2- \mu_{\mathrm{UV}}^2\right)^3},
\eq
where $P^{\alpha\beta}_{\mathrm{UV}}$ is a polynomial in $\bar{k}$ and $p$.
The explicit expression can be found in ref.~\cite{Becker:2010ng}.
With the choice $\mu_{\mathrm{UV}}=\mu$ the ultraviolet subtraction
term integrates to
\bq
 -i \Sigma^{\alpha\beta}_{\mathrm{UV}}\left(p,\mu^2,\mu^2\right)
 & = &
 - i 
 p\!\!\!/^{\alpha\beta}
 {\mathcal Z}_{\;q,\mathrm{UV}}^{(1)}
 +
 {\mathcal O}\left(\eps\right),
\eq
with
\bq
 {\mathcal Z}_{\;q,\mathrm{UV}}^{(1)}
 & = &
 \frac{\alpha_s}{4 \pi}
 C_F
 \left(- \frac{1}{\eps_{\mathrm{UV}}} \right).
\eq
In analogy with the gluon self-energy let us consider the quantity
\bq
\label{def_Sigma_hat}
 \hat{\Sigma}_{\alpha\beta}\left(p^0,\vec{p}\right)
 & = &
 p\!\!\!/_{\alpha\gamma} \left( -i \Sigma^{\gamma\delta} + i \Sigma^{\gamma\delta}_{\mathrm{UV}} \right)
 \frac{i p\!\!\!/_{\delta\beta}}{p^2}.
\eq
For $p^2=0$ we have
\bq
 \mbox{Re}\; \left(
                         \left.{\mathcal A}^{\alpha \; (0)}\right.^\ast 
                         \hat{\Sigma}_{\alpha\beta}\left(p^0,\vec{p}\right)
                         {\mathcal A}^{\beta \; (0)} 
                  \right)
 & = &
 \left( {\mathcal Z}_q^{(1)} - {\mathcal Z}_{\;q,\mathrm{UV}}^{(1)} \right) \left| {\mathcal A}^{(0)} \right|^2
 + \mathcal{O}\left(\eps\right).
\eq
For large $|p^0|$ the quantity $\hat{\Sigma}_{\alpha\beta}$ grows (up to logarithminc corrections)
linearly with $|p^0|$. As in the gluon case we will therefore use a dispersion relation with a subtraction. 
Let us further denote by $P^{\alpha\beta}(k,p,\mu^2)$ the numerator of the integrand of the quark self-energy,
i.e.
\bq
 - i \Sigma^{\alpha\beta} 
 & = & 
 \int \frac{d^Dk}{(2\pi)^Di} \frac{P^{\alpha\beta}\left(k,p,\mu^2\right)}{k_1^2 k_2^2},
 \nonumber \\
 P^{\alpha\beta}\left(k,p,\mu^2\right)
 & = &
 - i g^2 C_F S_\eps^{-1} \mu^{4-D} \left[ -2 \left(1-\eps\right) k\!\!\!/_1^{\alpha\beta} \right].
\eq
and define $\hat{N}_{\alpha\beta}(p,k)$ in analogy with eq.~(\ref{def_Sigma_hat}):
\bq
 \hat{N}_{\alpha\beta}\left(p,k\right)
 & = &
 p\!\!\!/_{\alpha\gamma} P^{\gamma\delta}\left(k,p,\mu^2\right) 
 \frac{i p\!\!\!/_{\delta\beta}}{p^2}.
\eq
With the help of a dispersion relation we may re-write $\hat{\Sigma}_{\alpha\beta}$ as
an expression which we may evaluate at $p^2=0$:
\bq
\lefteqn{
 \left. \hat{\Sigma}_{\alpha\beta}\left(p^0,\vec{p}\right) \right|_{p^2=0}
 = } & &
 \\
 & &
 \frac{1}{\left(2\pi\right)^{D-1}} 
 \int d^D k_1
 \int d^D k_2
 \delta_+\left(k_1^2\right)
 \delta_-\left(k_2^2\right)
 \delta^{D-1}\left(\vec{p}-\vec{k}_1+\vec{k}_2\right)
 \frac{\hat{N}_{\alpha\beta}\left(\tilde{p},k\right)}{\left(\tilde{p}^0-p^0\right) \left(1-\frac{\tilde{p}^2}{\mu_{\mathrm{DI}}^2}\right)} 
 \nonumber \\
 & &
 +
 \frac{1}{\left(2\pi\right)^{D-1}} 
 \int d^D k_1
 \int d^D k_2
 \delta_-\left(k_1^2\right)
 \delta_+\left(k_2^2\right)
 \delta^{D-1}\left(\vec{p}-\vec{k}_1+\vec{k}_2\right)
 \frac{\hat{N}_{\alpha\beta}\left(\tilde{p},k\right)}{\left(\tilde{p}^0-p^0\right) \left(1-\frac{\tilde{p}^2}{\mu_{\mathrm{DI}}^2}\right)} 
 \nonumber \\
 & &
 +
 \left. \frac{\mu_{\mathrm{DI}}^2 \hat{\Sigma}_{\alpha\beta}\left(\tilde{p}^0,\vec{p}\right)}{2 \tilde{p}^0 \left( \tilde{p}^0 - p^0 \right)} \right|_{\tilde{p}^0=\sqrt{|\vec{p}|^2+\mu_{\mathrm{DI}}^2}}
 +
 \left. \frac{\mu_{\mathrm{DI}}^2 \hat{\Sigma}_{\alpha\beta}\left(\tilde{p}^0,\vec{p}\right)}{2 \tilde{p}^0 \left( \tilde{p}^0 - p^0 \right)} \right|_{\tilde{p}^0=-\sqrt{|\vec{p}|^2+\mu_{\mathrm{DI}}^2}},
 \nonumber 
\eq
with $\tilde{p}=(k_1^0-k_2^0, \vec{p})$ 
and
$k = 1/2 \; (k_1+k_2)$.

Finally, using the loop-tree duality for the loop integrals in the last two terms, we may re-write
all terms as an integration over $d^{D-1}k_2$:
\bq
 \left. \hat{\Sigma}_{\alpha\beta}\left(p^0,\vec{p}\right) \right|_{p^2=0}
 & = &
 g^2 S_\eps^{-1} \mu^{2\eps}
 \int \frac{d^{D-1}k_2}{\left(2\pi\right)^{D-1}}
 \;
 \left[ X_q\left(p,k_2\right) \right]_{\alpha\beta}.
\eq
This defines $[ X_q(p,k_2) ]_{\alpha\beta}$.

\subsection{The massive quark self-energy}

The self-energy for a massive quark is given by
\bq
 - i \Sigma^{\alpha\beta} 
 & = & 
 - i g^2 C_F S_\eps^{-1} \mu^{4-D} \int \frac{d^Dk}{(2\pi)^Di} \frac{\left[ -2 \left(1-\eps \right) k\!\!\!/_1^{\alpha\beta} + 4 \left(1-\frac{1}{2} \eps \right) m \delta^{\alpha\beta} \right]}{\left(k_1^2-m^2 \right) k_2^2}.
\eq
One expands the self-energy around $p^2=m^2$:
\bq
 - i \Sigma^{\alpha\beta} 
 & = & 
 - i 
 \left[ \delta^{\alpha\beta} \Sigma\left(m\right)
  + \left( p\!\!\!/^{\alpha\beta} - m \delta^{\alpha\beta} \right) \Sigma'\left(m\right)
 + ...
 \right].
\eq
Then
\bq 
 Z_{m,\mathrm{on-shell}}^{(1)}
 \;\; = \;\;
 - \frac{1}{m} \Sigma\left(m\right),
 & &
 {\mathcal Z}_{\;Q}^{(1)}
 \;\; = \;\;
 \Sigma'\left(m\right),
\eq
and one finds
\bq
 Z_{m,\mathrm{on-shell}}^{(1)}
 & = &
 \frac{\alpha_s}{4 \pi} C_F 
 \left( - \frac{3}{\eps_{\mathrm{UV}}} - 4 + 3 \ln\frac{m^2}{\mu^2} \right),
 \nonumber \\
 {\mathcal Z}_{\;Q}^{(1)}
 & = &
 \frac{\alpha_s}{4 \pi} C_F 
 \left( - \frac{1}{\eps_{\mathrm{UV}}} - \frac{2}{\eps_{\mathrm{IR}}} - 4 + 3 \ln\frac{m^2}{\mu^2} \right).
\eq
Let us denote by $\Sigma^{\alpha\beta}_{\mathrm{UV}}(p,\mu^2,\mu_{\mathrm{UV}}^2)$ an ultraviolet approximation term to the one-loop self-energy.
$\Sigma^{\alpha\beta}_{\mathrm{UV}}$ has the integral representation
\bq
 - i \Sigma^{\alpha\beta}_{\mathrm{UV}}\left(p,\mu^2,\mu_{\mathrm{UV}}^2\right)
 & = &
 - i g^2 S_\eps^{-1} \mu^{4-D} \int \frac{d^Dk}{(2\pi)^Di} \frac{P^{\alpha\beta}_{\mathrm{UV}}\left(\bar{k},p,\mu_{\mathrm{UV}}^2\right)}{\left(\bar{k}^2- \mu_{\mathrm{UV}}^2\right)^3},
\eq
where $P^{\alpha\beta}_{\mathrm{UV}}$ is a polynomial in $\bar{k}$ and $p$.
With the choice $\mu_{\mathrm{UV}}=\mu$ and by adding a suitable chosen finite term we can ensure
that $\Sigma^{\alpha\beta}_{\mathrm{UV}}$
takes into account the ultraviolet divergence and the finite part due to the on-shell 
mass renormalisation.
The explicit expression is
\bq
 - i \Sigma^{\alpha\beta}_{\mathrm{UV}}\left(p,\mu^2,\mu^2\right)
 & = &
 i g^2 C_F \Sigma^{\mathrm{CT},\mathrm{UV},\alpha\beta}\left(\mu^2\right),
\eq
where $\Sigma^{\mathrm{CT},\mathrm{UV}}$ is given in eq.~(\ref{def_Sigma_CT_UV}).
The ultraviolet subtraction
term integrates to
\bq
 -i \Sigma^{\alpha\beta}_{\mathrm{UV}}\left(p,\mu^2,\mu^2\right)
 & = &
 - i \left[ {\mathcal Z}_{\;Q,\mathrm{UV}}^{(1)} p\!\!\!/ - \left( {\mathcal Z}_{\;Q,\mathrm{UV}}^{(1)} + Z_{m,\mathrm{on-shell}}^{(1)} \right) m \right],
\eq
with
\bq
 {\mathcal Z}_{\;Q,\mathrm{UV}}^{(1)}
 & = &
 \frac{\alpha_s}{4 \pi}
 C_F
 \left(- \frac{1}{\eps_{\mathrm{UV}}} \right),
 \nonumber \\
 Z_{m,\mathrm{on-shell}}^{(1)}
 & = &
 \frac{\alpha_s}{4 \pi} C_F 
 \left( - \frac{3}{\eps_{\mathrm{UV}}} - 4 + 3 \ln\frac{m^2}{\mu^2} \right).
\eq
Let us consider the quantity
\bq
\label{def_Sigma_hat_massive}
 \hat{\Sigma}_{\alpha\beta}\left(p^0,\vec{p}\right)
 & = &
 \left( p\!\!\!/_{\alpha\gamma} + m \delta_{\alpha\gamma} \right) 
 \left( -i \Sigma^{\gamma\delta} + i \Sigma^{\gamma\delta}_{\mathrm{UV}} \right)
 \frac{i \left( p\!\!\!/_{\delta\beta} + m \delta_{\delta\beta}\right)}{p^2-m^2}.
\eq
For $p^2=m^2$ we have
\bq
 \mbox{Re}\; \left(
                         \left.{\mathcal A}^{\alpha \; (0)}\right.^\ast 
                         \hat{\Sigma}_{\alpha\beta}\left(p^0,\vec{p}\right)
                         {\mathcal A}^{\beta \; (0)} 
                  \right)
 & = &
 \left( {\mathcal Z}_Q^{(1)} - {\mathcal Z}_{\;Q,\mathrm{UV}}^{(1)} \right) \left| {\mathcal A}^{(0)} \right|^2
 + \mathcal{O}\left(\eps\right).
\eq
Let us further denote by $P^{\alpha\beta}(k,p,\mu^2)$ the numerator of the integrand of the quark self-energy,
i.e.
\bq
 - i \Sigma^{\alpha\beta} 
 & = & 
 \int \frac{d^Dk}{(2\pi)^Di} \frac{P^{\alpha\beta}\left(k,p,\mu^2\right)}{k_1^2 k_2^2},
 \nonumber \\
 P^{\alpha\beta}\left(k,p,\mu^2\right)
 & = &
 - i g^2 C_F S_\eps^{-1} \mu^{4-D} \left[ -2 \left(1-\eps \right) k\!\!\!/_1^{\alpha\beta} + 4 \left(1-\frac{1}{2} \eps \right) m \delta^{\alpha\beta} \right].
\eq
and define $\hat{N}_{\alpha\beta}(p,k)$ in analogy with eq.~(\ref{def_Sigma_hat_massive}):
\bq
 \hat{N}_{\alpha\beta}\left(p,k\right)
 & = &
 \left( p\!\!\!/_{\alpha\gamma} + m \delta_{\alpha\gamma} \right) 
 P^{\gamma\delta}\left(k,p,\mu^2\right)
 \frac{i \left( p\!\!\!/_{\delta\beta} + m \delta_{\delta\beta}\right)}{p^2-m^2}.
\eq
With the help of a dispersion relation we may re-write $\hat{\Sigma}_{\alpha\beta}$ as
an expression which we may evaluate at $p^2=m^2$:
\bq
\lefteqn{
 \left. \hat{\Sigma}_{\alpha\beta}\left(p^0,\vec{p}\right) \right|_{p^2=m^2}
 = } & &
 \\
 & &
 \frac{1}{\left(2\pi\right)^{D-1}} 
 \int d^D k_1
 \int d^D k_2
 \delta_+\left(k_1^2-m^2\right)
 \delta_-\left(k_2^2\right)
 \delta^{D-1}\left(\vec{p}-\vec{k}_1+\vec{k}_2\right)
 \frac{\hat{N}_{\alpha\beta}\left(\tilde{p},k\right)}{\left(\tilde{p}^0-p^0\right) \left(1-\frac{\tilde{p}^2}{\mu_{\mathrm{DI}}^2}\right)} 
 \nonumber \\
 & &
 +
 \frac{1}{\left(2\pi\right)^{D-1}} 
 \int d^D k_1
 \int d^D k_2
 \delta_-\left(k_1^2-m^2\right)
 \delta_+\left(k_2^2\right)
 \delta^{D-1}\left(\vec{p}-\vec{k}_1+\vec{k}_2\right)
 \frac{\hat{N}_{\alpha\beta}\left(\tilde{p},k\right)}{\left(\tilde{p}^0-p^0\right) \left(1-\frac{\tilde{p}^2}{\mu_{\mathrm{DI}}^2}\right)} 
 \nonumber \\
 & &
 +
 \left. \frac{\mu_{\mathrm{DI}}^2 \hat{\Sigma}_{\alpha\beta}\left(\tilde{p}^0,\vec{p}\right)}{2 \tilde{p}^0 \left( \tilde{p}^0 - p^0 \right)} \right|_{\tilde{p}^0=\sqrt{|\vec{p}|^2+\mu_{\mathrm{DI}}^2}}
 +
 \left. \frac{\mu_{\mathrm{DI}}^2 \hat{\Sigma}_{\alpha\beta}\left(\tilde{p}^0,\vec{p}\right)}{2 \tilde{p}^0 \left( \tilde{p}^0 - p^0 \right)} \right|_{\tilde{p}^0=-\sqrt{|\vec{p}|^2+\mu_{\mathrm{DI}}^2}},
 \nonumber 
\eq
with $\tilde{p}=(k_1^0-k_2^0, \vec{p})$ 
and
$k = 1/2 \; (k_1+k_2)$.

Finally, using the loop-tree duality for the loop integrals in the last two terms, we may re-write
all terms as an integration over $d^{D-1}k_2$:
\bq
 \left. \hat{\Sigma}_{\alpha\beta}\left(p^0,\vec{p}\right) \right|_{p^2=m^2}
 & = &
 g^2 S_\eps^{-1} \mu^{2\eps}
 \int \frac{d^{D-1}k_2}{\left(2\pi\right)^{D-1}}
 \;
 \left[ X_Q\left(p,k_2\right) \right]_{\alpha\beta}.
\eq
This defines $[ X_Q(p,k_2) ]_{\alpha\beta}$. 

\section{The momenta mapping from the virtual to the real space}
\label{appendix:virtual_to_real}

In this appendix we determine the constants $\alpha$ and $\beta$ in the mapping of eq.~(\ref{map_virtual_to_real_massive}):
\bq
 p_k' & = & \alpha \left( p_k - \frac{Q\cdot p_k}{Q^2} Q \right) + \beta Q,
 \nonumber \\
 p_j' & = & - k_i,
 \nonumber \\
 p_i' & = & Q + k_i - p_k',
\eq
with $Q=p_i+p_k$.
We require that
\bq
 p_i'{}^2 \; = \; m_i'{}^2,
 & &
 p_k'{}^2 \; = \; m_k'{}^2 \; = \; m_k^2.
\eq
We may use the first equation to eliminate $\alpha$, doing so gives us a quadratic equation for $\beta$:
\bq
 a \beta^2 + b \beta + c & = & 0,
\eq
with
\bq
 a & = &
 4 Q^4 \left\{ 4 m_k^2 Q^4
              + \left[ 4 m_k^2 \left(2 k_i Q \right) + \left( 2 k_i p_k \right)^2 - \left( 2 p_k Q \right)^2 \right] Q^2
 \right. \nonumber \\
 & & \left.
              + 2 k_i Q \left[ m_k^2 \left( 2 k_i Q \right) - \left( 2 p_k Q \right)^2 - \left( 2 p_k Q \right) \left( 2 k_i p_k \right) \right]
       \right\},
 \nonumber \\
 b & = &
 2 Q^2 \left[ 2 Q^2 + 2 k_i Q \right) \left[ \left( 2 p_k Q \right)^2 - 4 m_k^2 Q^2 \right]
 \left[ Q^2 + 2 k_i Q - m_i'{}^2 + m_j'{}^2 + m_k^2 \right],
 \nonumber \\
 c & = &
 4 m_k^2 Q^4 \left[ \left( Q^2 + 2 k_i Q - m_i'{}^2 + m_j'{}^2 + m_k^2 \right)^2 - \left(2 k_i p_k\right)^2 \right]
 +
 4 m_k^2 Q^2 \left(2 k_i p_k \right) \left( 2 k_i Q \right) \left( 2 p_k Q \right)
 \nonumber \\
 & &
 -
 \left\{
 Q^6 
 + 2 \left[ 2 k_i Q - m_i'{}^2 + m_j'{}^2 + m_k^2 \right] Q^4
 - \left[ 2 k_i Q - m_i'{}^2 + m_j'{}^2 + m_k^2 \right]^2 Q^2
 + m_k^2 \left(2 k_i Q \right)^2
 \right\}
 \nonumber \\
 & &
 \times
 \left( 2 p_k Q \right)^2.
\eq
We thus have
\bq
 \beta & = &
 - \frac{b+\sqrt{b^2-4ac}}{2a},
\eq
where the sign of the root is fixed by matching on the massless limit.
The constant $\alpha$ is then given by
\bq
 \alpha
 & = &
 2 Q^2 \frac{\left[Q^2 + 2 k_i Q - m_i'{}^2 + m_j'{}^2 + m_k^2 - \left( 2 Q^2 + 2 k_i Q \right) \beta \right]}{2 Q^2 \left(2 k_i p_k \right) - \left(2 p_k Q \right)\left(2 k_i Q \right)}.
\eq

\end{appendix}

\bibliography{/home/stefanw/notes/biblio}
\bibliographystyle{/home/stefanw/latex-style/h-physrev5}

\end{document}